%% file: paper.tex
\documentclass[useAMS,usenatbib]{mnras}
\usepackage{amsmath}
\usepackage{amssymb}
\usepackage{graphicx}
\usepackage{float}
\usepackage{hyperref}
\usepackage{morefloats}
\usepackage{microtype}
\input{journals.tex}

\usepackage[dvipsnames]{xcolor}

\usepackage[normalem]{ulem}

\renewcommand{\vec}[1]{ {\bf #1} }
\usepackage{txfonts}

\newcommand{\ci}{\mathrm{i}}
\newcommand{\dd}{\mathrm{d}}

\newcommand*{\defeq}{\mathrel{\vcenter{\baselineskip0.5ex \lineskiplimit0pt
                     \hbox{\scriptsize.}\hbox{\scriptsize.}}}%
                     =}
\newcommand*{\eqdef}{= \mathrel{\vcenter{\baselineskip0.5ex \lineskiplimit0pt
                     \hbox{\scriptsize.}\hbox{\scriptsize.}}}%
                     }

\newcommand{\bvec}[1]{\ensuremath{\boldsymbol{#1}}}

\title[Intrinsic Alignments in IllustrisTNG]
{Intrinsic Alignments in IllustrisTNG and their implications for weak lensing: Tidal shearing and tidal torquing mechanisms put to the test}
\author[J. Zjupa, B.M. Sch{\"a}fer, and O. Hahn]{
Jolanta Zjupa$^{1}$\thanks{E-mail: jolanta.zjupa@oca.eu},
Bj{\"o}rn Malte Sch{\"a}fer$^{2}$,
and Oliver Hahn$^{1,3,4}$
\vspace*{0.2cm}\\
$^{1}$Laboratoire Lagrange, Universit\'e C\^ote d'Azur, Observatoire de la C\^ote d'Azur, CNRS, Blvd de l'Observatoire, CS 34229, 06304 Nice, France\\
$^{2}$Zentrum f\"ur Astronomie der Universit\"at Heidelberg, Astronomisches Rechen-Institut, Philosophenweg 12, 69120 Heidelberg, Germany\\
$^{3}$Department of Astrophysics, University of Vienna, T\"urkenschanzstraße 17, 1180 Vienna, Austria\\
$^{4}$Department of Mathematics, University of Vienna, Oskar-Morgenstern-Platz 1, 1090 Vienna, Austria\\
}

\begin{document}

\date{Accepted, Received; in original form}
\pubyear{2020}

\pagerange{\pageref{firstpage}--\pageref{lastpage}} 

\maketitle

\label{firstpage}

\begin{abstract}
Accurate measurements of the cosmic shear signal require a separation of the true weak gravitational lensing signal from intrinsic shape correlations of galaxies.  These `intrinsic alignments' of galaxies originate from galaxy formation processes and are expected to be correlated with the gravitational field through tidal processes affecting the galaxies, such as tidal shearing for elliptical galaxies and tidal torquing for spiral galaxies. In this study, we use morphologically selected samples of elliptical and spiral galaxies from the {\sc IllustrisTNG} simulation at $z=0$ and $z=1$ to test the commonly employed linear (tidal shearing) and quadratic (tidal torquing) models for intrinsic alignments. We obtain local measurements of the linear and quadratic alignment parameters, including corrections for large-scale anisotropies of the cosmologically small simulation volume, and study their dependence on galaxy and environmental properties. We find a significant alignment signal for elliptical galaxies (linear model), that increases with mass and redshift. Spiral galaxies (quadratic model) on the other hand exhibit a significant signal only for the most massive objects at $z=1$. We show the quadratic model for spiral galaxies to break down at its fundamental assumptions, and simultaneously obtain a significant signal of spiral galaxies to align according to the linear model. We use the derived alignment parameters to compute intrinsic alignment spectra and estimate the expected contamination in the weak lensing signal obtained by {\sc Euclid}.
\end{abstract}

\begin{keywords}
cosmology: theory -- galaxies: formation -- methods: numerical.
\end{keywords}

\section{Introduction}\label{sect_intro}

Weak gravitational lensing is one of the primary probes used to constrain cosmological parameters. The {\sc Euclid} satellite, to be launched in 2022, will observe a weak gravitational lensing signal from $\sim 1.5\times 10^9$ background galaxies in the redshift range $0\lesssim z \lesssim 2$, peaking around $z\simeq 1$ \citep{Laureijs11} in order to put below $5\%$ level constrains on the dark energy equation of state (\href{https://www.euclid-ec.org}{euclid-ec.org}), while the Large Synoptic Survey Telescope ({\sc LSST}, \href{https://www.lsst.org}{lsst.org}) survey is expected to observe $\sim 2\times 10^{10}$ galaxies over a span of 10 years, starting 2023. Currently, The Dark Energy Survey ({\sc DES}, \href{https://www.darkenergysurvey.org}{darkenergysurvey.org}) is collecting data from $\sim 3 \times 10^8$ source galaxies in four tomographic bins, centred between $z \simeq 0.3$ and $z \simeq 1$ \citep{Hoyle18}, while \citet{Hildebrandt20} and \citet{Asgari20} employ the Kilo Degree Survey ({\sc KiDS}) and {\sc VIKING} survey to analyse shear data in the redshift range $0.1 \lesssim z \lesssim 1.2$ from almost $1.2 \times 10^7$ ({\sc KV450}) and $\sim 10^8$ galaxies ({\sc KiDS-1000}), respectively.

All this rich data of weak gravitational lensing induced correlations in observed galaxy ellipticities, referred to as cosmic shear, is contaminated by an inherent correlation of galaxy shapes originating from galaxy formation processes. The inherent shape correlations of neighbouring galaxies, called intrinsic alignments (IAs), introduce, along photometric redshift uncertainties (\citealt{Guglielmo20}; see also \citealt{Desprez20}), the largest systematic bias in the shear auto-correlation function measured by aforementioned surveys. In order to obtain unaffected measurements, it is therefore crucial to either have reliable and robust models for intrinsic alignments that are included in the analysis pipeline, or rely on `nulling' techniques that suppress the impact of IAs at the expense of reduced constraining power \citep{Joachimi08}.

Intrinsic alignments of galaxies can be caused by a range of mechanisms related to the galaxies' formation process or through gravitational interaction with the surrounding large-scale structure, depending on the galaxy type. First works by \citet{Heavens00} and \citet{CroftMetzler00} have used collisionless $N$-body simulations to measure the intrinsic alignment signal for galaxies of distinct morphological type based on properties of their dark matter haloes. Thereby, \citet{Heavens00} have focused on extracting the IA signal for spiral galaxies, assuming that the orientation of a disk galaxy is set by the angular momentum vector of its host halo. \citet{CroftMetzler00} on the other hand focused on deriving the intrinsic alignment of elliptical galaxies, whose shape they assumed was inherited from their dark matter halo. Both studies found a similar magnitude of the intrinsic alignment signal at $\sim 10 \%$ level of the shear signal obtained from {\sc SDSS}-like surveys at $z\simeq 1$.

These early studies have prompted the formulation of theoretical models for intrinsic alignments for the two dominant galaxy types. Building on tidal torque theory \citep{Hoyle49,Doroshkevich70,White84} that sets the angular momentum vector of haloes during their early collapse through tidal torques from the surrounding large-scale structure, spiral galaxies can be modelled as thin discs inhereting their host halo's specific angular momentum. \citet{LeePen00,LeePen01} used this ansatz to correlate the expectation value of the galaxies' angular momentum with the gravitational tidal field. \citet{Crittenden01} extended this formalism to a correlation directly between the observed galaxy ellipticity and the tidal field, known as the quadratic alignment model for spiral galaxies. For elliptical, velocity dispersion supported galaxies, angular momentum is not a meaningful concept when it comes to defining the galaxy shape, which is rather set by the response of the galaxies' dynamical state to a non-spherical gravitational potential \citep{Hirata04,Piras18}. The latter is again characterised by the local tidal shear field giving rise to a correlation between the observed ellipticity of an elliptical galaxy and the gravitational tidal field, encompassed in the so called linear alignment model. The model names reflect the fact that intrinsic alignments of elliptical and spiral galaxies can be traced back to a respectively linear and quadratic dependence of the observed galaxy ellipticity on the tidal field generated by the large-scale matter distribution. 

The linear alignment model for elliptical galaxies has been further extended by \cite{BridleKing07} to mildly non-linear scales through substitution of the linear by the non-linear matter power spectrum, giving this model its name. \cite{Blazek19} showed that applying the non-linear model, widely used to account for the IA contamination in observational data, to a mock {\sc LSST} survey (not discriminating by galaxy type) results in a significant bias in the estimated cosmological parameters, including the dark energy equation of state. Instead, those authors propose a perturbative approach that includes both the tidal shearing and tidal torquing alignment mechanism (see also \citealt{Blazek15}), that recently was also applied to simulation data from {\sc MassiveBlack-II} and {\sc IllustrisTNG} by \citet{Samuroff20}. Complementary, \citet{Tugendhat18} showed that not taking into account the tidal shearing (ellipticals) and tidal torquing (spiral galaxies) induced intrinsic alignments, can lead to a bias in the deduced cosmological parameters of several times the statistical error expected for {\sc Euclid}. These studies motivate us to employ the {\sc IllustrisTNG} galaxy formation simulation to directly verify the applicability of the linear and quadratic alignment model to the respective galaxy types.

In recent years, analytical studies have been complemented by a variety of direct measurements of shape (and position) correlations of galaxies taken from cosmological galaxy formation simulations. Galaxy formation simulations have become a primary tool for IA studies not only because they naturally follow the full non-linear evolution, but particularly due the implementation of successful subgrid models for galaxy formation physics which eliminate the need to rely on restricting theoretical assumptions inherent to $N$-body studies, such as the perfect alignment between a central galaxy and its host halo. On the other hand, observables derived from state-of-the-art galaxy formation simulation are (if those are not modelled explicitly) inherently free of distortion due to gravitational lensing, such that all measured galaxy alignments are of `intrinsic' nature. To our knowledge, the first hydrodynamical simulations to study the alignment of disc galaxies with the large-scale tidal field (which then translates into alignments of nearby spirals) by \cite{Navarro2004} and \cite{Hahn10} found some hints for alignments at low redshift, based on small samples of galaxies. Employing the {\sc HorizonAGN} simulation 
at $z=1.2$, \citet{Codis15} studied the alignment of galaxy shapes given by their spin vector (for all galaxy types) with the large-scale structure geometry, while \citet{Chisari15} focused on galaxy correlations at $z\simeq 0.5$, and found elliptical galaxies to be radially aligned towards massive, central ellipticals, while spiral galaxies show a tangential alignment w.r.t. centrals with reduced amplitude compared to ellipticals. The redshift evolution of these effects has further been studied in \citet{Chisari16}. \citet{Velliscig15ias} studied the alignment of galaxies independent of type taken from the {\sc EAGLE} \citep{Schaye15eagle} and cosmo-{\sc OWLS} \citep{LeBrun14} simulations at $z=0$, and similarly found satellite galaxies to be preferentially elongated towards the massive central, as well as the satellite distribution to be correlated with the shape of the central galaxy, increasingly so with higher satellite as well as central mass. \citet{Tenneti20} further focused on small scale alignments at low redshift between the shape of a central galaxy and the shape traced by its satellite system in {\sc MassiveBlack-II} \citep{Khandai15} and {\sc IllustrisTNG}, and found a correlation between shapes on the projected sky that is stronger for more massive central galaxies, as well as for quenched, red, spheroidals than for star-forming, blue, disc galaxies, consistent between the two simulations. \citet{Bate20} and \citet{Bhowmick20} study the alignment of galaxy shapes with the underlying density field for a morphological mix of galaxies as a function of redshift in {\sc HorizonAGN} and {\sc MassiveBlack-II}, respectively. 

Complementary to these more geometric approaches, \citet{Hilbert17} used the {\sc Illustris} simulation \citep{Vogelsberger14Nature} to extract the gravitational shear - intrinsic ellipticity ($GI$) and intrinsic ellipticity - intrinsic ellipticity ($II$) angular correlation functions of galaxies with different stellar mass, colour, and apparent magnitude at various redshifts between $z=0$ and $z=1$. \cite{Shi20} used the TNG300 simulation from {\sc IllustrisTNG} to derive 3D intrinsic alignment power spectra from ellipticities that, motivated by the linear and quadratic model, respectively, where derived from the (reduced) tensor of inertia for kinematically classified elliptical galaxies, and from the angular momentum vector for spiral galaxies. The obtained alignment power spectra in the redshift range $0.3<z<2$ serve to derive alignment parameters within the (non-)linear alignment model framework, designed for ellipticals, applied to all morphological types (as also done in recent theoretical work on TNG300 by \citet{Samuroff20}, and observational work by \citet{Yao20a}), and study their dependence on galaxy stellar mass, morphology, central-satellite distinction, and redshift. Note, however, that TNG300 was run at a lower spatial resolution than TNG100, such that some basic galaxy properties in TNG300, like e.g. the galaxy stellar mass and as a consequence derivatives from it, are not fully converged below $M_{\rm tot} \sim 10^{12} M_\odot$ (see figure A2 in \citealt{Pillepich18}), and cannot be straightforwardly used without applying recalibration techniques. 
 
In this work we employ the TNG100 hydrodynamical simulation from the {\sc IllustrisTNG} simulation suite to directly test the validity of the quadratic model for spiral galaxies and linear model for elliptical galaxies. After a brief overview of the {\sc IllustrisTNG} simulation, as well as of our galaxy sample selection and the derivation of the relevant properties in section~\ref{sect_simulation}, we give a summary of the two primary alignment models in section~\ref{sect_ia_lensing}. We verify the linear and quadratic model on samples of elliptical and spiral galaxies from {\sc IllustrisTNG} in sections~\ref{sect_e} and~\ref{sect_s}, respectively. We directly measure the IA strengths and study their scaling behaviour with mass, environment, and redshift. In section~\ref{sect_lens}, we use the derived parameters to compare predictions for weak lensing ellipticity spectra for {\sc Euclid} with idealised models. We summarise our results in section~\ref{sect_disc}.

\section{Simulation and post-processing}\label{sect_simulation}

\subsection{IllustrisTNG cosmological simulation}
For our analysis we employ the galaxy formation simulation TNG100 from the {\sc IllustrisTNG} simulation suite \citep{fTNGNelson18, fTNGPillepich18, fTNGMarinacci18, fTNGNaiman18, fTNGSpringel18}, which evolves a $75$~Mpc$/h$ wide periodic cosmological box initialised with the best-fit cosmology reported by \citet{Planck15} using the moving-mesh code {\sc AREPO} \citep{Springel2010Arepo} to $z = 0$. The TNG100 simulation cube contains $1820^3$ dark matter particles and $1820^3$ initial gas cells, corresponding to a mass resolution of $7.46 \times 10^6 M_\odot$ in dark matter and $\sim 1.39 \times 10^6 M_\odot$ in baryonic matter, respectively. {\sc IllustrisTNG} employs a set of subgrid models for galaxy formation physics, including primordial and metal-line cooling, star formation and stellar evolution, chemical enrichment by nine elements, stellar feedback, black hole formation and growth, as well as feedback by active galactic nuclei (AGN), and magnetic fields. The {\sc IllustrisTNG} galaxy formation physics reproduces well many late time properties, such as (on the predictive side) the sizes of star-forming and quenched galaxies \citep{Genel18} and the star-formation rate (SFR) stellar mass relation \citep{Donnari19} at $0 \lesssim z \lesssim 2 $, the galaxy colour bimodality at $z=0$ \citep{fTNGNelson18}, and optical morphologies \citep{RodriguezGomez19}. For a full description of the subgrid models and parameters adopted in the {\sc IllustrisTNG} simulation see \cite{Pillepich18}. 

Furthermore, the absence of grid locking effects in simulations produced with {\sc AREPO}, which lead to some spurious alignment of galaxy orientations with the cartesian axes of the simulation volume in eulerian AMR codes \citep[see e.g.][]{Hahn10,Chisari15,Codis15}, as well as excellent angular momentum conservation properties \citep{Pakmor16}, make {\sc IllustrisTNG} a particularly suitable simulation to study intrinisc alignments.

\subsection{Galaxy sample selection}
\label{sec:galaxy_selection}
We employ the TNG100 galaxy catalogue to extract a sample of elliptical galaxies and a sample of spiral galaxies based on properties of their stellar population and their kinematics. 

\paragraph*{Selection of well-resolved galaxies.} \ For our study, we only consider galaxies from TNG100 that are (1) resolved with at least 1000 stellar particles, and (2) have a total mass, i.e. the combined dark matter and baryonic mass of the entire (sub)halo, of at least $10^{10}M_\odot$. Prior studies found that at least 300 particles are needed for reliable measurements of galaxy shapes \citep{Velliscig15alg,Chisari15} and halo angular momentum \citep{Bett07}, but not baryonic galaxy spin. Thus we adopt a more conservative choice of at least 1000 stellar particles for well-resolved galaxies, as also done in \citet{Tenneti20}.
Furthermore, with the choice of 1000 stellar particles we are selecting objects above the {\sc Euclid} sensitivity limit of 24.5 mag at $z\simeq 1$ \citep{Laureijs11}, the peak redshift of the {\sc Euclid} galaxy distribution and thus of the intrinsic alignment contamination. In order to match the {\sc Euclid} sensitivity at $z=0.3$, \citet{Hilbert17} adopted a limit of 100 stellar particles. At $z\simeq 1$, with a roughly three times larger luminosity distance, the apparent magnitude of 24.5 mag requires a ten times higher luminosity, corresponding to our choice of 1000 stellar particles. This way, we achieve consistency with the selection by \citet{Hilbert17}, as well as reliable shape and spin estimates. We apply our selection criteria to galaxies defined as `subhaloes' in the TNG database, which are gravitationally self-bound structures identified by the {\sc SUBFIND} group finder \citep{Springel01}. This selection results in a total of $\sim 2\times 10^4$ well-resolved galaxies.

\begin{figure*} 
\centering
\includegraphics[width=\textwidth,trim= 0 15 0 5,clip]{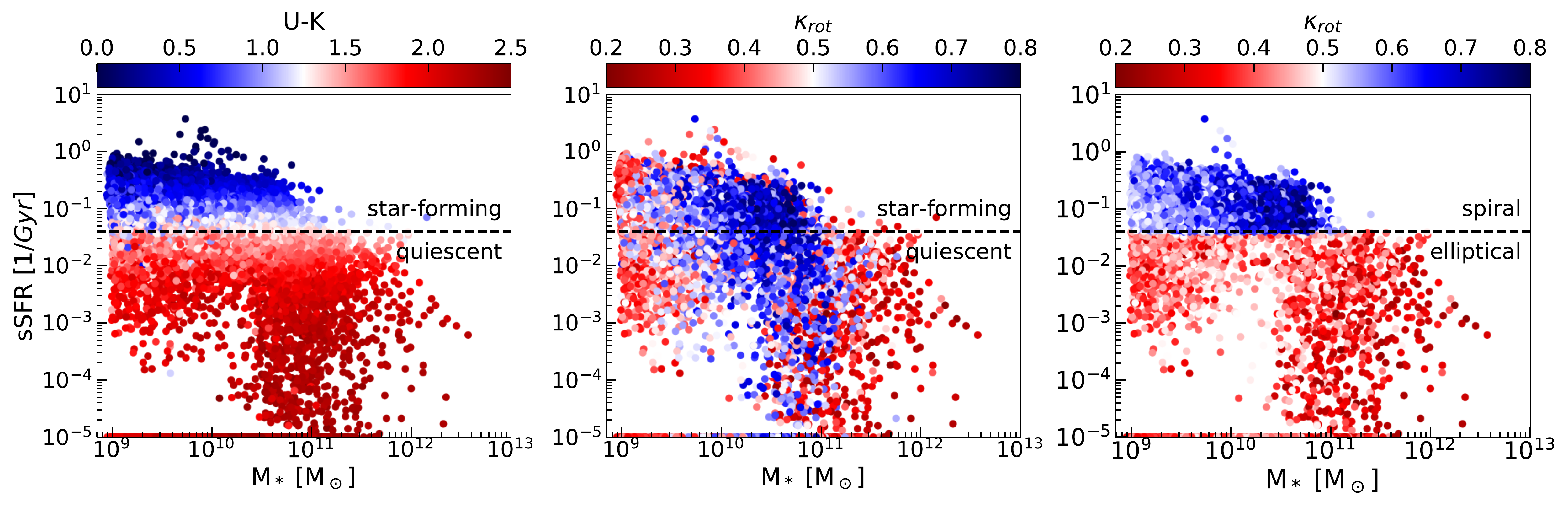}
\caption{\label{fig_samp}
Well-resolved galaxies from TNG100 in the sSFR-stellar mass plane, colour-coded by their $U$-$K$ colour (\textit{left panel}), and by the $\kappa_\mathrm{rot}$ parameter, which quantifies the rotational support of each galaxy (\textit{middle panel}). Galaxy colour is tightly correlated with sSFR which allows us to devide our sample in blue star-forming and red quiescent galaxies based on a threshold specific star-formation rate of sSFR $= 0.04\ Gyr^{-1}$ (dashed line). For the kinematic split we follow \citet{RodriguezGomez17} and classify galaxies with $\kappa_\mathrm{rot} > 0.5$ as spirals, and with $\kappa_\mathrm{rot} < 0.5$ as ellipticals. Rotationally supported galaxies with high $\kappa_\mathrm{rot}$ (discs) can be predominantly found at high sSFR/ blue colour as well as at Milky Way stellar mass. Our morphological classification of galaxies based on both star formation and kinematics yields 7905 spiral galaxies and 4616 elliptical galaxies at $z=0$ in TNG100 (\textit{right panel}).}
\end{figure*}

\paragraph*{Star-formation properties and galaxy colours.} \ In the left panel of Fig.~\ref{fig_samp}, we show the specific star-formation rate (sSFR) of all well-resolved galaxies versus their stellar mass, colour-coded by the $U$-$K$ colour of the galaxy. Galaxies with no (i.e. zero instantaneous) star formation are assigned a default sSFR of $10^{-5} Gyr^{-1}$ in order to be visible in the figure. The $U$-$K$ colour is calculated from the galaxy luminosities in the respective bands. {\sc IllustrisTNG} provides galaxy and stellar luminosities in 8 bands: $U$, $V$, $B$, $K$, $g$, $r$, $i$, and $z$ that are obtained through a model that accounts for stellar age, mass, and metallicity (\citealt{Vogelsberger13}, based on \citealt{BruzualCharlot03}), whereas the galaxy luminosity is calculated as a sum of the luminosities of individual stellar particles constituting the galaxy. The resulting galaxy colours in {\sc IllustrisTNG} are in excellent agreement with observations from the Sloan Digital Sky Survey (SDSS, DR12) at low redshift \citep{fTNGNelson18}. Furthermore, Fig.~\ref{fig_samp} clearly shows a tight correlation between galaxy colour and sSFR in {\sc IllustrisTNG}. We thus divide our sample into blue star-forming and red quiescent galaxies based on a threshold specific star-formation rate of sSFR $= 0.04\ Gyr^{-1}$ that is shown as a dashed black line in Fig.~\ref{fig_samp}, thereby effectively applying a colour cut.

\paragraph*{Kinematic split into spirals and ellipticals.} \ In order to distinguish between spiral and elliptical galaxies, we compute the amount of rotational support of the stellar component. Following \cite{Sales12}, we define 
\begin{equation} \label{kappa}
\kappa_\mathrm{rot} \defeq 
\frac{E_\mathrm{rot}}{E_\mathrm{kin}} \defeq 
\sum_n \frac{1}{2}m_n\left(\frac{j_{z,n}}{r_n}\right)^2 / \ \sum_n \frac{1}{2}m_n \bvec{\upsilon}_n^2,
\end{equation}
where $E_\mathrm{kin}$ and $E_\mathrm{rot}$ denote the total kinetic and rotational energy, respectively, $m_n$ the mass of stellar particle $n$ (in given subhalo which we use to define a galaxy), $\bvec{\upsilon}_{n}$ its velocity in the galaxy's centre of mass frame, $j_{z,n}$ its angular momentum in the direction of the total angular momentum $\bvec{J}_*$ of the stellar component, and $r_n$ its distance from the galaxy centre projected on the plane perpendicular to $\bvec{J}_*$. With this definition, galaxies with small $\kappa_\mathrm{rot}$ values are predominantly stabilised by their velocity dispersion, whereas galaxies with large $\kappa_\mathrm{rot}$ values are rotationally supported. We refer to \citet{RodriguezGomez17} for a more detailed discussion of the $\kappa_\mathrm{rot}$ parameter.

In the middle panel of Fig.~\ref{fig_samp}, we display for all our galaxies the relationship between the sSFR and stellar mass, this time colour-coded by their $\kappa_\mathrm{rot}$ parameter. Again, galaxies with no star formation are assigned a default value of sSFR $= 10^{-5} Gyr^{-1}$. This figure reveals two interesting properties. First, as expected, galaxies with high rotational support (discs) show predominantely high specific star-formation rates and have as a consequence a blue colour. Second, rotationally supported galaxies can mostly be found in a very specific stellar mass range that coincides with the stellar mass of typical Milky Way-like galaxies, while lower and higher mass galaxies are predominately velocity dispersion supported. This is also fully consistent with results from \citet{Chisari15} who employ the ratio of squared rotational velocity $\bvec{\upsilon}_\mathrm{rot}$ and velocity dispersion $\sigma^2$ of the stellar component as a kinematic criterion for morphological classification, and find $\bvec{\upsilon}_\mathrm{rot}^2/\sigma^2$ to peak at $M_* \sim 10^{10.5} M_\odot$. Following \citet{RodriguezGomez17}, we regard galaxies with $\kappa_\mathrm{rot} > 0.5$ to be rotationally supported and galaxies with $\kappa_\mathrm{rot} < 0.5$ to be velocity dispersion dominated. 

Applying both the colour cut as well as the kinematic classification to all well-resolved TNG100 galaxies yields a sample of 7905 blue spiral galaxies (sSFR $> 0.04\ Gyr^{-1}$, $\kappa_\mathrm{rot} > 0.5$) and a sample of 4616 red elliptical galaxies (sSFR $< 0.04\ Gyr^{-1}$, $\kappa_\mathrm{rot} < 0.5$) at $z=0$, which we simply will refer to as spirals and ellipticals in the following. The two final samples are displayed in the right panel of Fig.~\ref{fig_samp}. Applying the same selection at $z=1$ yields a sample of 8092 spiral and 969 elliptical galaxies.

\subsection{Projected galaxy shapes}
We analyse the projected shape of the simulated galaxies in terms of the moments of their light distribution. For all stellar particles $n = 1\dots N$ belonging to a galaxy, let $L^{(b)}_n$ be the luminosity of star $n$ in the photometric band $b$, and $\bvec{x}_n$ its three-dimensional position. We then define the centred second moment tensor of the light distribution \citep{Valdes83,MiraldaEscude91} as
\begin{equation}
q_{ij}^{(b)} \defeq  \frac{1}{L^{(b)}}\sum_n  \left(\bvec{x}_n-\overline{\bvec{x}}^{(b)}\right)_i\,\left(\bvec{x}_n-\overline{\bvec{x}}^{(b)}\right)_j \, L_n^{(b)}, \label{eq_def_qij}
\end{equation}
where $\overline{\bvec{x}}^{(b)}$ is the galaxy centre defined by its potential minimum, and $L^{(b)}\defeq\sum_n L_n^{(b)}$ is the total luminosity of the galaxy in band $b$. From the quadrupole moment $q_{ij}^{(b)}$ of the brightness distribution one commonly defines the ellipticity $\epsilon$, on which gravitational lensing operates. Assuming for simplicity, and without loss of generality, a projection of the light distribution along the cartesian $z$-axis, one has, using the components $q_{ij}$ appropriate for the projection,
\begin{equation} \label{eq_def_eps}
\epsilon \defeq \epsilon_+ + \ci\, \epsilon_\times \defeq 
\frac{q_{xx} -q_{yy}}{q_{xx} + q_{yy}} + \ci\,\frac{2q_{xy}}{q_{xx} + q_{yy}},
\end{equation}
where we have omitted the `$(b)$' for notational simplicity. Using eqs.~(\ref{eq_def_qij}) and~(\ref{eq_def_eps}), we calculate the real and imaginary components of ellipticity, $\epsilon_+$ and $\epsilon_\times$, for all our sample galaxies. Note, that though {\sc IllustrisTNG} is remarkably successful in reproducing galaxy sizes \citep{Genel18}, star-formation rates \citep{Donnari19}, and colours \citep{fTNGNelson18}, the resulting observed galaxy ellipticities constitute a prediction that can be affected by choices in the subgrid models. A verification based on observational data is not straight forward precisely because of the presence of lensing effects in real data. 
Also, the derived ellipticities ultimately depend on the selection of stellar particles constituing a galaxy, which in literature can range from all gravitationally bound stars belonging to a subhalo, to stars within the stellar half mass (light) radius, down to a fixed apperture, whereby \citet{Velliscig15ias} have shown that imposing sphericity on the stellar particle selection reduces the obtained IA signal, such that sometimes upweigthing schemes for central stars (mimicking luminosity weighting) are used to counteract this effect \citep[see][]{Tenneti15}. We circumvent such complications by weighting with the stellar luminosity, instead of with stellar mass, and focusing on the physical mechanism generating IAs (and not trying to reproduce specific observations) employ the subhalo definiton. In addition, observed ellipticities are affected by the presence of dust along the line-of-sight, which is not taken into account in the majority of previous IA studies, and is currently beyond the scope of our analysis. We comment on the impact of dust below.

Note, that $\epsilon_+$ and $\epsilon_\times$ are not invariant and depend on the coordinate system chosen in the projection plane. They transform like tensors of spin two, $\epsilon\rightarrow\exp(2\mathrm{i}\varphi)\epsilon$, under rotation of the coordinate frame by an angle $\varphi$. This implies that for small cosmological volumes, such as the TNG100 volume, cosmic variance causes them to deviate from isotropy when averaged over the entire galaxy sample. To circumvent this problem, we will consider random rotations around the $z$-axis of the $(x,y)$-plane for every galaxy and its ambient tidal field (see section~\ref{sec:measurement_parameters}). 

\begin{figure*}
\centering
\includegraphics[width=\textwidth,trim= 0 15 0 0,clip]{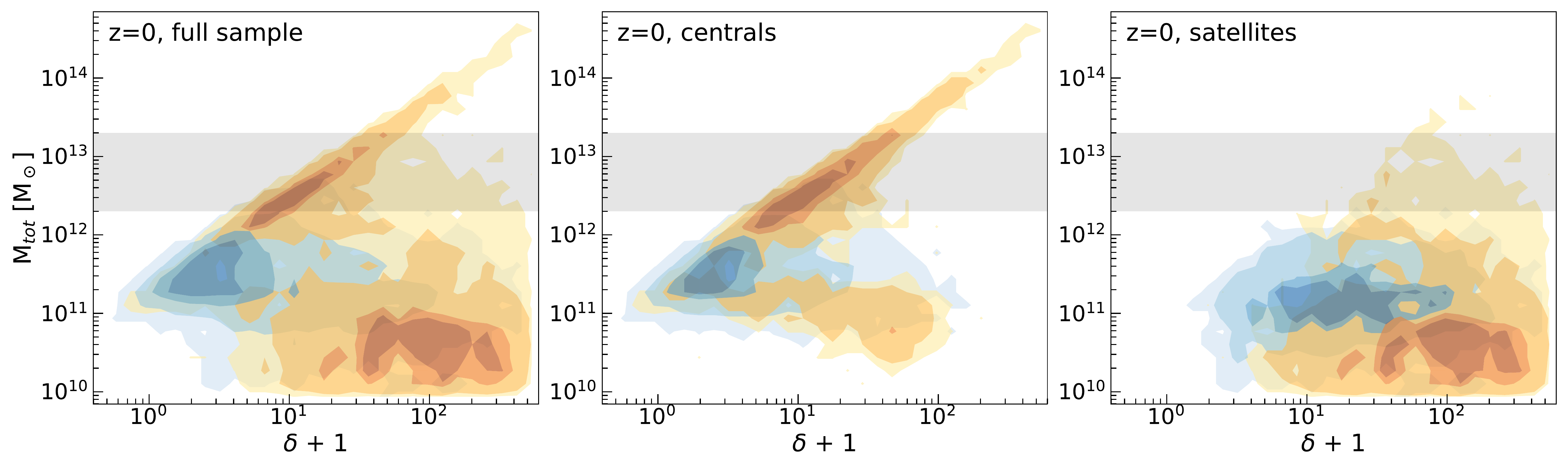}
\caption{\label{fig_cntsat} 2D-histogram of spiral (\textit{blue contours}) and elliptical (\textit{orange contours}) galaxies in the halo mass vs. local overdensity $\delta$ (smoothed at 1 Mpc) plane at $z=0$. Contours with increasing intensity are drawn around $100\%, 90\%, 50\%,$ and $10\%$ of galaxies. The three panels display from \textit{left} to \textit{right}: all galaxies, centrals only, and satellites only. Two distinct populations of elliptical galaxies become visible: central ellipticals with mass close to maximal possible at given overdensity, and satellite ellipticals with typically an order of magnitude lower mass found at high overdensities. The grey shaded band highlights the selected mass range for Fig.~\ref{fig_D_aniso} where elliptical galaxies exhibit a particularly high discrepancy in the alignment strength $D$ derived directly from TNG100 without applying anisotropy corrections (see section~\ref{sec:measurement_parameters} and appendix~\ref{sec:app_anisotropy_lss}).}
\end{figure*}

\subsection{The local gravitational tidal field}
\label{sec:tidal_field}
Intrinsic alignments are concerned with local correlations of galaxy shapes and the shape of the gravitational potential. To quantify the latter, we extract from the simulation the tidal shear tensor
\begin{equation}
 \Phi_{,ij}(\bvec{x}) = \frac{\partial^2\Phi(\bvec{x})}{\partial x_i \partial x_j}
\end{equation}
given by the Hessian of the gravitational potential $\Phi(\bvec{x})$ at the location of each galaxy in our sample.
To obtain $\Phi_{,ij}(\bvec{x})$, we first determine the three-dimensional overdensity field $\delta(\bvec{x})$ for the full simulation volume on a $1024^3$ grid using cloud-in-cell interpolation \citep[CIC, cf.][]{Hockney:1981}. We then use a discrete Fourier transform to solve Poisson's equation and obtain the Hessian of the potential algebraically in Fourier space. Due to an intrinsic uncertainty on which scale the effective tidal field relevant for intrinsic alignment correlations should be evaluated (see our discussion in sections~\ref{sect_e_props} and~\ref{sect_s_props}), we allow for an additional smoothing scale $\lambda_s$ (implemented by a Gaussian filter). We then have 
\begin{equation} \label{eq_posf}
\Phi_{,ij}(\bvec{x}) = \frac{3\Omega_m}{2\chi_H^2a}\, \mathcal{F}^{-1}\left[ \frac{k_i k_j}{|\bvec{k}|^2}\, \exp\left( -\frac{1}{2}|\bvec{k}|^2 \lambda_s^2\right) \, \mathcal{F}\left[ \delta(\bvec{x})\right] \right],
\end{equation}
where $\bvec{k}$ is the comoving wave vector, $\chi_H \defeq c/H_0$ is the Hubble-distance and $a$ the cosmic scale factor, so that $\Phi_{,ij}(\bvec{x})$ is dimensionless. $\mathcal{F}$ and $\mathcal{F}^{-1}$ denote discrete Fourier- and inverse Fourier-transforms, respectively. For the Gaussian smoothing $\lambda_s$, we consider here five different values: 250~kpc, 500~kpc, 1~Mpc, 2~Mpc, and 5~Mpc. 1~Mpc corresponds roughly to the mass scale of $M_{\rm tot} \sim 10^{12} M_\odot$ of Milky Way-like galaxies (estimated using the relation $M = 4\pi/3\:\Omega_m\rho_\mathrm{crit}\lambda_s^3$ with $\rho_\mathrm{crit} = 3H_0^2/(8\pi G)$), whereas 250~kpc corresponds to galaxies of $\sim 10^{10} M_\odot$ total mass. In a final step, we obtain the tidal shear field at the actual galaxy positions by applying an inverse CIC interpolation. The effective smoothing scale is therefore slightly larger than $\lambda_s$ by about one grid cell.

\subsection{Galaxy environment: large-scale overdensity and the central-satellite distinction}
\label{sec:environment}
In this work, we generally do not distinguish between central and satellite galaxies. As a consequence, dividing our sample into ellipticals and spirals, the satellite fraction varies with average galaxy mass and the environmental density (which is in some sense an observational proxy for the central-satellite distinction). We show in Fig.~\ref{fig_cntsat}, where centrals and satellites for the two galaxy morphologies reside in the space spanned by total galaxy (subhalo) mass and environmental overdensity (smoothed on a scale of 1~Mpc). Since halo mass is strongly correlated with the overdensity $\delta$ measured on a fixed scale, centrals are preferentially located on a tight relation in the $1+\delta$ versus $M_{\rm tot}$ plane, whereby central ellipticals constitute the most massive galaxies located at higher overdensities than central spirals. Satellites are scattered off the density-mass relation for centrals to higher densities (by definition due to the presence of a more massive host). More interestingly, low mass spiral satellites occupy intermediate densities and are located in slightly more massive subhaloes than elliptical satellites that occupy higher densities and have slightly lower mass. Such correlations between morphological mix and large-scale overdensity are also known from observations \citep[e.g.][]{Dressler:1980,Kauffmann:2004}. Regarding our full sample $\sim 70\%$ of elliptical galaxies at $z=0$ are satellites, whereas this is the case for only $\sim 25\%$ of spirals. Bearing this rather clear distinction of galaxy type with mass in mind (especially for elliptical galaxies), the trends of the alignment strength as a function of galaxy mass that we present in section~\ref{sect_e_props} and~\ref{sect_s_props} can also be interpreted in terms of a central-satellite distinction.

\section{Intrinsic alignments and weak lensing} \label{sect_ia_lensing}
In this section we provide a concise overview over the linear and quadratic alignment models for elliptical and spiral galaxies, respectively. We discuss the implications for lensing and bridge the gap to angular ellipticity spectra, both for lensing and intrinsic alignments, as they would appear in {\sc Euclid} data.

\subsection{Alignments of elliptical galaxies} \label{sect_e_theory}
We start our discussion of intrinsic alignment models with elliptical galaxies, which are assumed to be virialised, velocity-dispersion stabilised systems. In the linear alignment model \citep{Catelan01, Hirata04} presented in eq.~(\ref{eq_D}), an elliptical galaxy is assumed to be spherically symmetric in isolation. The presence of a cosmic tidal field, however, breaks spherical symmetry, leading to a correlation between the galaxy shape and the deforming tidal field which is at the same time responsible for gravitational lensing of background galaxies. As neighbouring galaxies are exposed to similar tidal fields, their shapes become correlated with each other.

\paragraph*{Response of the galaxy to large-scale tidal fields.} \ Let $\sigma^2(r)$ be the isotropic velocity dispersion of stars in the unperturbed, spherically symmetric galaxy, and $\bvec{r}\defeq\bvec{x}-\bvec{x}_0$, $r\defeq|\bvec{r}|$, be the radial coordinate w.r.t. the galaxy centre $\bvec{x}_0$. In hydrostatic equilibrium one has 
\begin{equation}
\frac{1}{\rho}\frac{\mathrm{d}(\sigma^2\rho)}{\mathrm{d} r} = - \frac{\mathrm{d}\Phi}{\mathrm{d} r},
\label{eq_jeans}
\end{equation} 
with density $\rho(r)$ and gravitational potential $\Phi(r)$.  The hydrostatic density profile is therefore
\begin{equation}
\rho(r) = \rho_0 \cdot \exp\left(-\frac{\Phi(r)}{\sigma^2}\right),
\label{eq_jeans_slv}
\end{equation}
with a constant $\rho_0$, if the velocity dispersion $\sigma^2$ is constant and does not depend on radius (i.e. if isothermality is given). 

An anisotropic gravitational tidal field, that surrounds the galaxy, introduces perturbations to this spherically symmetric state through tidal interactions. One expects that the galaxy responds by re-adjusting to a new dynamical equilibrium on the free-fall time scale $\propto 1/\sqrt{G\rho}$. A suitable upper limit for this time scale can be obtained with $\rho = 200\Omega_m\rho_\mathrm{crit}$ and $\rho_\mathrm{crit} = 3H_0^2/(8\pi G)$, giving a fraction $\sqrt{8\pi/3/200\Omega_m}$ of the Hubble time $1/H_0$. The new shape will reflect the orientation of the local tidal field. A Taylor expansion of the large-scale gravitational potential to second order at the galaxy location $\bvec{r}_0$ yields
\begin{equation}
\Phi(\bvec{r}) = \Phi(\bvec{r}_0) +  \Phi_{,a}(\bvec{r}_0) r_a + \frac{1}{2}\Phi_{,ab}(\bvec{r}_0)r_a r_b + \ldots
\end{equation}
The first derivative accelerates the galaxy as a whole into the direction of the gradient of the potential and gives rise to a non-zero peculiar velocity. Changes in the shape of a galaxy can only be evoked if there is a differential acceleration between the different parts of the galaxy, necessitating second or higher order derivatives. Assuming that the galaxy is indeed small compared to the curvature scale $S$,
\begin{equation}
\frac{1}{S^2} \defeq \left|\frac{\mathrm{d}^2}{\mathrm{d}r^2}\frac{\Phi}{\sigma^2}\right|, 
\end{equation}
and that $\sigma^2$ remains unchanged, the perturbed (stellar) density $\tilde{\rho}(\bvec{r})$ then is related to the unperturbed density $\rho(r)$ as
\begin{equation} \label{eq_rho_pert}
\tilde{\rho}(\bvec{r}) 
= \rho_0 \cdot \exp \left( -\frac{\Phi(\bvec{r})}{\sigma^2} \right) 
\simeq \rho(r)\, \left( 1 -\frac{\Phi_{,ab}(\bvec{r}_0)r_ar_b}{2\sigma^2}  \right),
\end{equation}
where the second step is a first order approximation. Assuming a mass-to-light conversion $L(\bvec{r})\propto \rho(\bvec{r})$, the first three even moments of the brightness distribution are defined as
\begin{subequations} 
\begin{alignat}{3}
L_0 &\defeq  \int \dd^3 r\: L(\bvec{r}), \\
q_{ij} &\defeq  \frac{1}{L_0} \int {\rm d}^3 r\: L(\bvec{r})\,r_i r_j, \\
s_{ijkl} &\defeq \frac{1}{L_0} \int \dd^3r\: L(\bvec{r})\,r_i r_j r_k r_l, \label{eq_sijkl}
\end{alignat}
\end{subequations}
whereas previously we have computed the 2D projected $q_{ij}$ in discretized form, as given in eq.~(\ref{eq_def_qij}). The perturbation $\Delta q_{ij}$ of the second moment due to large-scale tides is then computed to be
\begin{align}  \label{eq_qij_pert}
\Delta q_{ij} &\defeq \tilde{q}_{ij}-q_{ij} = -\frac{ 1}{2\sigma^2} \Phi_{,ab}(\bvec r_0)  s_{abij}
\end{align}
\citep{Piras18}, where $\tilde{q}_{ij}$ is obtained using the perturbed luminosity $\tilde{L}(\bvec{r})\propto\tilde{\rho}(\bvec{r})$ given by eq.~(\ref{eq_rho_pert}). 

As we assume the unperturbed galaxy to be spherically symmetric, the unperturbed $q_{ij}$ has effectively only a scalar degree of freedom $q_{ij}=\frac{q}{2} \delta_{ij}$, such that $s_{ijkl}$ is given by the isotropic tensor  $s_{ijkl}=c_{ijkl}+q_{ij}q_{kl}+q_{ik}q_{jl}+q_{il}q_{jk} = \left(c+ \frac{q^2}{4}\right)\left( \delta_{ij}\delta_{kl}+\delta_{ik}\delta_{jl}+\delta_{il}\delta_{jk}\right)$, where $c_{ijkl}$ is the 4th cumulant of the light distribution, which is again a scalar, $c$, for a point-symmetric case. A qualitatively similar result has been obtained by \cite{Ghosh20} through other means. Note, that the prefactor of $(c+q^2/4)$ inherits the $\sim R^4$ scaling from $s_{ijkl}$, that is apparent in eq.~(\ref{eq_sijkl}). Also, by symmetry, eq.~(\ref{eq_qij_pert}) must hold up to factors even if we had assumed a more general unperturbed $q_{ij}$. A small anisotropy in the velocity dispersion would appear as a higher order correction. 

\paragraph*{Observed ellipticity-tidal field correlation.} \ Given the perturbed second brightness moments $\tilde q_{ij}$, the observed ellipticity in projection along the cartesian $z$-axis is
\begin{equation} \label{eq_def_eps_tilde}
\tilde\epsilon = 
\frac{\tilde q_{xx} - \tilde q_{yy}}{\tilde q} + \ci\frac{2\tilde q_{xy}}{\tilde q} ,
\end{equation}
where $\tilde q \defeq \tilde q_{xx} + \tilde q_{yy}$ is the galaxy size, which we assume to be unchanged at leading order, i.e. $\tilde q = q\defeq q_{xx}+q_{yy}$. Inserting the tidal shape perturbation from eq.~(\ref{eq_qij_pert}) the observed ellipticity becomes
\begin{equation} \label{eq_eps_zw}
\tilde \epsilon = \epsilon_0 -\frac{1 }{2\sigma^2q}  \sum_{i,j\in\{x,y\}}\Phi_{,ij} \left(s_{i j x x} - s_{i j y y} + 2 \ci\,s_{i j x y} \right).
\end{equation}
It thus becomes apparent that the observed ellipticity has an intrinsic contribution $\epsilon_0$, which reflects for instance the shape of the protogalactic cloud from which the galaxy forms and other intrinsic formation processes, and a contribution due to tidal interaction with the environment.
If the unperturbed galaxy is perturbatively isotropic, i.e. if we can assume $q_{\ij}\approx \frac{q}{2} \delta_{ij}+\eta_{ij}$ and $\eta_{ij}$ is a second order term, then we find at leading order
\begin{equation}
\tilde \epsilon = \epsilon_0 - D \left( \Phi_{,xx}-\Phi_{,yy} + 2 \ci  \Phi_{,xy} \right) \eqdef \epsilon_0 - D\ (T_+ +\ci T_\times),
\end{equation}
where we have defined $T_+ \defeq \Phi_{,xx} - \Phi_{,yy}$ and $T_\times \defeq 2\Phi_{,xy}$, and where $D\defeq \frac{c/q + q/4}{\sigma^2}$ absorbs the galactic properties, i.e. the velocity dispersion $\sigma^2$ of the galaxy, its apparent area $q$, and its peakiness or concentration $c$, and scales as $\sim R^2/ \sigma^2$.

In a measurement it will be the case that the intrinsic ellipticity $\epsilon_0$ will be a random variable with zero mean and some amount of dispersion. In a statistical sample of galaxies, one then expects a mean relation (dropping the tilde for simplicity since the unperturbed ellipticity is not observable)
\begin{equation} \label{eq_D}
\epsilon = \epsilon_+ + \ci \epsilon_\times \simeq -D \ (T_+ + \ci T_\times), 
\end{equation}
which is referred to as the linear alignment model for elliptical galaxies. Note the sign convention chosen for $D$ to be strictly positive and reflect the alignment strength, as well as that we quote $D$ in units of $c^2$, as we use the dimensionless gravitational potential $\Phi/c^2$. Since we do not know which scales in tidal shear tensor are the dynamically most relevant for establish the correlation, we allowed for a tuneable scale parameter $\lambda_s$ in our computation. Comparing to gravitational lensing it is thus apparent that tidal fields in both cases are able to distort shapes in a completely analogous way. Physically, the intrinsic alignment effect is governed by the second derivatives of $\Phi/\sigma^2$ while lensing by the second derivatives of $2\Phi/c^2$. Clearly, there are no astrophysical effects to include in lensing as it is completely determined by relativity alone.  It is clear that introducing a scale dependence through $\lambda_s$ will also make $D$ scale dependent. 

\subsection{Alignments of spiral galaxies} \label{sect_s_theory}
The picture for spiral galaxies is that their observed shape is fully determined by the direction of their angular momentum vector which is oriented perpendicularly to a thin, circular galactic disc. Angular momentum is imprinted on a halo and ultimately on its central galaxy through torques from the surrounding large-scale matter distribution during early stages of gravitational collapse. As neighbouring spiral galaxies are surrounded by the same large-scale structure, they are subject to similar tidal forces, thus acquiring angular momentum vectors pointing in similar directions. This results in a correlation between the observed galaxy ellipticity and the surrounding tidal field presented in eq.~(\ref{eq_A}) referred to as the quadratic alignment model.

In more detail, the quantity that determines the ellipticity of a spiral galaxy is the angle of inclination under which the stellar disc is viewed, i.e. the angle between the symmetry axis of the galactic disc and the line-of-sight. Assuming that this symmetry axis reflects the angular momentum direction $\hat{\bvec{L}} = \bvec{L}/|\bvec{L}|$ of the stellar component, the ellipticity $\epsilon = \epsilon_+ + \ci\epsilon_\times$ for an infinitely thin disc observed along the $z$-axis is given by
\begin{equation} \label{eq_eL}
\epsilon = \frac{\hat L_y^2 - \hat L_x^2}{1 + \hat L_z^2} + \ci\frac{2 \hat L_y \hat L_x}{1 + \hat L_z^2}
\end{equation}
\citep{Crittenden01}, such that the absolute value $\left|\epsilon\right| = \sqrt{\epsilon\epsilon^*} = \sqrt{\epsilon_+^2+\epsilon_\times^2}$ of the ellipticity scales according to
\begin{equation} \label{eq_eLz}
\left|\epsilon\right| = \frac{1 - \hat L_z^2}{1 + \hat L_z^2}.
\end{equation}
These relationships can be weakened by introducing a constant of proportionality less than one, for describing a disc of finite thickness where the orientation effect is less pronounced. \citet{Crittenden01} quote that the observed ellipticity of spiral galaxies is on average suppressed by a factor of $\simeq 0.85$. If the symmetry axis of the stellar component is parallel to the angular momentum direction, and if the stellar angular momentum direction coincides with that of the host halo, the apparent ellipticity can be traced back to the tidal gravitational field, which is responsible for exerting torques onto the halo and its central galaxy, and for building up its angular momentum. 

Fundamental models for angular momentum generation by tidal torquing \citep{Hoyle49, White84, CatelanTheuns96} assume a first-order process in Lagrangian perturbation theory and express the angular momentum $\bvec{L}$ as 
\begin{equation}
L_i(t) = -a^2(t) \dot{D}(t) \, \epsilon_{ijk}\, I_{jl}\, \Phi_{,lk},
\end{equation}
where $a(t)$ is the scale factor, $\dot{D}(t)$ the linear growth rate, $I_{jl}$ the moment of inertia tensor of the proto-galaxy before collapse, and $\epsilon_{ijk}$ the antisymmetric Levi-Civita symbol. Using the most general second order relationship for the variance of the angular momentum directions as a function of the tidal shear tensor, 
\begin{equation}
\left< \hat{L}_i \hat{L}_j \ \middle|\tilde{\Phi}_{} \right> = \frac{1+A}{3} \delta_{ij} + A\tilde \Phi_{,ik} \tilde \Phi_{,kj},
\label{eq_llt_hat}
\end{equation}
\citet{LeePen01} have found an effective description of the random process $p(\hat{L}_i|\Phi_{,ij})$ conditional on the tidal shear $\Phi_{,ij}$, which enters the above relation as the traceless, unit normalised shear $\tilde{\Phi}_{,ij} = \hat \Phi_{,ij} / (\hat \Phi_{,ij}\hat \Phi_{,ij})^{1/2}$, where $\hat{\Phi}_{,ij}\defeq\Phi_{,ij}-\frac{1}{3}\Phi_{,kk}\delta_{ij}$ is the traceless part of the tidal shear tensor. It has the properties $\tilde{\Phi}_{,ii} = 0$ and $\tilde{\Phi}_{,ij}\tilde{\Phi}_{,ji} = 1$. The constant of proportionality $A$ in this model determines how random the angular momentum field can be in the presence of a tidal shear field, and the probability distribution $p(\hat{L}_i|\Phi_{,ij})$ accounts for the range of angular momenta that are compatible for a given tidal shear field for the full range of inertia tensors, assuming those are uncorrelated with the tidal field. In this framework, $A$ can reach a maximum value of $A = 3/5$ in case the alignment is fully set by the linear theory prediction and preserved at full strength at all redshifts, whereas $A = 0$ would imply completely random angular momentum directions. \citet{LeePen00} have verified eq.~(\ref{eq_llt_hat}) in $N$-body simulations and derived the free model parameter to be $A=0.237\pm0.023$.

Combining the ansatz from \citet{LeePen01} with the relation between observed ellipticity and angular momentum direction given in eq.~(\ref{eq_eL}), \citet{Crittenden01} derive an expression for the complex ellipticity $\epsilon$ as a function of the tidal shear field, which was revised in \citet{SchaeferMerkel15}. 
In our work, we work with the following version of the quadratic model,
\begin{equation} 
\epsilon = \frac{A}{2} \ (\tilde \Phi_{,xi} \tilde\Phi_{,ix} - \tilde\Phi_{,yi} \tilde\Phi_{,iy} + 2\ci \tilde \Phi_{,xi} \tilde \Phi_{,iy}) \eqdef A\,(Q_+ + \ci Q_\times)
\label{eq_A}
\end{equation} 
where we have defined $Q_+ \defeq 1/2\ (\tilde \Phi_{,xi} \tilde\Phi_{,ix} - \tilde\Phi_{,yi} \tilde\Phi_{,iy})$ and $Q_\times \defeq \tilde \Phi_{,xi} \tilde \Phi_{,iy}$. Note, that the version given in eq.~(\ref{eq_A}) differs from literature in the signs of the real and/or imaginary part. We adopt this version of the quadratic model as the correct default, as it yields identical values of $A$ derived from the real and imaginary part independently, as well as produces a significant signal for specific sub-populations of galaxies. We leave a revision of the model for future theoretical studies, if deemed to be of importance after inspecting our results in section~\ref{sect_s}. 
Theoretical predictions of ellipticity correlation functions in the literature are unaffected by different sign conventions, as the direct comparison between tidal shear and resulting ellipticity needs consistency in the real and imaginary parts separately. Note, that the parameter $A$ combines the description of the randomness of the angular momentum field, misalignments between galactic disc and angular momentum, as well as the effect of a stellar disc of finite thickness, which further reduces the observed ellipticity with increasing inclination angle. Physical limitations of the tidal torquing model are those of first-order Lagrangian perturbation theory, but also other processes such as anisotropic accretion or dissipative mechanisms that change or reorient the angular momentum and are not accounted for. 

\subsection{Measurement of alignment parameters}
\label{sec:measurement_parameters}
Both the linear alignment model, 
\begin{subequations}
\begin{equation}
\epsilon_++\ci \epsilon_\times = D \, (T_+ +\ci T_\times),
\end{equation}
and the quadratic alignment model
\begin{equation}
\epsilon_++\ci \epsilon_\times =  A \, ( Q_+ + \ci Q_\times), 
\end{equation}
\end{subequations}
depend on effective couplings, or `alignment' parameters, $D$ and $A$, that we will determine empirically in the remainder of this paper for various galaxy samples at different redshifts, and investigate their dependence on scale and galaxy environment. Note that both alignment models are linear w.r.t. these parameters, while the linear (quadratic) model is linear (quadratic) in its relation to the tidal field.

\begin{figure*} 
\centering
\includegraphics[width=0.49\textwidth,trim= 10 10 10 10,clip]{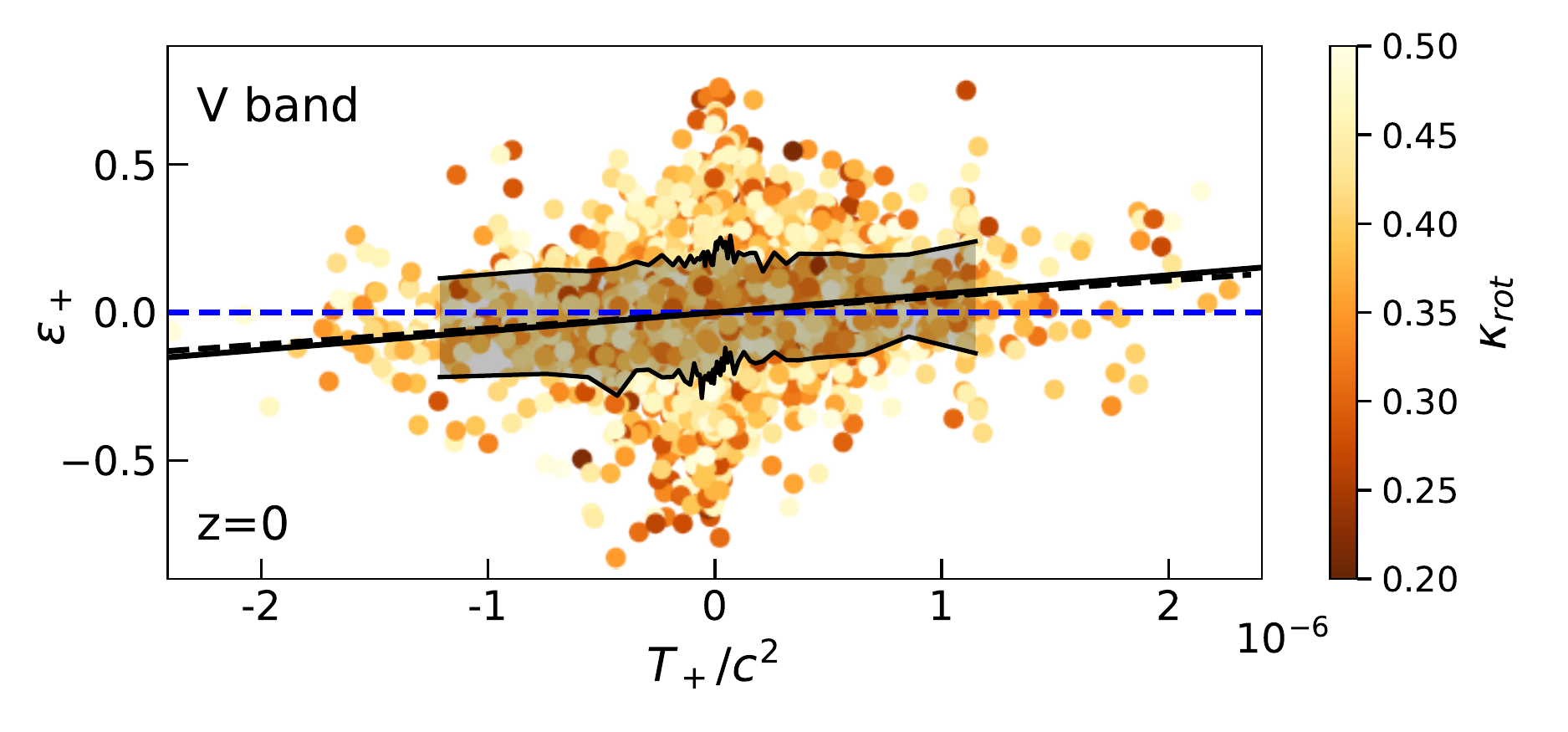}
\includegraphics[width=0.49\textwidth,trim= 10 10 10 10,clip]{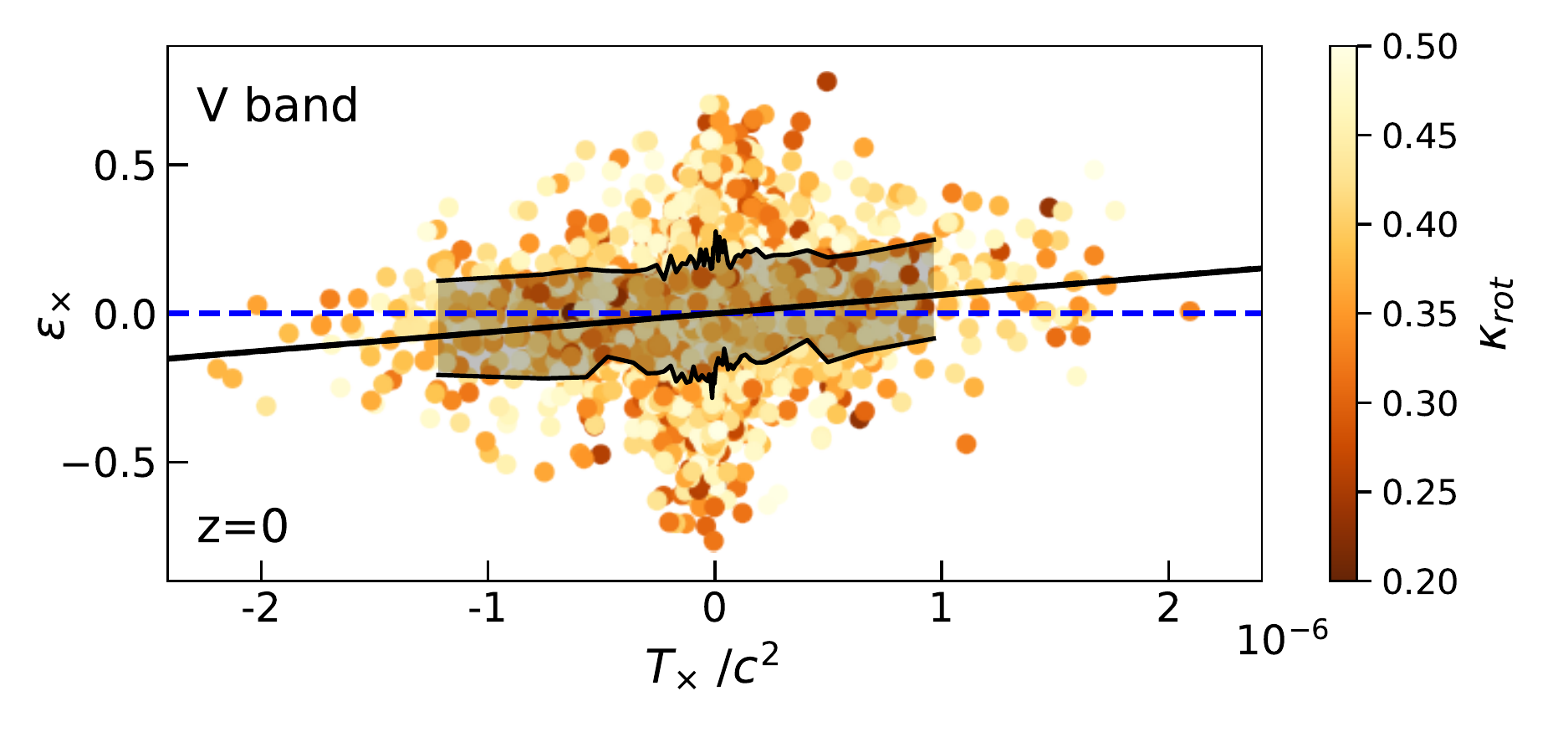}
\caption{Correlation of the real (\textit{left panel}) and imaginary part (\textit{right panel}) of the observed galaxy ellipticity of elliptical galaxies with the respective tidal field components $T_{+,\times}$ (smoothed on a scale of 1 Mpc) according to eq.~(\ref{eq_D}) in the $V$-band at $z=0$. Each point represents one galaxy colour-coded by its $\kappa_\mathrm{rot}$ value. No correlation is indicated by a blue dashed line at $\epsilon_{+,\times} = 0$. Black dashed line correponds to a linear fit to the binned data points, the error in each bin is given by the standard deviation and is displayed as a grey band that shows where $68.3\%$ of all galaxies come to lie. Black solid line shows the fit to the anisotropy corrected data that yields a slope of $D_{z=0} = (6.53 \pm 0.58) \times 10^{-7} c^2$, and thus a significant intrinsic alignment signal according to the linear model. \label{fig_D_z0}}
\vspace{0.5cm}
\includegraphics[width=0.49\textwidth,trim= 10 10 10 10,clip]{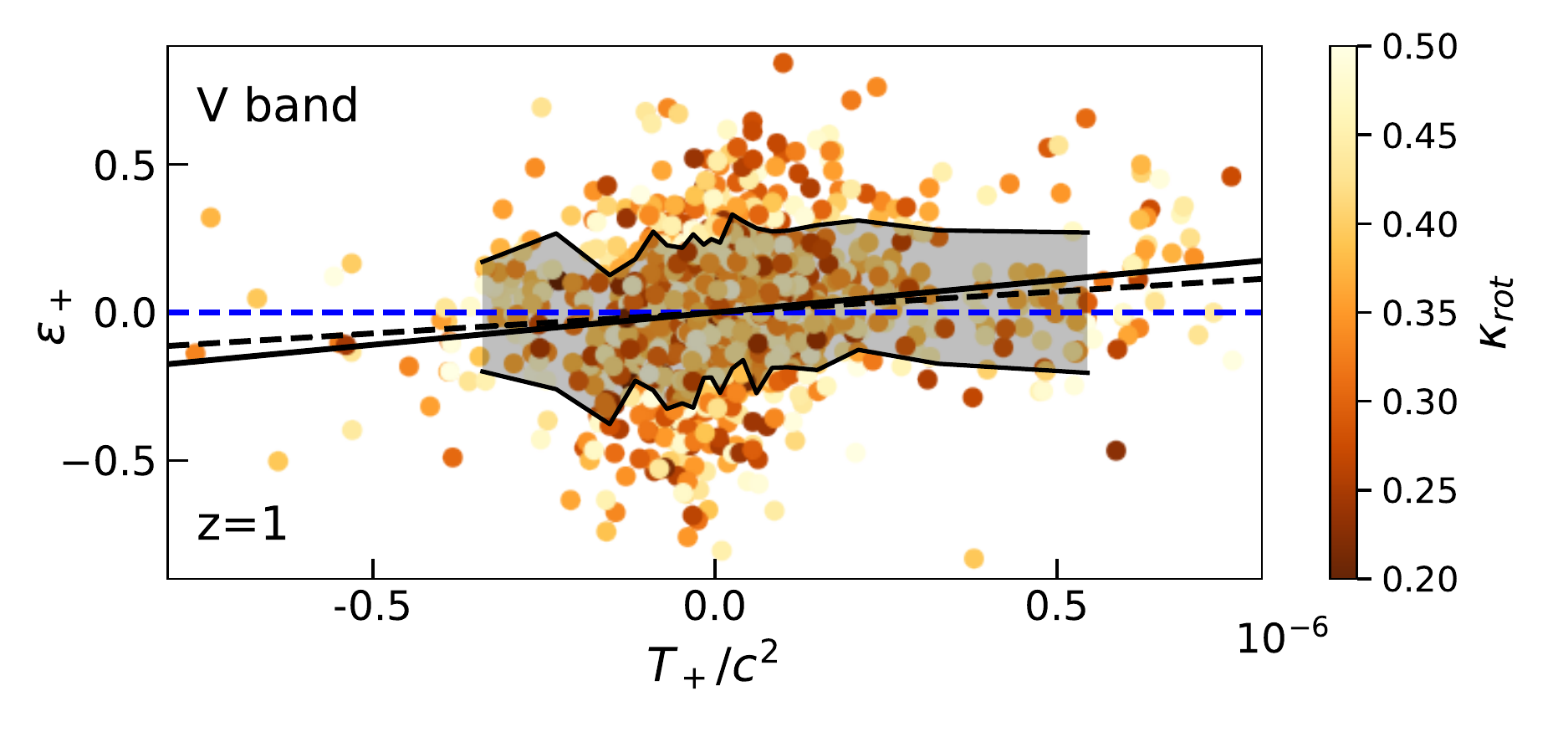}
\includegraphics[width=0.49\textwidth,trim= 10 10 10 10,clip]{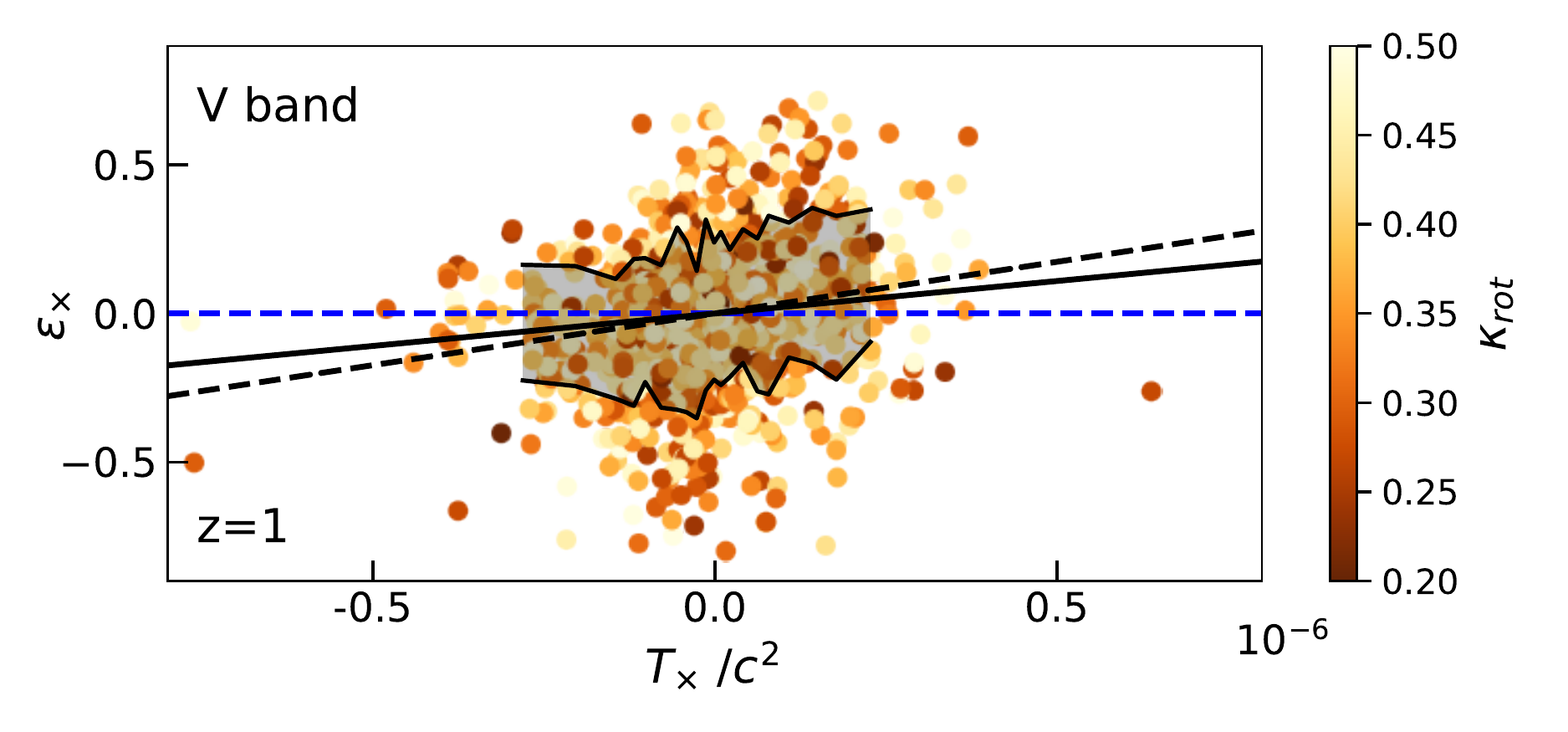}
\caption{\label{fig_D_z1} Same as Fig.~\ref{fig_D_z0}, but for elliptical galaxies at $z=1$. At higher redshift we find the direct measurement from TNG100 to deviate stronger from the anisotropy corrected result, that yields an alignment parameter of $D_{z=1} = (2.43 \pm 0.36) \times 10^{-6} c^2$, which is significantly larger than the value at $z=0$. Also, elliptical galaxies at $z=1$, selected based on the same criteria as at $z=0$, exhibit lower $\kappa_{\rm rot}$ values.}
\end{figure*}

\paragraph*{Galaxy ellipticity and dust.} \ We first determine the complex ellipticity $\epsilon$ for each galaxy, elliptical or spiral, from its light distribution. {\sc IllustrisTNG} provides stellar luminosities in the $U$, $B$, $V$, and $K$ broad bands, as well as $g$, $r$, $i$, and $z$ narrow bands which all yield observed ellipticities that we found statistically indistinguishable from the ones shown in Fig.~\ref{fig_D_z0} (ellipticals) and Fig.~\ref{fig_A_z0} (spirals) for the $V$-band. Stellar luminosities in {\sc IllustrisTNG} have, however, been derived without taking into account dust attenuation \citep{Vogelsberger13}, which can impact the observed ellipticities especially in the $U$-band and in the infrared. We find that the differences in age and metallicity alone do not significantly alter the apparent shape of galaxies, but this is expected to be different once dust is accounted for. A study of the effect of dust on apparent galaxy shapes is however beyond the scope of this study, but certainly an aspect that should be included in future work. In this study we thus will present results always in the $V$-band, expected to be least affected by dust. 

\paragraph*{Fitting procedure.} \ In order to measure the alignment parameters, we bin our various galaxy samples into multiple bins in $T_{+,\times}$ and $Q_{+,\times}$ respectively (both obtained from the smoothed tidal field at each galaxy's position, as described in section~\ref{sec:tidal_field}). We keep the number of galaxies in each bin constant. We then fit a linear model to the mean $\epsilon_{+,\times}$ in each bin, taking the error on the mean (i.e. the scatter in $\epsilon$ in the bin, divided by $\sqrt{N-1}$, where $N$ is the number of galaxies in each bin). Note, that while we will display the standard deviation in each bin as a grey band in the corresponding figures, the error $\pm \Delta D$ and $\pm \Delta A$ on the alignment parameters $D$ and $A$ is obtained from the factor of $\sqrt{N-1}$ smaller standard error in each bin. Per default we will fit to tidal field values smoothed on a 1 Mpc scale, which is the correspnding length scale of a steriotypical $10^{12}M_\odot$ Milky Way-like spiral galaxy. For direct comparability we adopt the same smoothing scale for elliptical galaxies, but stress that ultimately the scale for smoothing is a purely numerical choice. It is worth noting that the two ellipticity components $\epsilon_+$ and $\epsilon_\times$ in this case constitute mutually independent measurements and can be converted into each other by a rotation of the coordinate frame.

\paragraph*{Isotropisation of ellipticity frame.} \ A complication of our analysis is that $\epsilon_+$ and $\epsilon_\times$ transform as components of a tensor and are therefore sensitive to the absolute orientation of structure. Since the TNG100 cosmological volume is too small to contain isotropic cosmic large-scale structure, one observes statistical differences in the parameters derived from the two ellipticity components $\epsilon_+$ and $\epsilon_\times$ (see appendix~\ref{sec:app_anisotropy_lss}). To eliminate this effect, we randomise the frame in which we measure the alignment parameters by performing a rotation of the local frame in which ellipticity and shear are given, effectively decorrelating these frames between galaxies, i.e. we rotate each galaxy and its corresponding tidal field values by the same random angle between 0 and $\pi$, which removes any preferred directions in the simulation box, while retaining the unmitigated correlation between tidal field strength and galaxy ellipticity. We apply a given number of such random rotations (`randomisations') to the orientations of all galaxies in a chosen sample and their tidal field values, thus deriving an equal number of independent realisations of TNG100 without anisotropic contamination, each resulting in one measurements for an alignment parameter. The average of those measurements then yields the `true' value for the alignment strength $D$ or $A$. We have performed convergence tests with $10^2$, $10^3$, $10^4$, and $10^5$ such random rotations, and we find that both the mean alignment parameters, as well as their standard error agree well after about $10^3$ randomisations with the results from $10^4$ and $10^5$ randomisations. Therefore, in the rest of this study we present for the alignment parameters $D$ and $A$ the mean value derived from $10^3$ realisations of TNG100 with randomised galaxy and tidal field orientations.

\paragraph*{Non-parametric alignment test.} \ The alignment parameters are typically small compared to the scatter in the ellipticity-tide relations, and depend explicitly on the magnitude of the quantities entering the models. For this reason, their absolute magnitude depends non-trivially on other quantities such as the mass, or the smoothing scale applied to the tidal field (see section~\ref{sect_e_props} below). To aid physical intuition, we therefore supplement our analysis also with Spearman rank correlation coefficients, which are insensitive to magnitudes and non-linear (monotonic) transformations of the data.

\subsection{Gravitational lensing}
Intrinsic shape correlations of galaxies, either caused by correlated angular momenta or correlated tidal gravitational shear fields, are a serious contaminant of weak gravitational lensing on small scales. These effects do not only possess the same fundamental observable, namely ellipticity correlation functions or angular ellipticity spectra, but depend as well on the same fundamental fields, e.g. the traceless tidal gravitational shear. The change in ellipticity in the weak lensing limit is given by the mapping $\epsilon\rightarrow\epsilon+\gamma$ of the complex ellipticity $\epsilon=\epsilon_++\ci\epsilon_\times$ with the complex shear $\gamma=\gamma_++\ci\gamma_\times$. Those can be obtained from the Hessian tensor $\psi_{ab}$,
\begin{equation}
\psi_{A,ab} = \partial_a\partial_b\psi_A
\end{equation}
of the lensing potential $\psi_A$, by using $\gamma_+ = (\psi_{A,xx}-\psi_{A,yy})/2$ and $\gamma_\times = \psi_{A,xy}$. The lensing potential is given by
\begin{equation}
\psi_A = 2\int\dd\chi\:\frac{G_A(\chi)}{\chi}\frac{D_+}{a}\Phi = \int\dd\chi\:W_{\psi,A}(\chi)\ \Phi,
\end{equation}
where $\chi$ is the comoving distance, $a$ the scale factor, $D_+$ the growth factor, and $\Phi$ the tidal field, for a redshift $z$ distribution $p_A(\chi)\dd\chi$ of lensed galaxies which is subdivided into tomographic bins $A$, which defines the lensing efficiency,
\begin{equation}
G_A(\chi) = \int_{\mathrm{max}(\chi,\chi_A)}^{\chi_{A+1}}\dd\chi^\prime\:p(\chi^\prime)\frac{\dd z}{\dd\chi^\prime}\left(1-\frac{\chi^\prime}{\chi}\right).
\end{equation}
Carrying out a Limber-projection yields the angular spectrum of weak lensing shear as a function of multipole moments $\ell$,
\begin{equation}
C^{\gamma\gamma}_{AB}(\ell) = 
\ell^4\:\int\frac{\dd\chi}{\chi^2}\:W_{\psi,A}(\chi)W_{\psi,B}(\chi)\:P_{\Phi\Phi}(k = \ell/\chi),
\end{equation}
and an analogous expression for the ellipticity correlations due to intrinsic alignments for the linear interaction model, thought to be applicable for elliptical galaxies,
\begin{equation}
C^{\epsilon\epsilon}_{AB}(\ell) = 
\ell^4\:\int\frac{\dd\chi}{\chi^2}\:W_{\varphi,A}(\chi)W_{\varphi,B}(\chi)\:P_{\Phi\Phi}(k = \ell/\chi),
\end{equation}
and finally for their cross-correlation,
\begin{equation}
C^{\gamma\epsilon}_{AB}(\ell) =
\ell^4\:\int\frac{\dd\chi}{\chi^2}\:W_{\varphi,A}(\chi)W_{\psi,B}(\chi)\:P_{\Phi\Phi}(k = \ell/\chi),
\end{equation}
with a suitable definition of the line-of-sight weighting function for the intrinsic ellipticities, through the projected physical tidal field
\begin{equation}
\varphi_A = D\int\dd\chi\:p_A(z(\chi))\frac{\dd z}{\dd\chi}\frac{D_+}{a}\:\Phi = \int\dd\chi\:W_{\varphi,A}(\chi)\:\Phi,
\end{equation}
with the alignment parameter $D$ relevant for elliptical galaxies. From the projected potentials $\varphi_{A}$ and $\psi_{A}$ one obtains the projected tidal field by angular differentiation, $\varphi_{A,ab} = \partial_a\partial_b\varphi_A$ and $\psi_{A,ab} = \partial_a\partial_b\psi_A$. As both gravitational lensing and the alignment model for elliptical galaxies depend linearly on the traceless tidal shear as the fundamental field, they have to be in fact cross-correlated. For a cosmic tidal shear field with Gaussian statistics, there should be no cross-correlations between intrinsic ellipticities of elliptical galaxies and spiral galaxies, nor between gravitational lensing and spiral galaxies, as both correlation functions become proportional to a third moment of a Gaussian random variable, which is necessarily zero. An analogous treatment applies to the derivation of ellipticity spectra of spiral galaxies, which are modelled in configuration space and projected with the Limber-equation, before Fourier-transforming to obtain $E$-mode and $B$-mode spectra, details of this procedure can be found in \citet{Tugendhat18}.

\section{Elliptical galaxies} \label{sect_e}
In this section, we measure at $z=0$ and $1$ the alignment parameter $D$ of the linear model for elliptical galaxies, derived previously and presented in eq.~(\ref{eq_D}), followed by a more detailed investiagtion of the mass, scale, and environmental dependence of these results.

\subsection{Intrinsic alignments at $z=0$} \label{sect_e_z0}
Based on our kinematically and colour-selected sample of well-resolved elliptical galaxies from TNG100  (as described in section~\ref{sec:galaxy_selection}), we will calibrate the linear alignment model which predicts the linear relationship given in eq.~(\ref{eq_D}) between galaxy ellipticities and the gravitational tidal field. In Fig.~\ref{fig_D_z0}, we show the real and imaginary parts of the observed galaxy ellipticity at $z = 0$ in the SDSS $V$-band against the respective tidal field values. Every data point corresponds to one elliptical galaxy from our sample, colour-coded by the value of its $\kappa_\mathrm{rot}$ parameter, as defined in eq.~(\ref{kappa}). Here, the tidal field values have been smoothed on a scale of $\lambda_s = 1$~Mpc.

We measure the alignment parameter $D$ by determining the mean ellipticities $\epsilon_{+,\times}$ in 50 bins in $T_{+,\times}$, such that every bin contains $\sim 100$ galaxies. We then performe linear regressions of $\epsilon_{+}(T_+)$ and $\epsilon_{\times}(T_\times)$ for 1000 randomisations of the local reference frame in TNG100 to account for the anisotropy of the large-scale structure in the small TNG100 volume (see section~\ref{sec:measurement_parameters} for details), and show the results as a solid black line. The regression uses the standard error in defining the $\chi^2$-value, which in this case is $\sim 10$ times smaller than the standard deviation of the ellipticity values in each bin that is indicated by the grey band. The fit for the unrandomized data is shown as a black dashed line. The null hypothesis of no correlation is indicated by a blue dashed line.

At $z=0$, we measure in the $V$-band an alignment parameter of
\begin{equation}
D_V = (6.53 \pm 0.58) \times 10^{-7} c^2,
\end{equation}
which differs significantly from zero at a level of about $10\sigma$. The raw measurements from TNG100 data yield instead $D_{+,V}= (6.02 \pm 0.79) \times 10^{-7} c^2$ and $D_{\times,V}= (7.04 \pm 0.84) \times 10^{-7} c^2$, which differ by only $1.2\sigma$ and $1.3\sigma$, respectively, and which come out with very similar dispersions. Averaging the two values yields a combined alignment parameter $D_{V}= (6.53 \pm 0.58) \times 10^{-7} c^2$, equivalent to the measurement after the randomisation procedure. However, we strongly emphasise that this is not true in general, and that a direct measurement of ellipticity correlations from any cosmological simulation of similar or smaller volume than TNG100 is affected by the same problematic issue. 

\subsection{Intrinsic alignments at $z=1$} \label{sect_e_z1}
Many of the source galaxies for weak lensing studies are situated at high redshift. In case of {\sc Euclid} the projected redshift distribution peaks around $z\simeq 1$, such that we are particularly intrested in this redshift. In Fig.~\ref{fig_D_z1}, we show the relation between the ellipticity components $\epsilon_{+,\times}$ as a function of tidal shear $T_{+,\times}$, where each elliptical galaxy is again colour-coded by the value of its $\kappa_\mathrm{rot}$ parameter. We apply the same selection criteria as at $z = 0$, which at $z = 1$ yields a sample of only 969 elliptical galaxies. Thus, when fitting for the alignment parameter $D$, following the same procedure as in Fig.~\ref{fig_D_z0}, but at $z=1$, we use only 20 bins in $T_{+,\times}$, with $\sim 50$ galaxies per bin. 

Correcting for the bias introduced by anisotropies in the large-scale structure of TNG100, we find at $z=1$ an alignment parameter
\begin{equation}
D_V = (2.43 \pm 0.36) \times 10^{-6} c^2,
\end{equation}
shown as solid black line in Fig.~\ref{fig_D_z1}. This measurement is incompatible with zero at a level of about $7\sigma$, making the alignment effect highly significant. The measured alignment parameter at $z = 1$ is about 3.7 times larger compared to the value measured at $z = 0$. The raw alignment parameters are $D_{+,V}= ( 1.58 \pm 0.43) \times 10^{-6} c^2$ and $D_{\times,V}= ( 3.87 \pm 0.64) \times 10^{-6} c^2$, which are inconsistent with one another but average to $D_{V}= (2.73 \pm 0.39) \times 10^{-6} c^2$, which is consistent with the anisotropy corrected measurement. 

The value for the alignment parameter $D$ of order $\simeq 10^{-6}c^2$ is significantly smaller than the one obtained by \citet{Tugendhat18} with a value close to $D\simeq 10^{-4}$, and even though the fundamental models in both cases are identical, the way of determining the numerical value differs subtantially and merits a detailed comment. While we are determining $D$ as the proportionality constant in the linear relation between ellipticity and tidal field, \citet{Tugendhat18} estimate the value of $D$ by setting up a model for angular ellipticity correlations and by determining the amplitude such that the detection significance of intrinsic alignments matches that seen in CFHTLenS-data. To be exact, with the lensing signal dominating, the estimate of the alignment parameter originates from a fit to the $GI$-cross correlations, with some amount of cancellation due to the $II$-terms because of their opposite sign. Thereby, working with the lensing survey \citet{Tugendhat18} employ a mass scale of $10^{12}M_\odot/h$, while in our analysis we select all galaxies above a lower mass cut at $10^{10}M_\odot$, and it is in fact the difference in mass-cutoff of the spectrum that biases the alignment parameter $D$ towards higher values in analysing CFHTLenS-data. A mass-cutoff on a scale $M_\mathrm{cut}$ would correspond to an angular scale $R_\mathrm{cut}$ determined by $4\pi/3\rho_\mathrm{crit}\Omega_mR^3 = M$. That this issue is important can be seen with the following rough estimate. $M_\mathrm{cut}\sim 10^{12}M_\odot/h$ corresponds to a comoving scale of $R\sim 1\mathrm{Mpc}/h$. At a comoving distance of $\chi\sim 1\mathrm{Gpc}/h$ the subtended angle is then about 3 arcmin, equivalent to a multipole of $\ell = \chi/R = 10^3$, and because CFHTLenS probes ellipticity correlations on angular scales of between a few arcmin up to 100 arcmin, the choice of the cutoff scale matters. If the galaxies in the lensing survey sample are not limited to the same mass threshold, the model prediction falls short of the actual signal, which needs to be compensated by a higher value of $D$, biasing the fit high.

For computing the influence of a cutoff in the spectrum $P(k)$ which enters the prediction of the ellipticity correlation $C(\ell)$ through the Limber-equation,
\begin{equation}
C(\ell) = 
\int\frac{\mathrm{d}\chi}{\chi^2}\:W(\chi)^2P(k=\ell/\chi) = 
\frac{1}{\ell}\int\mathrm{d}k\:W(\ell/k)^2P(k)
\end{equation}
where the integration variable was switched from $\chi$ to $k$ in the last step. Replacing the cutoff in $P(k)$, which was chosen to be Gaussian in \citet{Tugendhat18}, by a step function shows that the contribution to the ellipticity correlation at high $\ell$ which is missed by the analytical model, can be determined to be
\begin{equation}
\Delta C(\ell) = \frac{1}{\ell}\int_{k_\mathrm{cut}}^\infty\mathrm{d}k\:W(\ell/k)^2P(k),
\end{equation}
and scales proporational to $M^{2/3\ldots 1/3}$, by the following argument. If on small scales $P(k)$ is approximated by a power law $P(k)\propto k^{-2\ldots -3}$ and if the weighting function is taken to be slowly varying, the integral evaluates to $k^{-2\ldots -3}$, and replacing $k$ using its inverse relation to spatial scale $R \sim 1/k$ and linking it to $M \sim R^3 \sim k^{-3}$ yields the final result. With the mass ratio of $100$ one then should expect a change in $D$ by a factor of $100^{2/3}$, which is about one and a half order of magnitude, where an actual variation in the weighting function would exacerbate things.

Additionally, due to the marginal detection in CFHTLenS, one can expect a Malmquist-biased value for the alignment parameter, yielding values that are biased high with respect to the real value, as well as uncertainties concerning the morphological mix of galaxies when predicting the angular spectrum and uncertainties in relation to the galaxy biasing model, as the ellipticity field is only sampled at positions, where elliptical galaxies reside, making a direct comparison between the two results difficult.

Furthermore, we find $D$ to be significantly larger at $z=1$ than at $z=0$. The trend of an increasing alignment strength with redshift is also observed by \citet{Tenneti15} for the non-linear alignment model in the {\sc MassiveBlack-II} simulation, as well as recently by \citet{Samuroff20} in both {\sc MassiveBlack-II} and the TNG300 simulation for tidal shearing induced alignments in their new framework. Furthermore, \cite{Yao20a} obtained a significant IA signal from observations of $\sim 23 \times 10^6$ galaxies from the Dark Enegy Camera Legacy Survey (DECaLS, DR3) in the reshift range $0.1\lesssim z\lesssim 0.9$, and similarly find an increasing alignment signal of elliptical galaxies with increasing redshift. The linear (and non-linear) alignment model do not have any explicit redshift dependence, however, the measured alignment parameter $D$ depends implicitly on several galaxies properties and other quantities. We discuss the emerging scaling behaviour at the end of the next section.

\subsection{Dependence on galaxy and environmental properties} \label{sect_e_props}
In the previous sections, we have provided single values for the linear alignment parameter $D$ for the entire sample of well-resolved TNG100 ellipticals at $z=0$ and $z=1$. The value of $D$ however ultimately depends on properties of the observed galaxy sample, as quantified through the survey selection function, as well as on environmental properties. In this section, we therefore investigate the dependence of the alignment parameter $D$ on galaxy and host halo mass, environment, smoothing scale, and ultimately redshift. 

\begin{figure}
\centering
\includegraphics[width=0.48\textwidth,trim= 10 10 0 5,clip]{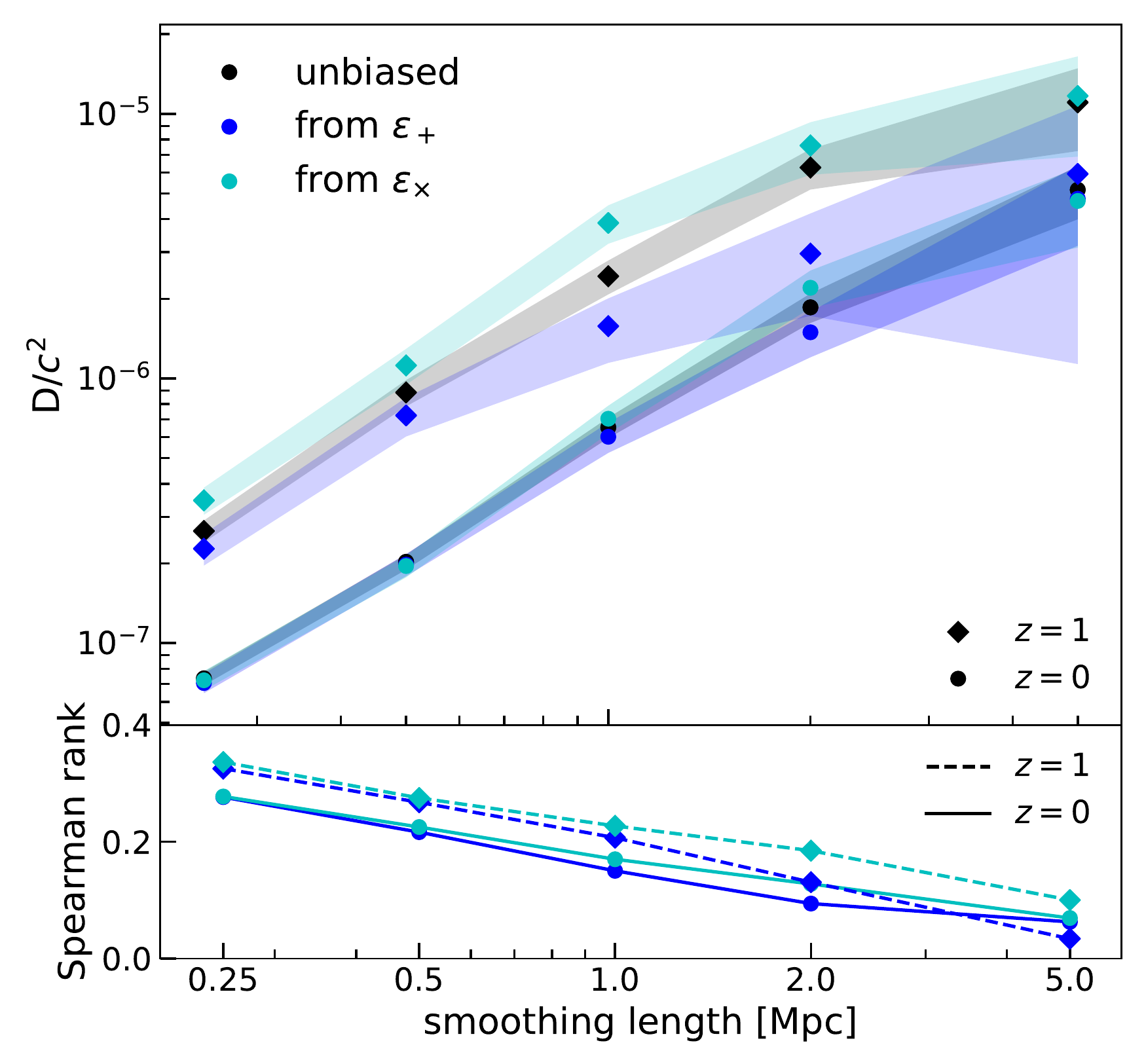}
\caption{\textit{Top panel}: Intrinsic alignment strength $D$ as a function of the tidal field smoothing scale at $z=0$ and $1$, with the standard error shown as shaded area. In black we show the anisotropy corrected values, and for comparison in blue and cyan the direct measurments from the real and imaginary part of the galaxy ellipticity in TNG100. \textit{Bottom panel}: Spearman's rank coefficient for the correlation between galaxy ellipticity and tidal field at given smoothing scale, derived from each elliptical galaxy taken as an individual data point. The best correlation can be found at the smallest smoothing scale.}
\label{fig_D_smooth}
\end{figure}

\paragraph*{Dependence on smoothing scale.} \ In the top panel of Fig.~\ref{fig_D_smooth}, we show the intrinsic alignment parameter $D$ for our full sample of elliptical galaxies obtained from correlating the observed ellipticity with the tidal field smoothed on five different scales: 250 kpc, 500 kpc, 1 Mpc, 2 Mpc, and 5 Mpc. The lower limit is set by the $10^{10}M_\odot$ lower halo mass cut on our sample galaxies, whereas the upper limit corresponds to scales larger than the biggest collapsed objects. In black we show the isotropised alignment parameter, and, for comparison, in blue and cyan the raw measurements for the real and imaginary part of the ellipticity, respectively.  We show results obtained at $z=0$ and $z=1$ as circle and diamonds, respectively. The coloured areas correspond to the $1\sigma$-error on $D$.

We find that at both $z=0$ and $z=1$, the alignment parameter $D$ increases strongly with increasing smoothing length, spanning almost two orders of magnitude between the smallest and the largest displayed scales. This behaviour arrises from the fact that the amplitude of the tidal field decreases when smoothed at larger scales, which directly translates into higher $D$ values. Note, that a higher $D$ value does not imply a stronger correlation between observed ellipticity and tidal field. In order to investigate whether there is indeed a preferred scale which maximises the physical correlation, we also plot the Spearman rank correlation coefficient $r$ between $\epsilon_{+,\times}$ and $T_{+,\times}$ in the bottom panel of Fig.~\ref{fig_D_smooth}. The Spearman rank correlation describes the evidence for a monotonic but not necessarily linear relationship between $\epsilon_{+,\times}$ and $T_{+,\times}$, ranging from 1 for a perfect correlation, to $-1$ for a perfect anti-correlation. Fig.~\ref{fig_D_smooth} showes that the correlation weakens with increasing smoothing scale, such that the best correlation is given at the smallest smoothing scale we probe. The decrease in correlation with increasing smoothing scale is furthermore reflected in the increasing standard error on $D$, shown as a coloured band in the upper panel of the figure. 

As this trend could be an artefact of our galaxy sample being dominated by ellipticals just above the lower halo mass cut of $10^{10} M_\odot$ (see Fig.~\ref{fig_pdf_mass} below), which have an associated physical scale of $\sim 250$ kpc, we have reproduced Fig.~\ref{fig_D_smooth} with elliptical galaxies in narrow bins of halo mass centred at $\sim 10^{10} M_\odot$, $\sim 10^{11} M_\odot$, and $\sim 10^{12.5} M_\odot$ (not explicitly shown). For all host halo mass bins we find exactly the same behaviour as displayed in Fig.~\ref{fig_D_smooth}, with both $D$ and its error increasing with increasing smoothing scale, as well as, more importantly, the Spearman's coefficient decreasing with increasing smoothing scale. In conclusion, we find that the tidal field is most strongly correlated with the galaxy shape at small distances from the galaxy, and the correlation steadily decreases with increasing distance, as probed by the smoothing scale. Nonetheless, the correlation is non-negligible at the level of $\sim0.2$ even for smoothing scales as large as 2 Mpc. 

Finally, we note that the systematic bias due to large-scale anisotropies in the directly observed alignment signal is visibly stronger at high redshift, and originates from the same selection criteria yielding a much smaller sample at $z=1$ with a higher fraction of central galaxies, $38\%$ of all ellipticals at $z=1$ compared to $30\%$ at $z=0$, that are predominately affected by large scale anisotropies (see appendix \ref{sec:app_anisotropy_lss}).

\begin{figure}
\centering
\includegraphics[width=0.48\textwidth,trim= 10 10 0 5,clip]{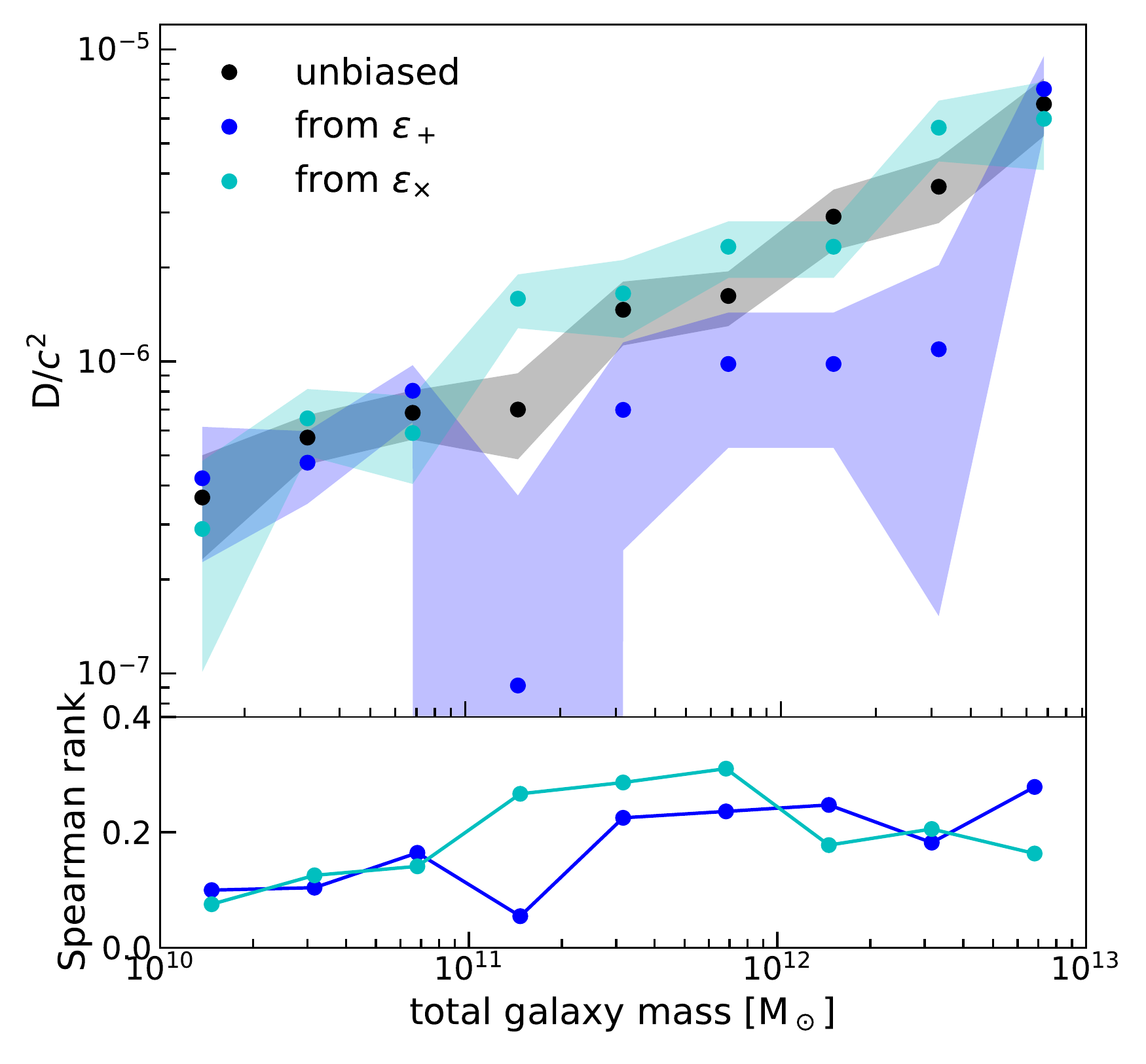}
\caption{\textit{Top panel}: Intrinsic alignment strength $D$ as a function of the total galaxy mass at $z=0$, with the standard error shown as shaded area. In blue and cyan we present the values measured directly from the real and imaginary part of the galaxy ellipticity, and in black the true, anisotropy corrected value.
\textit{Bottom panel}: Spearman's rank coefficient for the correlation between galaxy ellipticity and tidal field for elliptical galaxies at given mass. The best correlation is given for galaxies with total mass between $10^{11} M_\odot$ and $10^{12} M_\odot$.} \label{fig_D_mass}
\end{figure}

\begin{figure}
\centering
\includegraphics[width=0.48\textwidth,trim= 10 10 0 5,clip]{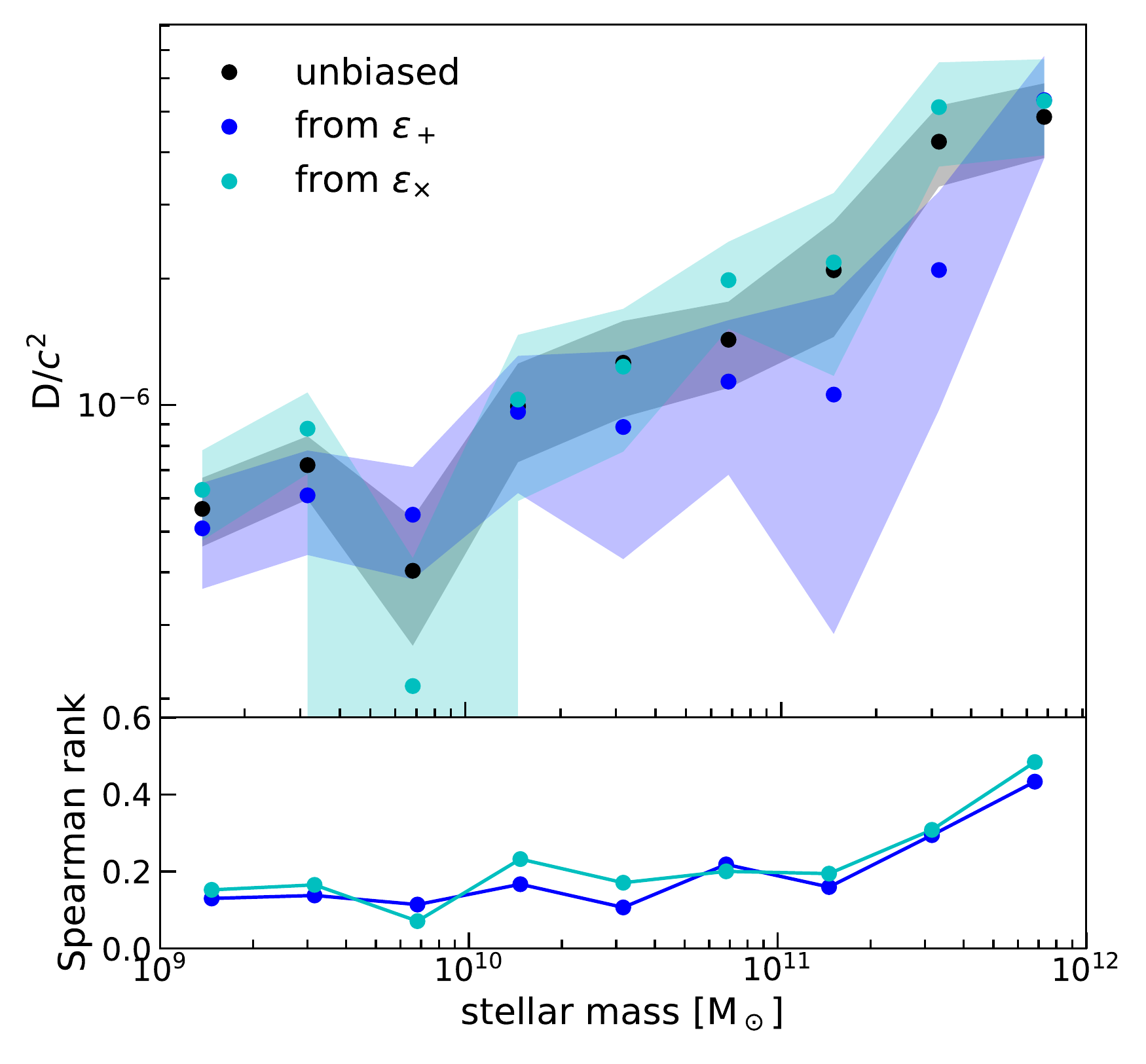}
\caption{Same as Fig.~\ref{fig_D_mass} with galaxy stellar mass instead of total mass. We find the same trend of larger alignment parameter $D$ with increasing mass, however with a much better correlation quantified by the Spearman's rank (note the larger y-axis range) for increasing stellar mass.}
\label{fig_D_stellar_mass}
\end{figure}

\paragraph*{Dependence on stellar and halo mass.} \ In Figs.~\ref{fig_D_mass} and~\ref{fig_D_stellar_mass} we show the dependence of $D$ on the total and on the stellar mass at $z=0$ in 9 logarithmically spaced mass bins in the range $10^{10} M_\odot - 10^{13} M_\odot$ for total galaxy mass, and $10^9 M_\odot - 10^{12} M_\odot$ for stellar mass. At each mass we determine $D$ based on 10 bins in tidal field strength, with the number of galaxies varying an order of magnitude between over 100 and only as few as $\sim 10$ galaxies per bin, depending on the mass. At $z=1$, the $\sim 10$ times smaller galaxy sample size does unfortunately not allow for this type of analysis in the TNG100 volume. As before, the anisotropy corrected measurement is shown in black, the raw measurements of the two ellipticity components $\epsilon_+$ and $\epsilon_\times$ in blue and cyan, shaded areas represent the $1\sigma$-error band.

For both the total galaxy mass and the stellar mass we find an increase in the value of $D$ as a function of increasing mass. This is in fact related to an increased alignment. A more massive halo or galaxy is associated with a stronger tidal field in its vicinity (at fixed smoothing), which cannot explain an increase in $D$ with increasing mass. This is also reflected in the Spearman rank correlation analysis shown in the bottom panels of Fig.~\ref{fig_D_mass} and \ref{fig_D_stellar_mass}, which clearly indicate a highly significant increase in correlation with increasing mass. These results are in perfect agreement with \citet{Chisari15} who study the correlation between the minor axis and separation vector of massive velocity dispersion supported galaxies, and find a larger negative correlation coefficient with increasing galaxy mass. When derived as a function of stellar mass the correlation is particularly high at the largest masses (note the different mass ranges in Figs.~\ref{fig_D_mass} and~\ref{fig_D_stellar_mass}), whereas in case of total galaxy mass the best correlation is found at masses between $10^{11} M_\odot$ and $10^{12} M_\odot$. These results are qualitatively consistent with previous studies by \citet{Tenneti15} who find a strong increase of the IA signal at $z=0.3$ with increasing mass for central galaxies (measured from the non-linear alignment model two-point correlation function, and by \citet{Velliscig15ias} who find a stronger correlation in the orientations of galaxy pairs derived from their 3D stellar distribution with increasing mass, that is also a decreasing function of galaxy separation, whereas however no morphological selection has been applied.

Note, that in Fig.~\ref{fig_D_mass} the raw measurements of $D$ closely resemble the anisotropy corrected value in the lowest three mass bins that are predominately populated by satellite ellipticals (compare to Fig.~\ref{fig_cntsat}). Only with a transition to higher masses the discrepancy between direct observation and corrected values becomes prominent, highlighting massive central ellipticals being mostly affected by the large-scale anisotropies in the TNG100 volume (see appendix~\ref{sec:app_anisotropy_lss}).

\begin{figure}
\centering
\includegraphics[width=0.48\textwidth,trim= 10 10 0 5,clip]{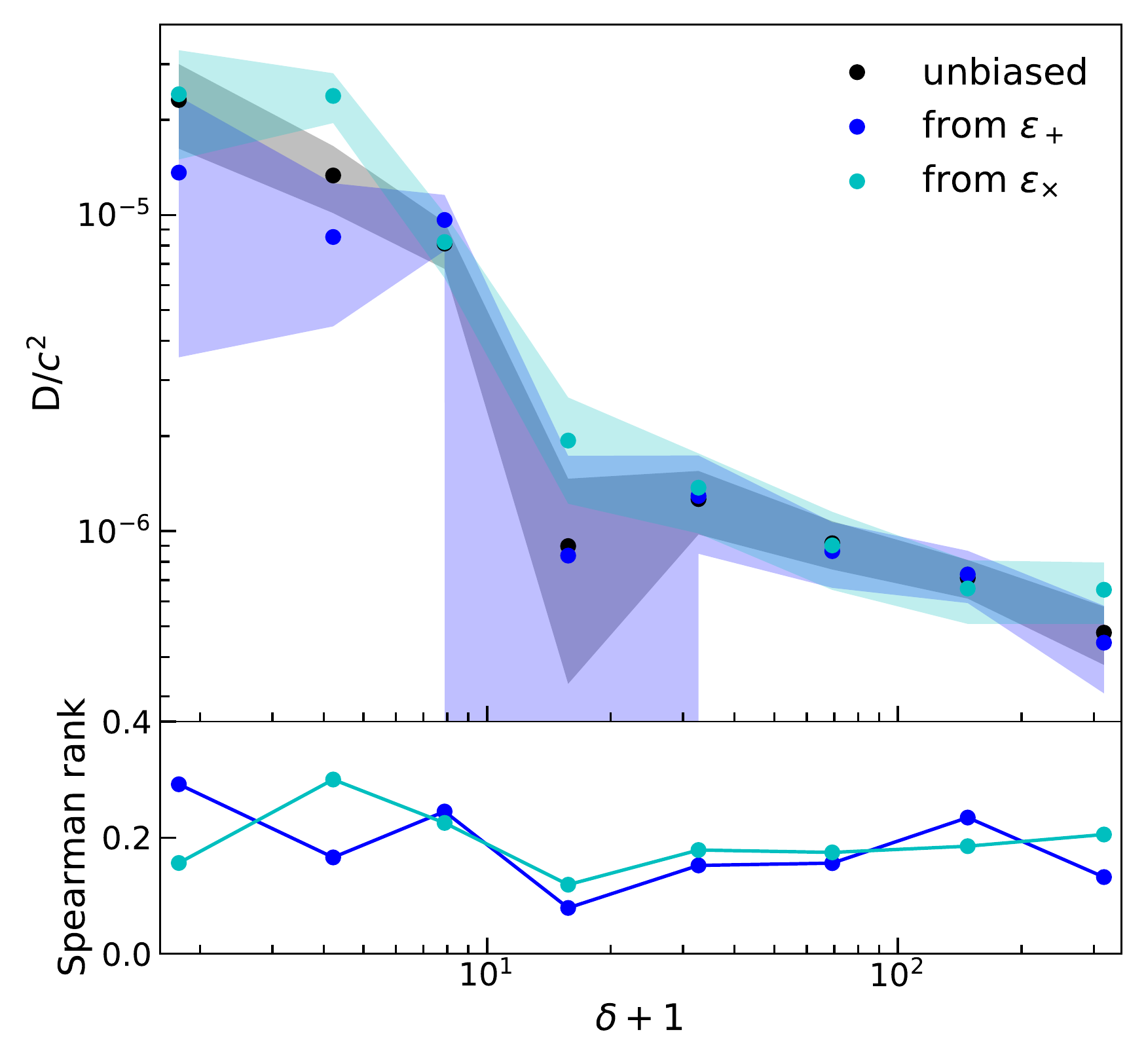}
\caption{\textit{Top panel}: Intrinsic alignment strength $D$ as a function of the local overdensity $\delta$ of elliptical galaxies at $z=0$, derived at a smoothing scale of 1~Mpc, with the $1\sigma$-error shown as shaded area. Black represents the anisotropy corrected values, while blue and cyan represent the raw measurements from the real and imaginary ellipticities in TNG100. \textit{Bottom panel}: Spearman's rank coefficient for the correlation between galaxy ellipticity and tidal field for subsamples of our elliptical galaxies at given overdensity $\delta$. While the alignment strength drops by almost two orders of magnitude with increasing $\delta$, the correlation strength is independent of environment.}
\label{fig_D_env}
\end{figure}

\paragraph*{Dependence on environmental density.} \ Finally, we investigate the dependence of alignment on environmental density. This is in some sense an indirect probe of the relative alignment of centrals and satellites (see Fig.~\ref{fig_cntsat}). In Fig.~\ref{fig_D_env} we show $D$ as a function of environmental overdensity $\delta$, measured on scales of 1~Mpc to match the tidal field smoothing scale. We bin all elliptical galaxies at $z=0$ in 8 logarithmic bins in $\delta$, ranging from $\delta + 1 = 0$ to $\delta + 1 \approx 450$. In each overdensity bin, we measure $D$ based on 10 bins in tidal field strength. Again, results are only shown for $z=0$, as the sample size at $z=1$ is too small for this type of analysis. Isotropised values of $D$ are shown in black, while cyan and blue are used for the raw measurements, the shaded area indicates the $1\sigma$-error.

The alignment strength $D$ exhibits particularly large values in underdense regions, and otherwise decreases with increasing overdensity. Naively, one might expect $D$ to increase with overdensity, equally to its increase with mass, as more massive galaxies reside in larger overdensities. However, hierarchical merging leads to a greater number of low mass satellites being accreted towards large overdensities (see Fig.~\ref{fig_cntsat}). Satellites outnumber massive galaxies in overdense environments and as such determine the measured alignment strength, leading to the observed trend. The variation of alignment strength across almost two orders of magnitude depending on environment is important to bear in mind when extracting the weak lensing signal from galaxies observed in the field versus in cluster environments.
In the bottom panel of Fig.~\ref{fig_D_env} we again show the Spearman's rank coefficient derived taking each galaxy in a given overdensity bin as an indvidual data point. We do not find any trend of the Spearman's rank with overdensity. This indicates that the change in $D$ as a function of $\delta$ is due to the change in amplitude of the tidal field alone.

\paragraph*{Scaling.} \ There are strong trends of the alignment parameter $D$ with total as well as stellar mass, and with properties of the environment such as tidal field smoothing scale and local overdensity, which the tidal alignment model has to be able to explain. As the ellipticity is a quantity of order unity, the scaling has to be generated by either internal properties of the galaxy such as its size and velocity dispersion, or by external properties such as the scale-dependence of the tidal gravitational field. Interestingly, the first two suspected variables cancel each other’s influence on the alignment parameter $D$, as shown by \citet{Piras18}: In the linear alignment model the ellipticity $\epsilon$ scales with $\partial^2\Phi/\sigma^2\cdot R^2$ (see section~\ref{sect_e_theory}), as a consequence of the quadrupolar term in the external gravitational potential. In virial equilibrium, the velocity dispersion $\sigma^2$ is proportional to the ratio $M/R$ between mass $M$ and size $R$ of the galaxy, such that $\sigma^2\sim R^2$ given $M\sim R^3$. With the ratio $R^2/\sigma^2$ being constant one recovers a direct proportionality between ellipticity and tidal field strength. Because the tidal shear field $\partial^2\Phi$ has the same statistical properties as $\Delta\Phi$, and by virtue of the Poisson-equation the same statistics as the overdensity field $\delta$, one can trace the scaling of the tidal shear to that of the density field itself.

Typical values for the variance of the tidal shear field can be computed as
\begin{equation}
\sigma^2_{\partial^2\Phi} = \int_0^{k_\mathrm{max}}\frac{k^2\dd k}{2\pi^2}\:P(k),
\end{equation}
where for simplicity of the argument we use a sharp cut-off in Fourier-space. The spectrum $P(k)$ can be approximated on small scales, relevant in our case, by a power laws $k^{-1\ldots-2}$, which yields a scaling of the variance $\sigma^2_{\partial^2\Phi} \sim k^{1\ldots 2}$, and typical tidal field values (obtained from the square root of the variance) to be proportional to $k^{0.5\ldots 1}$. Thus tidal field values exhibit an inverse scaling proportional to $\lambda^{-1\ldots -0.5}$ with physical scale $\lambda\sim 1/k$, which causes he alignment parameter $D$ to scale with $\lambda^{0.5\ldots 1}$ due to the inverse proportionality of ellipticity with tidal shear field. Correspondingly, we observe in Fig.~\ref{fig_D_smooth} an increase of the alignment parameter $D$ with smoothing scale $\lambda_s$ by a power law with index $\simeq 1$. The physical effect of smoothing is to wash out structures in the density field and consequently in the tidal shear field below a characteristic scale $\lambda_s$, such that larger values of the alignment parameter are needed to explain the same values for the ellipticity.
 
Furthermore, if the physical scale $\lambda$ is set equal to the galaxy size $R$, which is related to the galaxy mass by $M\sim R^3$, we obtain variances $\sigma^2_{\partial^2\Phi}$ which scale with $M^{-2/3\ldots -1/3}$ and consequently typical values of the tidal shear in the range $M^{-1/3\ldots -1/6}$. This results in an expected scaling of the alignment parameter $D$ with mass as $\propto M^{1/6\ldots 1/3}$. Figs.~\ref{fig_D_mass} and~\ref{fig_D_stellar_mass} show that indeed the alignment parameter $D$ increases with galaxy mass, both total and stellar, in an approximate power law with slope of $+1/3$. Such mass scaling can also be used to partially explain the observed redshift evolution of the alignment parameter. In Fig.~\ref{fig_pdf_mass}, we show the probability distribution function of the galaxy stellar mass of our sample of elliptical galaxies at $z=0$ and $1$, which can be viewed as a projection of Fig.~\ref{fig_cntsat} along the $x$-axis, such that the two peaks at low and high stellar masses can be identified to originate from the satellite and central ellipticals. Applying the same selection criteria at both redshifts results in the ellipitical galaxy sample at $z=1$ having on average a higher galaxy mass. The vertical dashed lines indicate the median stellar mass at each redshift, being $\sim 5\times 10^9 M_\odot$ at $z=0$ and $\sim 3\times 10^{10} M_\odot$ at $z=1$. This corresponds to a ratio of $\sim 6$ in median stellar mass, which given a scaling of $D\propto M^{1/3}$ is expected to yield a factor $\sim 2$ higher alignment parameter $D$ at $z=1$ compared to $z=0$ for our full samples. We measure, however, a factor $\sim 3.7$ higher value of $D$ at $z=1$, such that mass scaling is responsible for about half of the increase. A more profound understanding of the redshift evolution requires a detailed study of the dependecies of $D$ on further properties, such as the concentration $c$ (see section~\ref{sect_e_theory}) of the galaxy that will impact the responsivity of stellar dynamics to changes in the surrounding tidal field. We defer such analysis to future work.

Finally, Fig.~\ref{fig_D_env} shows a scaling of the alignment parameter $D$ with overdensity $\delta$ of power-law shape with an index of about $-1$. Recursing to the argument that tidal shear and density have similar scaling properties, the tidal shear in regions of low density becomes weaker, necessitating higher values for the alignment parameter. Making this argument more quantitative would use the direct inverse proportionality between tidal shear and alignment parameter to explain the decrease of $D$ by two orders of magnitude if $\delta$ increases by two orders of magnitude keeping in mind that our definition of mean overdensity at fixed smoothing scale of $1~\mathrm{Mpc}$ was applied.

\begin{figure}
\centering
\includegraphics[width=0.48\textwidth,trim= 10 15 0 10,clip]{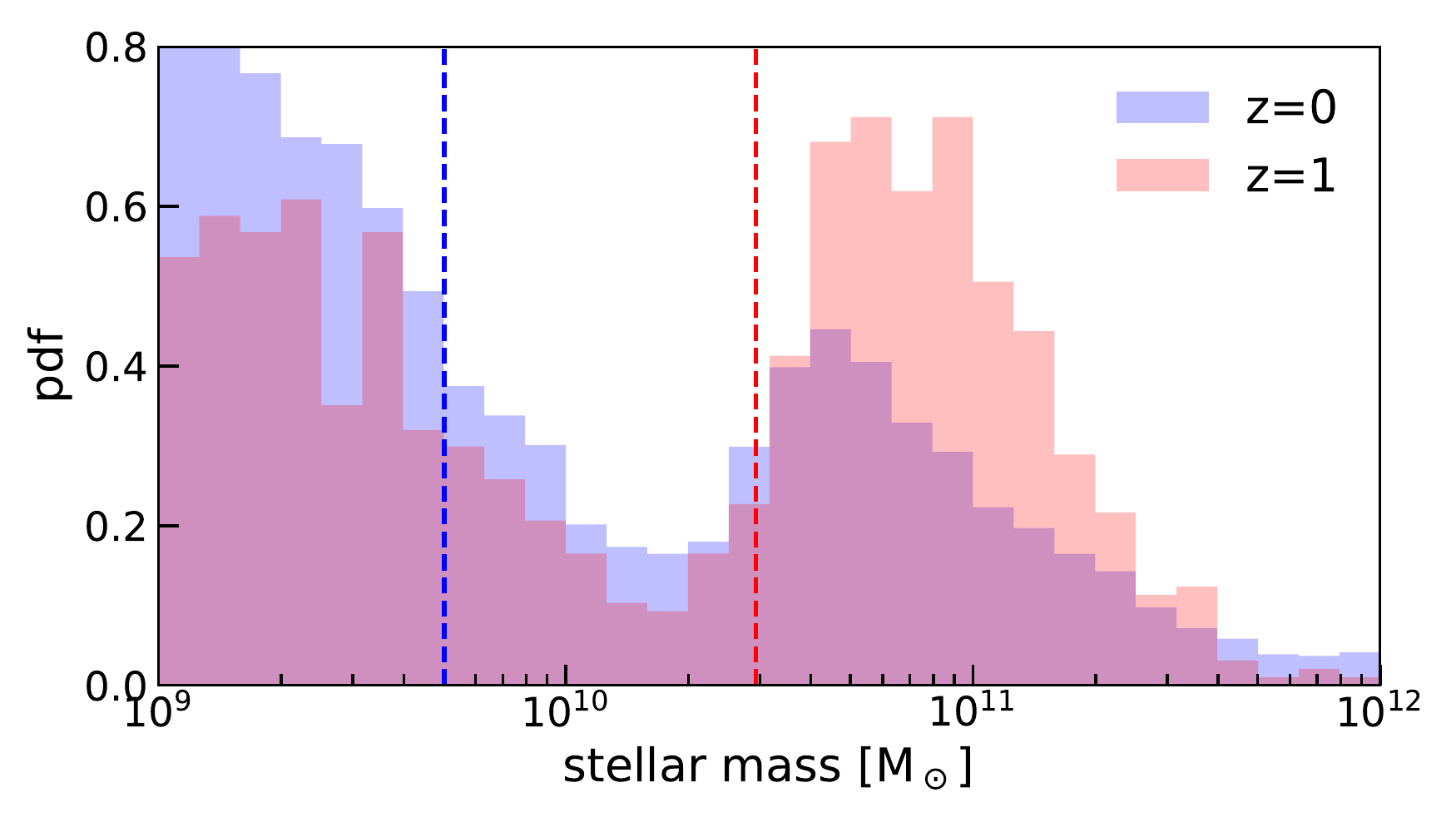}
\caption{Probability distribution function of the galaxy stellar mass for elliptical galaxies at $z=0$ and $1$. Applying the same selection criteria at both redshifts yields a higher fraction of massive galaxies at $z=1$, resulting in an overally higher alignment parameter $D$ at high redshift. Dashed lines indicate the median stellar mass of $\sim 5\times 10^9 M_\odot$ at $z=0$ and $\sim 3\times 10^{10} M_\odot$ at $z=1$. The ratio of $~6$ is expected to yield a factor $\sim 2$ higher alignment $D$ at $z=1$ compared to $z=0$, whereas we measure a $\sim 3.7$ times higher value.}
\label{fig_pdf_mass}
\end{figure}

\section{Spiral galaxies} \label{sect_s}
In this section, we examine the IA signal of spiral galaxies as predicted by the quadratic model. We measure the quadratic alignment parameter $A$ at $z=0$ and $1$, and study its scale, mass, and environment dependence. We revise the individual assumptions entering the quadratic model and compare with recent observational findings.

\begin{figure*} 
\centering
\includegraphics[width=0.49\textwidth,trim= 10 10 15 10,clip]{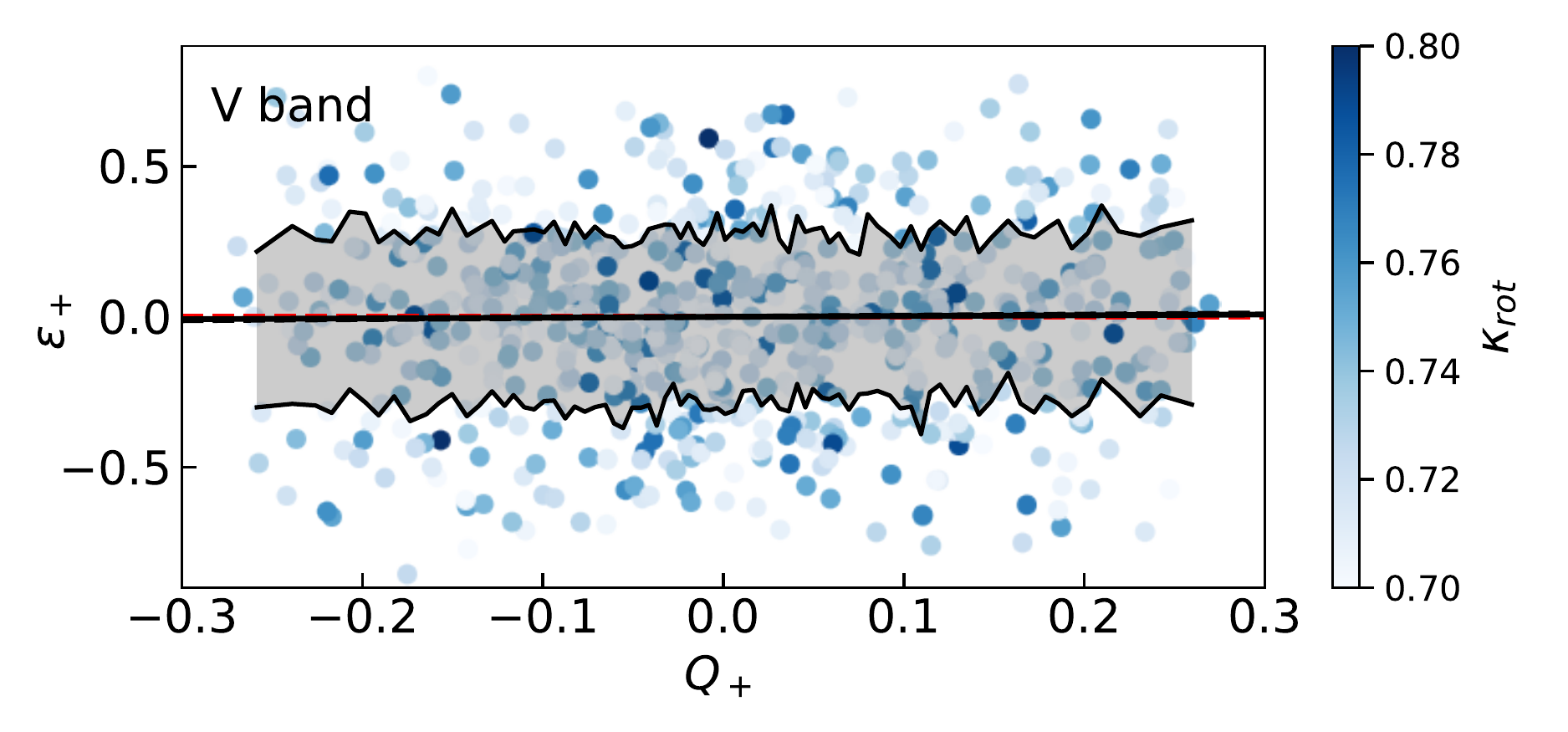}
\includegraphics[width=0.49\textwidth,trim= 10 10 15 10,clip]{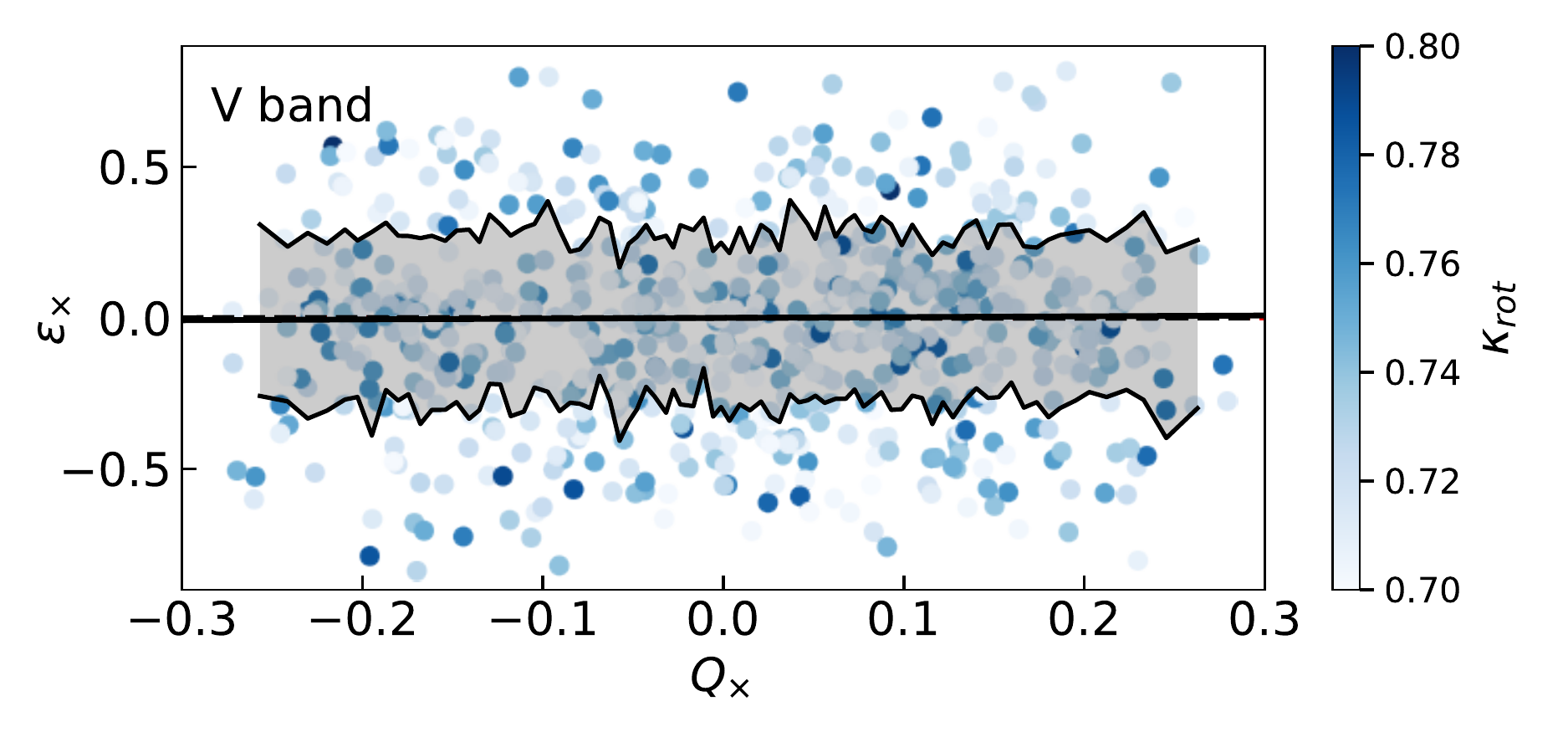}
\caption{$V$-band ellipticities $\epsilon_{+,\times}$ (\textit{left, right panel}) of spiral galaxies as a function of the respective tidal field components $Q_{+,\times}$ according to eq.~(\ref{eq_A}) at $z=0$. We show as individual dots only galaxies with high rotational support $\kappa_{\rm rot}>0.7$, representative of the full sample, colour-coded by their $\kappa_{\rm rot}$ parameter. No correlation is shown as a red dashed line, the direct fit to the observed spiral galaxy ellipticities as a black dashed, and the anisotropy corrected result as a black solid line, which are all indistinguishable. Results at $z=1$ closely resemble $z=0$ (thus not explicitly shown). The resulting alignment strength amounts to $A \simeq 0$, such that spiral galaxies in TNG100 do not exhibit a siginificant intrinsic alignment signal as predicted by the quadratic model neither at $z=0$ nor at $z=1$.} 
\label{fig_A_z0}
\end{figure*}

\subsection{Intrinsic alignments at $z=0$} \label{sect_s_z0}
To test the quadratic alignment model for spiral galaxies, we follow the same approach as for ellipticals. We measure the quadratic tidal field components $Q_+$ and $Q_\times$ given in eq.~(\ref{eq_A}) on a scale of 1~Mpc, and the observed ellipticities $\epsilon_+$ and $\epsilon_\times$ for the 7905 well-resolved spiral galaxies at $z=0$ and show the results as scatter plots in Fig.~\ref{fig_A_z0}. Due to the large sample size, we show as individual dots only spiral galaxies with $\kappa_{\rm rot}>0.7$ (and thus with the largest rotational support), representative of the full spiral galaxy sample ($\kappa_{\rm rot}>0.5$). Each galaxy is coloured by its $\kappa_{\rm rot}$ value, whereby a darker blue colour corresponds to a higher rotational support. We also show the linear regression of the alignment parameter $A_{+,\times}$ from $\epsilon_+(Q_+)$ and $\epsilon_\times(Q_\times)$, respectively, as a black dashed line. The regression was done using 80 bins in $Q_{+,\times}$, such that each bin contains approximately the same number of $\sim 100$ galaxies. The $1\sigma$-error in each bin is shown as a grey band. The null hypothesis of no correlation is indicated by a red dashed line. 

At $z=0$ we measure in the $V$-band an anisotropy corrected alignment parameter of
\begin{equation} \label{eq_A_z0}
A_{V}= 0.026 \pm 0.017,
\end{equation}
which is compatible with zero at a $1.5\sigma$ level. The parameters measured from the raw real and imaginary ellipticities are $A_{+,V} = 0.038 \pm 0.025$ and  $A_{\times,V} = 0.002 \pm 0.024$, respectively, and their average amounts to $A_{V}= 0.020 \pm 0.017$. 

Both the scatter plot and the regression yield no indication of appreciable alignment under the quadratic model. Also, the measured value is an order of magnitude smaller than the value of $A=0.237 \pm 0.023$ derived by \citet{LeePen00} from collisionless $N$-body simulations, even after accounting for a finite disc thickness that further reduces the expected alignment parameter by a factor of $\simeq 0.85$ \citep{Crittenden01} to $A \simeq 0.2$. In the underlying quadratic alignment model the correlation between ellipticity and tidal field is imprinted at high redshift during early stages of gravitational collapse \citep[at the turn-around time of the halo, cf.][]{Porciani:2002} and then is assumed to be preserved throughout cosmic time. However, non-linear evolution and astrophysical processes can be expected to have an effect in weakening any initial correlations, leading to the mismatch between the above two values for $A$. Following this logic we would expect to find a stronger correlation at higher redshift.

\subsection{Intrinsic alignments at $z=1$} \label{sect_s_z1}
In order to investigate whether spiral galaxies exhibit larger intrinsic alignments at higher redshift, we repeat our analysis with $8092$ spiral galaxies identified at $z=1$, which yield an anisotropy corrected alignment parameter of
\begin{equation} \label{eq_A_z1}
A_{V}= 0.039 \pm 0.016.
\end{equation}
This value is at $2.4\sigma$ different from zero. The raw measurements without isotropisation yield $A_{+,V} =  0.060 \pm 0.023$ and $A_{\times,V} = 0.017 \pm 0.024$, which average to $A_{V}= 0.039 \pm 0.016$, identical to the unbiased value. As naively expected the alignment strength $A$ is larger at high redshift than at $z=0$, as well as statistically less compatible with the null hypothesis, although the signal is sill weak at best. The $z=1$ results resemble very closely Fig.~\ref{fig_A_z0}, such that we refrain from showing them explicitly. For our full sample of spiral galaxies we thus do not find a significant intrinsic alignment signal neither at $z=0$ nor at $z=1$. A similar result was recently obtained from the TNG300 simulation by \citet{Samuroff20} through other means. The growing trend of both the value for $A$ as well as its significance with redshift, however, encourages a more detailed look into the dependence of the alignment parameter $A$ on additional properties of the spiral galaxy sample.

\begin{figure}
\centering
\includegraphics[width=0.48\textwidth,trim= 12 10 5 10,clip]{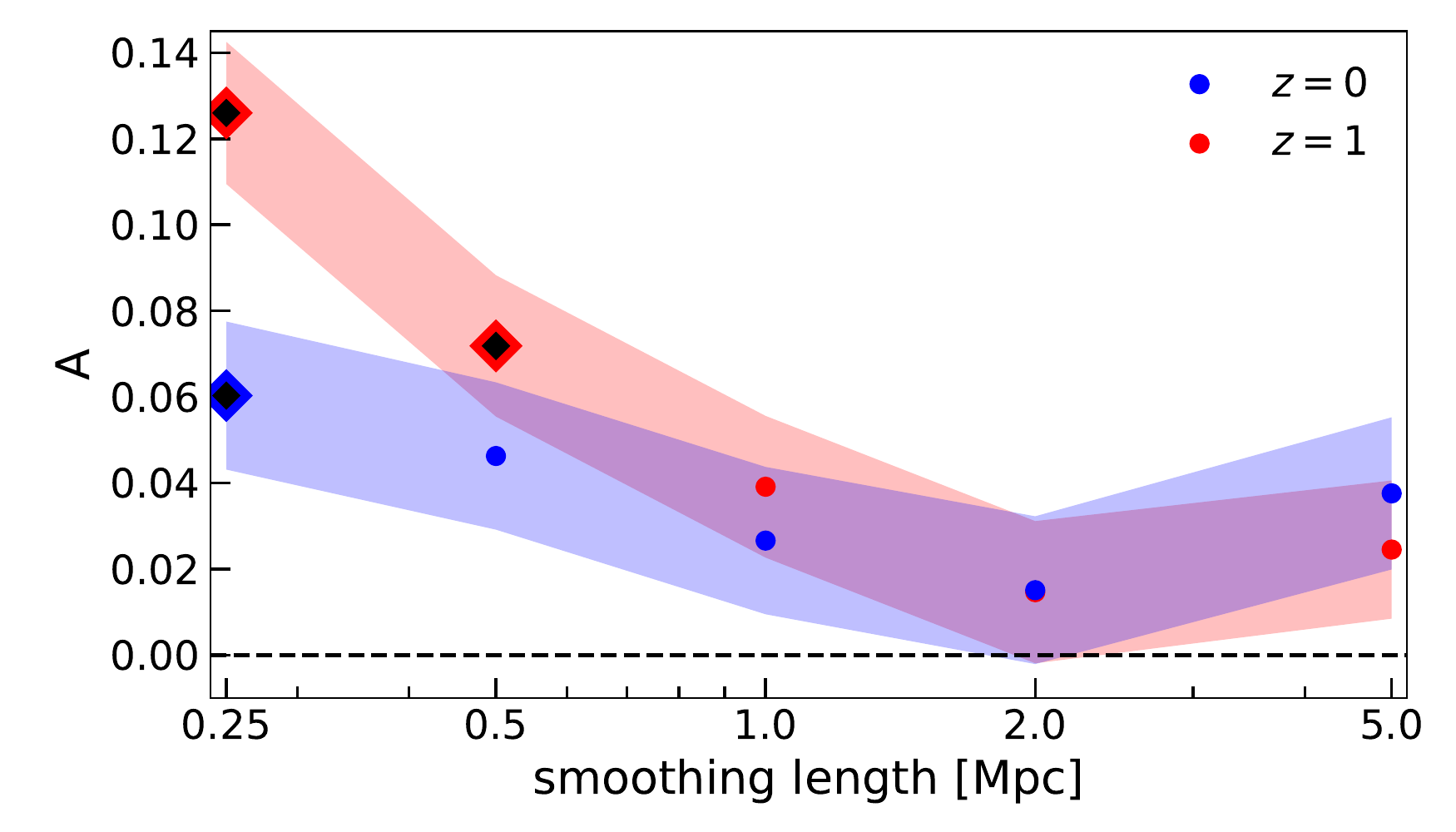}
\caption{Intrinsic alignment parameter $A$ as a function of tidal field smoothing scale at $z=0$ and $1$, with the standard error shown as shaded area. Diamond symbols highlight alignment parameters that are significantly larger than zero. No alignment is indicated by a black dashed line at $A=0$. Similarly to elliptical galaxies, we find the best correlation at the smallest smoothing scale.}
\label{fig_A_smooth}
\end{figure}

\subsection{Dependence on galaxy and environmental properties} \label{sect_s_props}
In this section, we study how the alignment parameter $A$ depends on properties of the galaxy sample at both $z=0$ and $1$. In particular, we investigate how $A$ changes as a function of average galaxy mass, host halo mass, environment, and tidal field smoothing scale which can provide insight on whether the intrinsic alignment signal is most prominently present in a specific class of spiral galaxies, and whether certain selection criteria can lead to a signal detection at a $3\sigma$ level. The large sample size of spirals at high redshift allows us to perform a detailed analysis also at $z=1$, whereas to keep figures clean we only show the anisotropy corrected values for $A$. Throughout this section, the $1\sigma$-error on $A$ is shown as a shaded area, and a black dashed line is drawn at $A=0$ to guide the eye. Alignment parameters that are significantly different from zero are highlighted by diamond symbols.

We would also like to point out that for all choices of parameters, we find very low Spearman rank coefficients below a value of $0.1$. The low Spearman rank is partially due to being derived from the directly observed, anisotropy biased data, which also applies to the values quoted previously for elliptical galaxies. It is not straight forward how to propagate the Spearman rank to the anisotropy corrected observables, only which reveal the intrinsic alignment signal, such that we refrain from showing the Spearman rank in this section.

\paragraph*{Dependence on smoothing scale.} \ We start by showing in Fig.~\ref{fig_A_smooth} the alignment parameter $A$  for the full sample of spiral galaxies at $z=0$ and $1$ as a function of the tidal field smoothing scale, which reveals a small decrease of $A$ with increasing smoothing scale, except for the largest scale of $5$~Mpc. This trend is somewhat stronger at $z=1$ than at $z=0$. A significant IA signal can be measured only at a scale of $250$~kpc at $z=0$, and of $250$~kpc and $500$~kpc at $z=1$. As the quadratic model makes use of the unit normalised tidal field values, there is no a priori dependence of the tidal field strength on smoothing scale, in contrast to the linear model for elliptical galaxies. Thus the trend in Fig.~\ref{fig_A_smooth} is indicative of the spiral disc orientation being stronger correlated with the tidal field on smallest measurable scales, and this effect being stronger at higher redshift. 

\begin{figure}
\centering
\includegraphics[width=0.49\textwidth,trim= 7 10 0 10,clip]{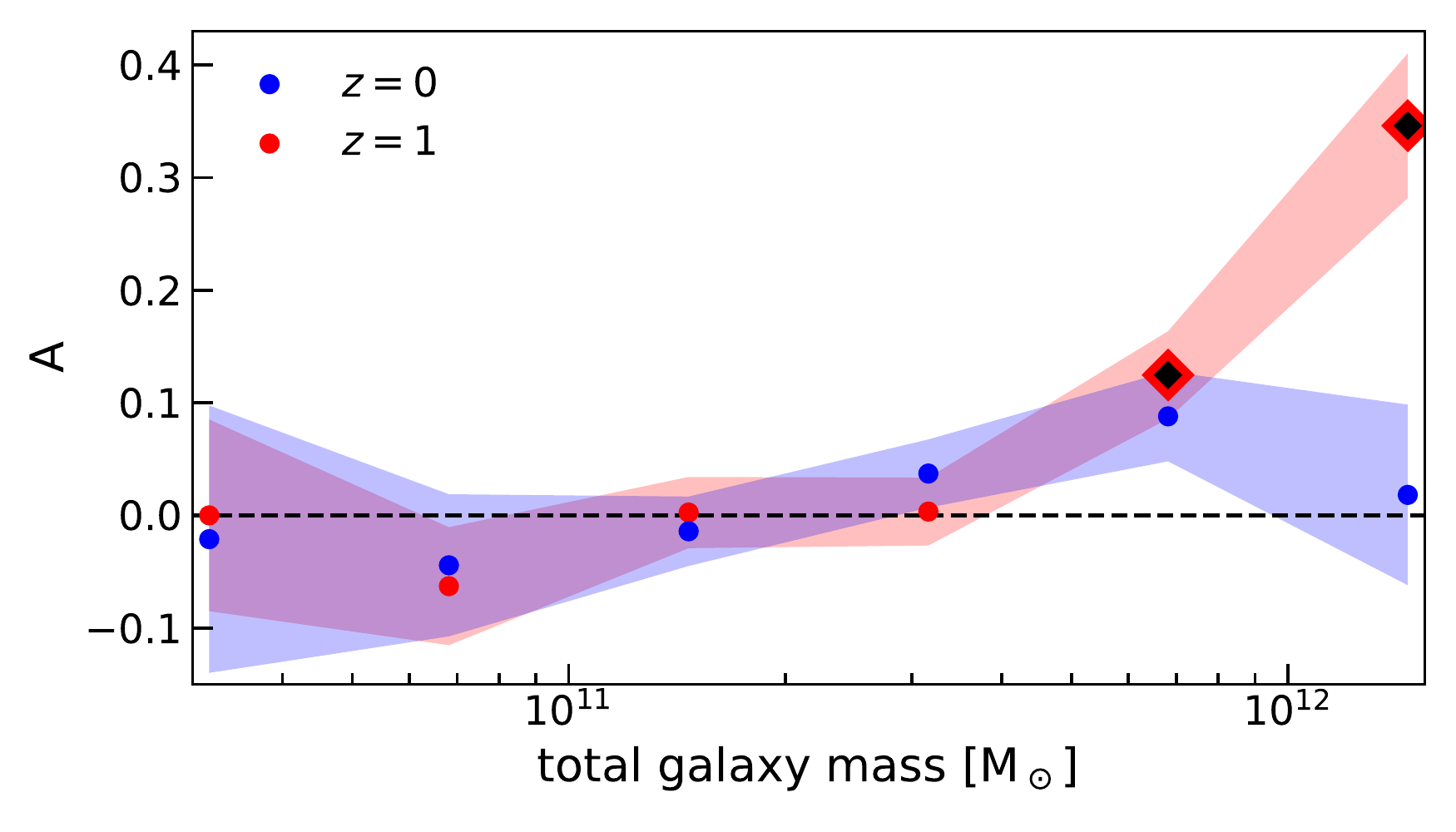}
\caption{Intrinsic alignment parameter $A$ as a function of the total galaxy mass at $z=0$ and $1$, with the standard error shown as shaded area. Diamonds highlight alignment parameters that differ from zero at more than a $3\sigma$ level. The IA signal is compatible with zero for all redshifts and masses, except for massive, $z=1$ spirals.}
\label{fig_A_mass}
\end{figure}

\begin{figure}
\centering
\includegraphics[width=0.49\textwidth,trim= 10 10 0 10,clip]{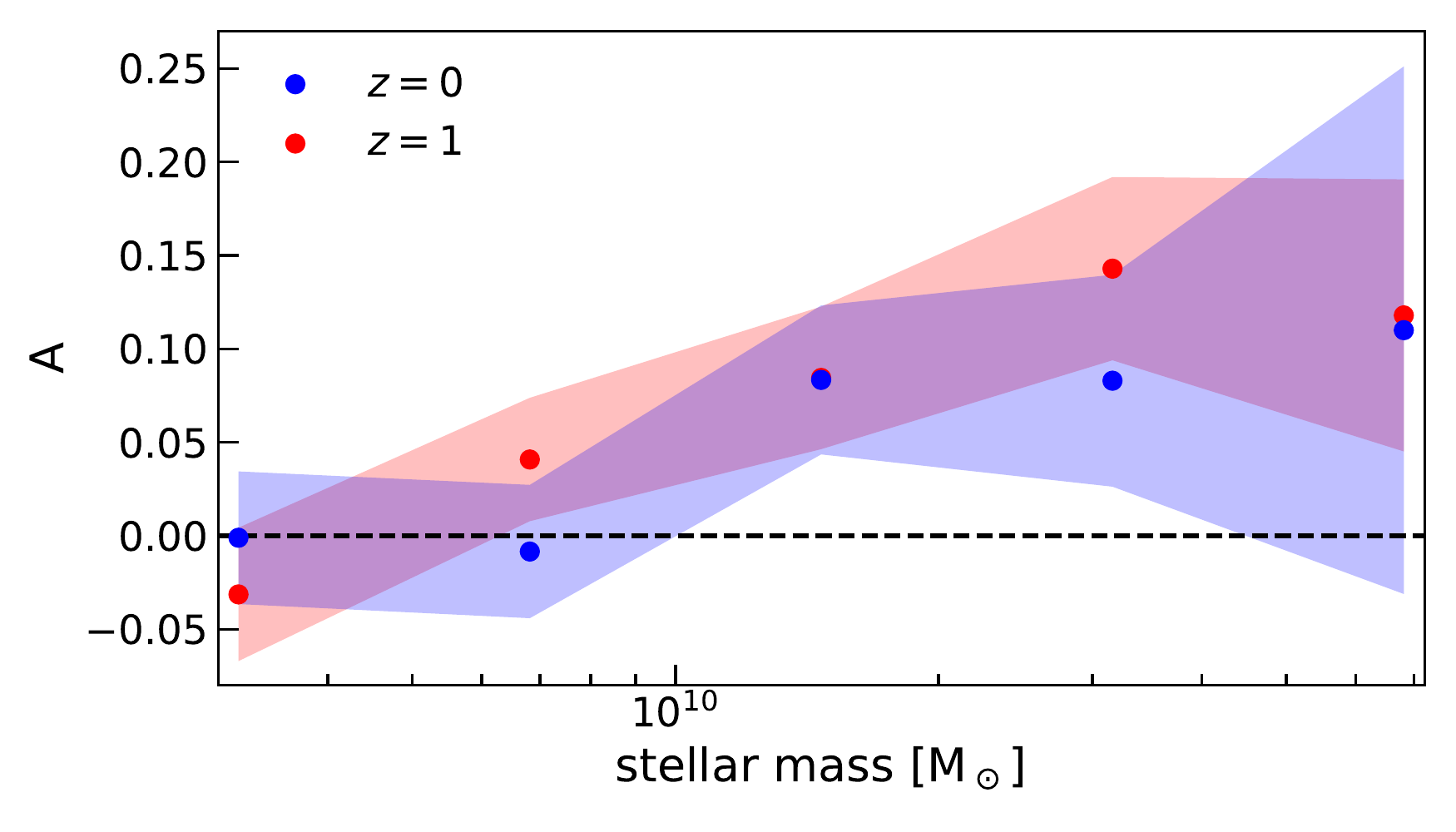}
\caption{Same as Fig.~\ref{fig_A_mass} but with galaxy stellar mass instead of total mass. In this case the alignment parameter $A$ shows a subtle increase with increasing stellar mass, however the signal remains compatible with zero for all stellar masses.}
\label{fig_A_stellar_mass}
\end{figure}

\begin{figure}
\centering
\includegraphics[width=0.48\textwidth,trim= 0 15 10 10,clip]{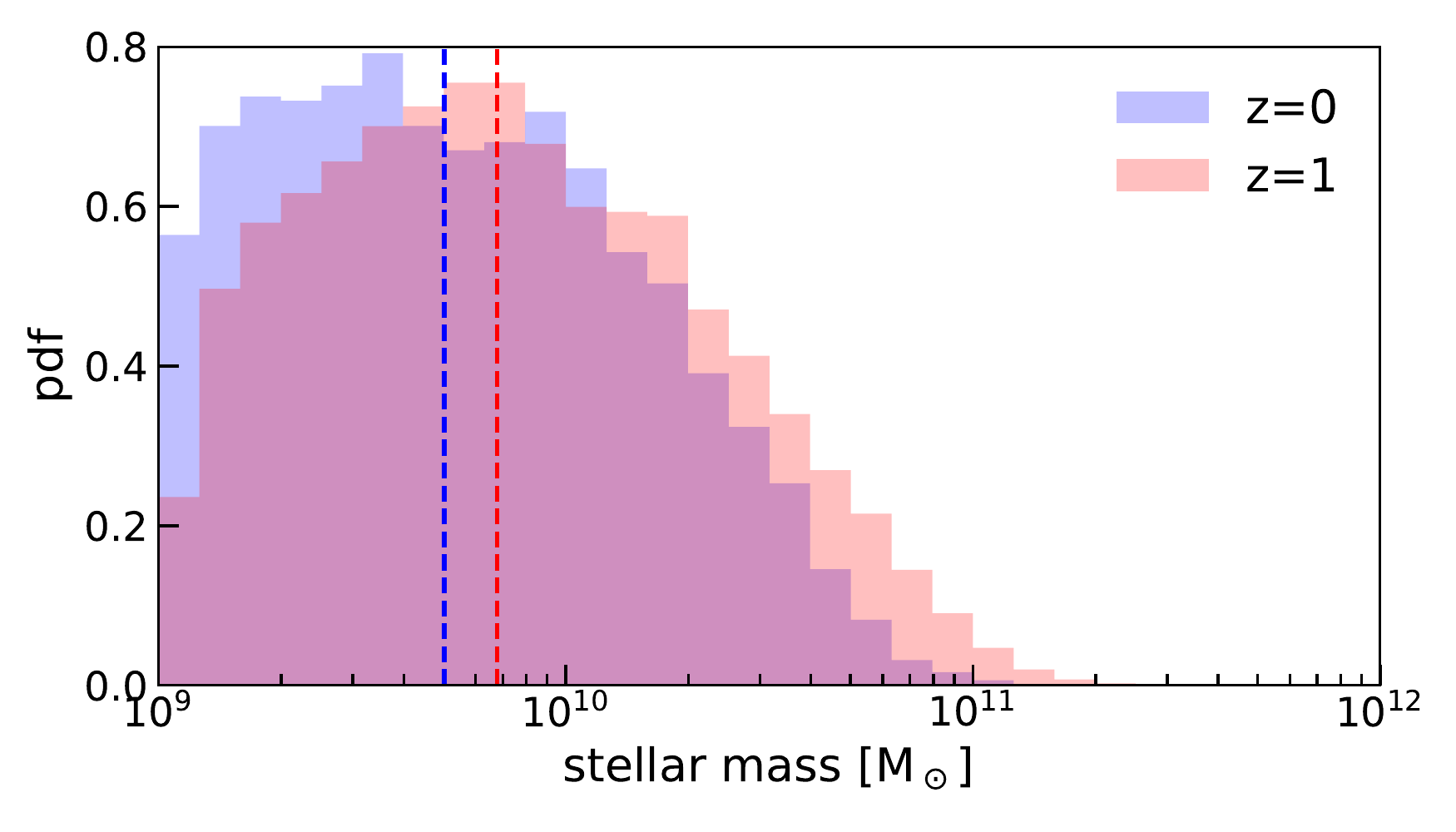}
\caption{Probabability distribution function of the galaxay stellar mass for spiral galaxies at $z=0$ and $1$. Both the distribution and the total number of spiral does not change considerably between redshifts, going from 8092 at $z=1$ to 7905 at $z=0$. This is also reflected in the median mass indicated by a dashed line for each disribution.}
\label{fig_pdf_mass_spirals}
\end{figure}

\paragraph*{Dependence on stellar and halo mass.} \ In Figs.~\ref{fig_A_mass} and~\ref{fig_A_stellar_mass}, we show how the alignment parameter $A$ depends on the total and on the stellar mass of galaxies at $z=0$ and $1$, respectively, with fixed smoothing scale at $1$~Mpc characteristic for typical Milky Way-type galaxies. In these measurements, we bin all galaxies in 9 equal mass bins in logarithmic space, in the mass range $10^{10} M_\odot - 10^{13} M_\odot$ for total galaxy mass, and $10^9 M_\odot - 10^{12} M_\odot$ for stellar mass. At each mass we derive the alignment parameter $A$ as before, binning all galaxies in 10 bins along the tidal field axis, with the number of galaxies varying by an order of magnitude from several hunderd to only a few dozen galaxies per bin, depending on the mass. We remove mass bins with less than 10 galaxies which we deem unreliable. This removes the lowest and the two highest total galaxy mass bins and the three highest stellar mass bins. The need to remove those bins can be seen best in Fig.~\ref{fig_pdf_mass_spirals}, where we show the probability distribution function for the stellar galaxy mass, with dashed lines indicating the median mass at each redshift. The mass distribution of spiral galaxies does not change considerably between the examined redshifts, similar to the total number of spiral galaxies in our sample, going from 8092 at $z=1$ to 7905 at $z=0$, which enables a fair comparison of the dependence of the alignment signal on secondary properties as a function of redshift. Note, that we have applied exactly the same sample selection criteria at both redshifts.

As expected from Fig.~\ref{fig_pdf_mass_spirals}, we do not find any redshift dependence of the alignment strength $A$ as a function of both total and stellar galaxy mass shown in Figs.~\ref{fig_A_mass} and~\ref{fig_A_stellar_mass}. When viewed as a function of total galaxy mass, however, one exception occures for the most massive spirals at $z=1$, who show a distinctive upward trend in the value of $A$ and exhibit a significant alignment signal. We have reproduced Fig.~\ref{fig_A_mass} with tidal fields smoothed at 500 kpc and 250 kpc, to investigate whether the particular mass scale above which the alignment signal at $z=1$ becomes significant is determined by measuring the tidal field strength smoothed on a scale that corresponds to the mass scale of the respective galaxies. If the significance of the signal is determined by the response of a galaxy to the tidal field on the characteristic length scale, reducing the smoothing scale would result into significant detections of an alignment signal at lower total galaxy mass. However, the choice of smoothing scale does not affect the results presented in Fig.~\ref{fig_A_mass}, such that the $3\sigma$ alignment signal for the most massive spiral galaxies at $z=1$ is not a consequence of our particular choice for how to measure the relevant properties. The intriguing question is now why does this significant signal appear for the most massive spirals at $z=1$ but not at $z=0$? We discuss this point further in the next section where we revise the quadratic alignment model. For now, we would like to point out that it is those high mass galaxies that drive the overall IA signal at $z=1$ to be $2\sigma$ above zero, and considerably larger than at $z=0$. Furthermore, this distinctive upward trend is not present when the alignment parameter is plotted as a function of stellar mass, as shown in Fig.~\ref{fig_A_stellar_mass}. In fact, for all stellar masses the alignment parameter $A$ is indistinguishable between $z=0$ and $1$, and compatible with zero, with a small but not significant increase in its value with increasing stellar mass. 

\begin{figure}
\centering
\includegraphics[width=0.49\textwidth,trim= 10 15 -15 8,clip]{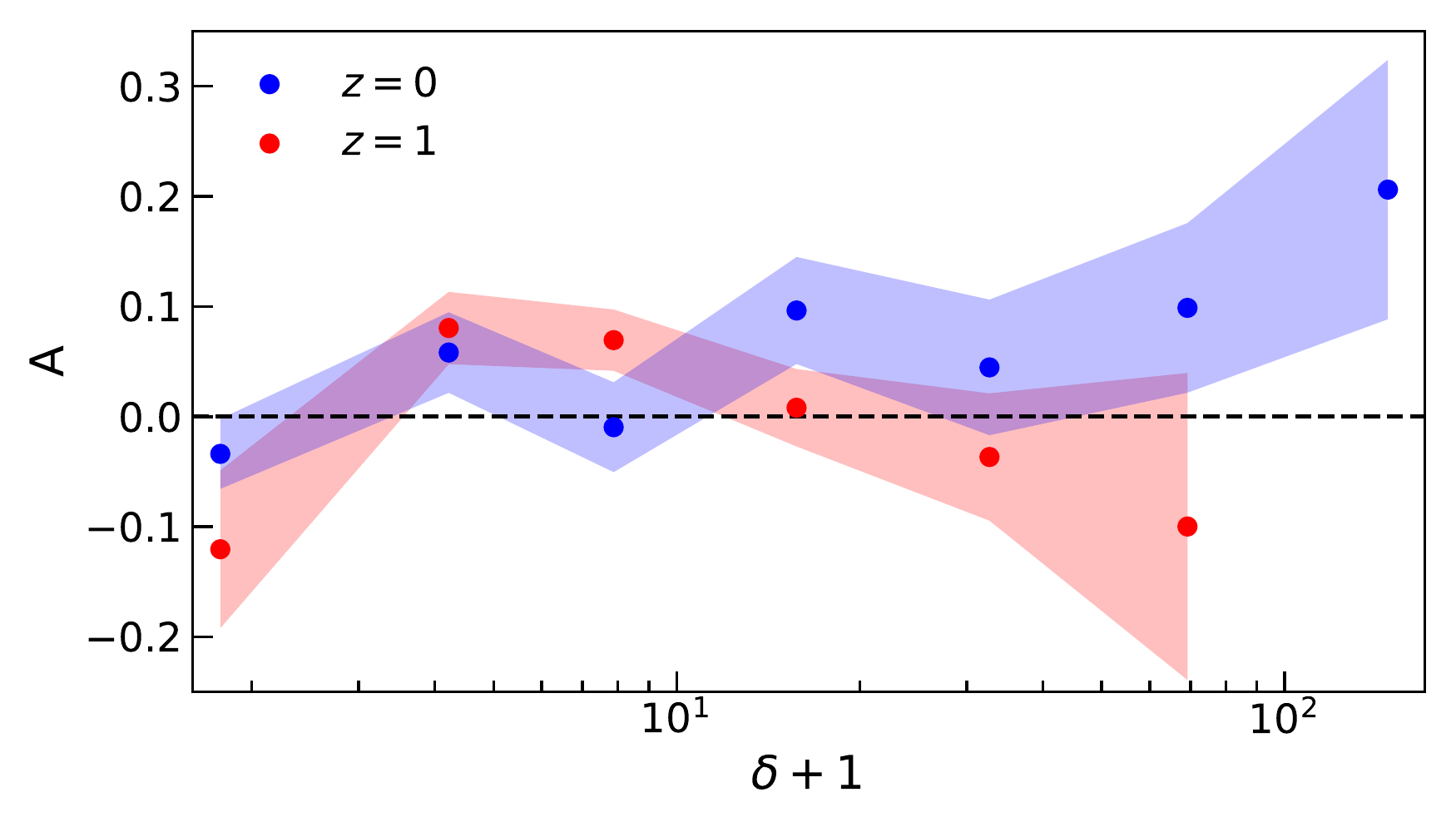}
\caption{Intrinsic alignment strength $A$ as a function of the local overdensity $\delta$, derived at a smoothing scale of 1 Mpc, at $z=0$ and $1$, with the standard error shown as shaded area. Within the accuracies of our measurement we do not see any trend of alignment parameter $A$ with overdensity.}
\label{fig_A_env}
\end{figure}

\paragraph*{Dependence on environmental density.} \ Finally, in Fig.~\ref{fig_A_env} we show the dependence of the alignment parameter $A$ on the local overdensity, smoothed at a scale of $1$~Mpc to match the tidal field smoothing. Again, we bin all spiral galaxies in 8 logarithmic bins according to their overdensity, ranging from $\delta + 1 = 0$ to $\delta + 1 \approx 450$. In each overdensity bin, we place all galaxies in 10 bins along the tidal field axis. Due to only few spiral galaxies populating very high overdensities, we dismiss the highest overdensity bin at $z=0$, and the two highest bins at $z=1$. We do not find any clear trend of intrinsic alignment strength $A$ as a function of local overdensity, but repeating this analysis for the more massive population of spirals in a larger cosmological volume could provide additional insight, since \citet{Hahn10} found potential hints for a density evolution in their simulation.

\begin{figure} 
\centering
\includegraphics[width=0.5\textwidth,trim= 10 0 0 10,clip]{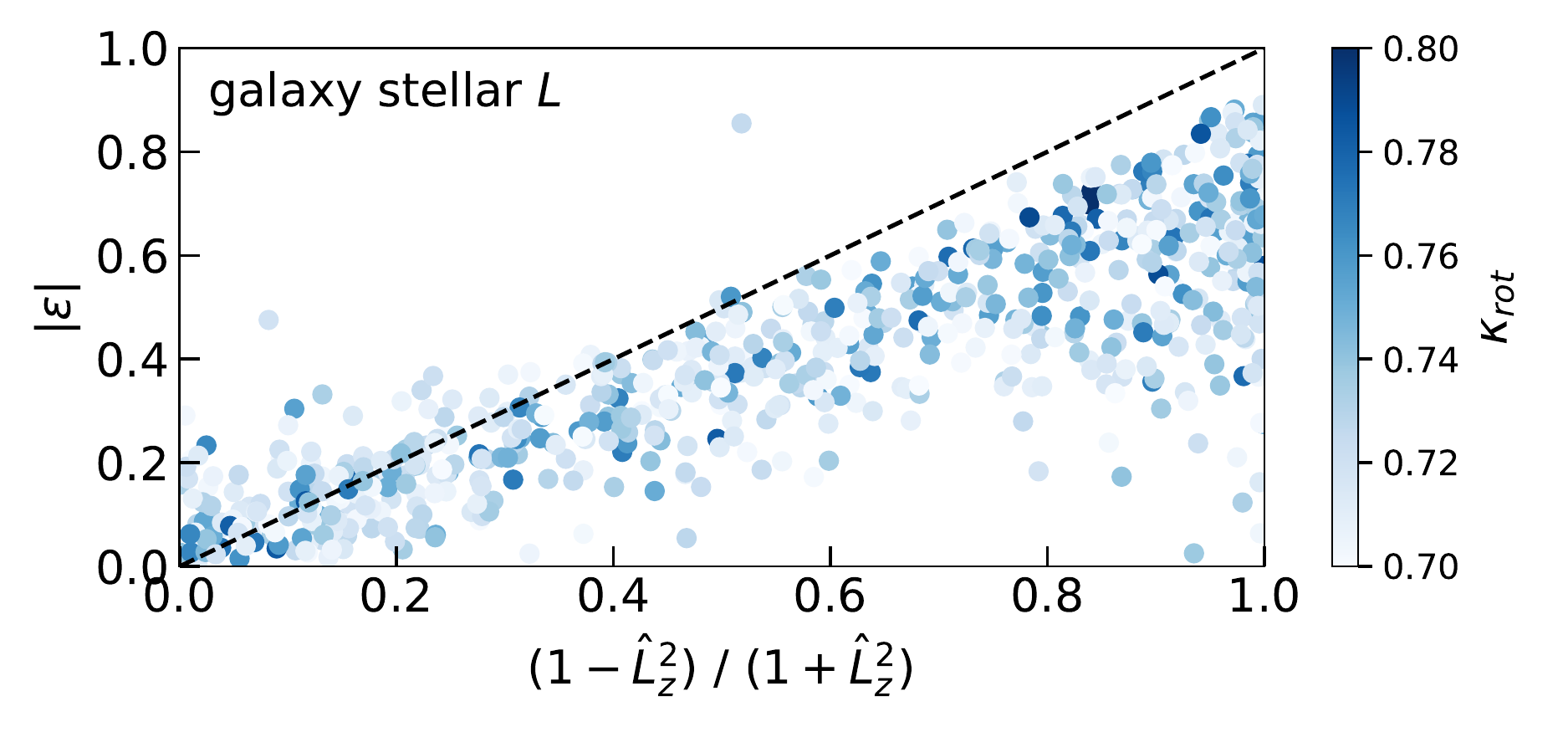}
\includegraphics[width=0.5\textwidth,trim= 10 10 0 10,clip]{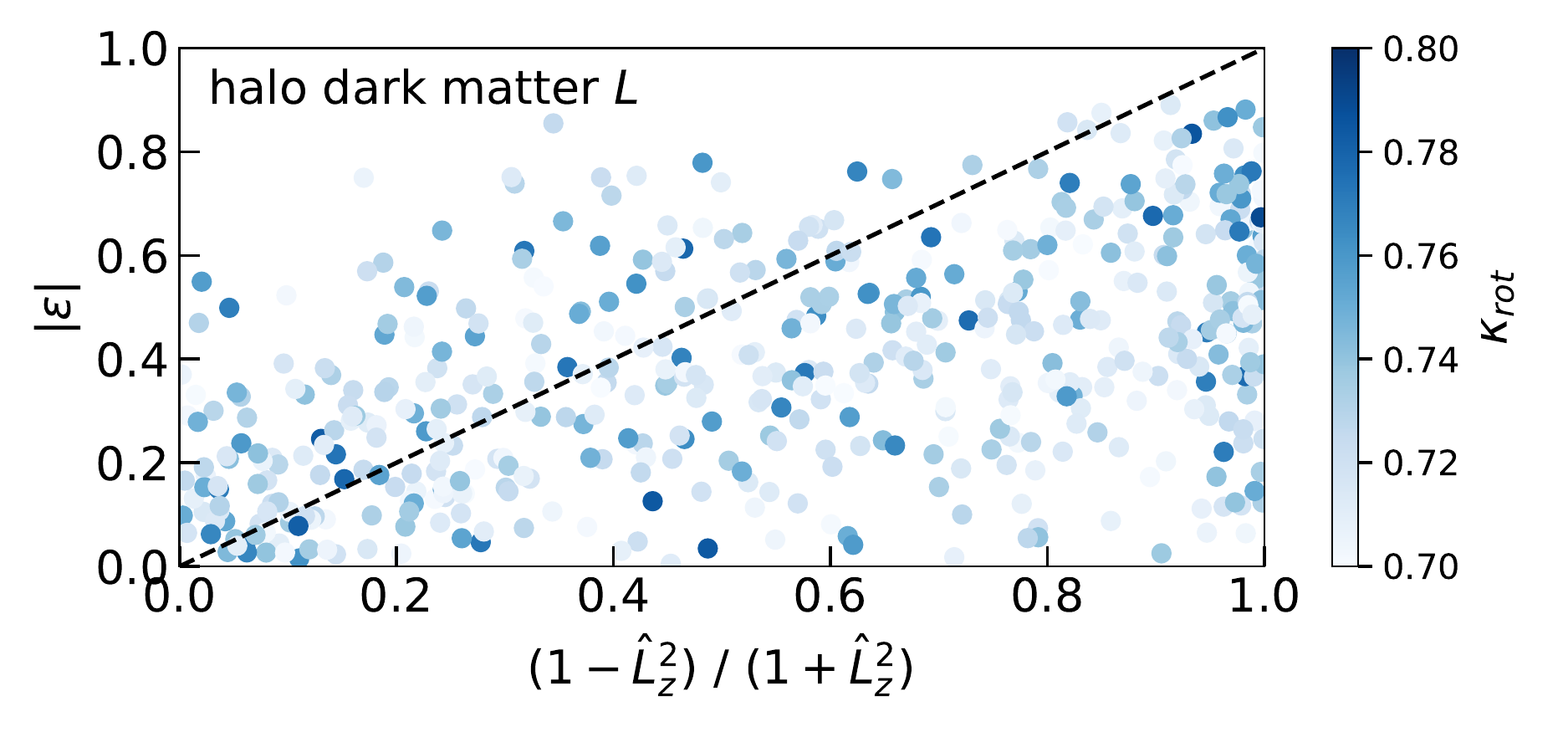}
\caption{Observed galaxy ellipticity of spiral galaxies with $\kappa_{\rm rot} > 0.7$ shown against the theory prediction from the angular momentum vector of the galactic stellar component (\textit{top panel}), and of the host dark matter halo (\textit{bottom panel}, for centrals only), assuming an infinitely thin, circular galactic disc. While the angular momentum vector of the stellar disc is well suited for estimating the observed ellipticity of spirals, with systematically lower observed values with increasing ellipticity due to a finit disc thickness, this is by no means the case for the host halo angular momentum.}
\label{fig_eL}
\end{figure}

\subsection{Quadratic model assumptions revised} \label{sect_s_why}
We have found that, apart from the case of massive discs at high $z$, the alignment parameter of the quadratic model for spiral galaxies is consistent with zero. The quadratic alignment model, described in section~\ref{sect_s_theory}, relies on a few fundamental assumptions: (1) the angular momentum $\vec L_\text{halo,dm}$ of the host halo being set by linear tidal torques during early stages of gravitational collapse, (2) the central galaxy inheriting its host halo's specific angular momentum vector, (3) the angular momentum direction defining the observed ellipticity (assuming an infinitely thin, perfectly regular stellar disc), and (4) the relations being preserved troughout cosmic time. In this section we revisit these assumptions to uncover where they conflict with the actual physical picture.

\paragraph*{Ellipticity prediction from angular momentum.} \ In order to deconstruct the quadratic model, we start by verifying assumption (3), namely that the observed galaxy ellipticity $\left|\epsilon\right| = \sqrt{\epsilon_+^2+\epsilon_\times^2}$ is given by the angular momentum direction through eq.~(\ref{eq_eLz}). In the top panel of Fig.~\ref{fig_eL}, we show the observed ellipticity at $z=0$ against the ellipticity predicted by the angular momentum vector $\vec L_\text{galaxy,stars}$ of the galaxy's stellar component, assuming an infinitely thin, round galactic disc. Due to the large sample size, we display only spiral galaxies with high rotational support, $\kappa_{\rm rot}>0.7$, as individual dots colour-coded by their $\kappa_{\rm rot}$ values. The black dashed line indicates perfect agreement between theory prediction and observed ellipticity. The top panel of Fig.~\ref{fig_eL} shows that the observed ellipticities follow nicely the theoretical prediction, with some scatter, and a characteristic suppression of the observed ellipticity compared to the theoretically expected with increasing $\left|\epsilon\right|$. The latter is a consequence of the finite disc thickness of real galaxies, which is absorbed into and reduces the alignment parameter $A$ (see section~\ref{sect_s_theory}). However, if instead of $\vec L_\text{galaxy,stars}$ we use for the theory predicition $\vec L_\text{halo,dm}$, whose direction is primarily set by interactions with the surrounding tidal field at high redshift, we obtain the correlation shown in the bottom panel of Fig.~\ref{fig_eL}. Here, we only include spiral galaxies that are central objects of their host haloes, reducing the total sample size to $5905$, and compute $\vec L_\text{halo,dm}$ from all dark matter particles within a spherical overdensity of $200 \cdot \rho_{\rm crit}$. Using the halo angular momentum direction to predict the galactic stellar angular momentum and subsequently the observed galaxy ellipticity thus erases the correlation between angular momentum direction and observed ellipticity.

\begin{figure} 
\centering
\includegraphics[width=0.49\textwidth,trim= 10 10 0 10,clip]{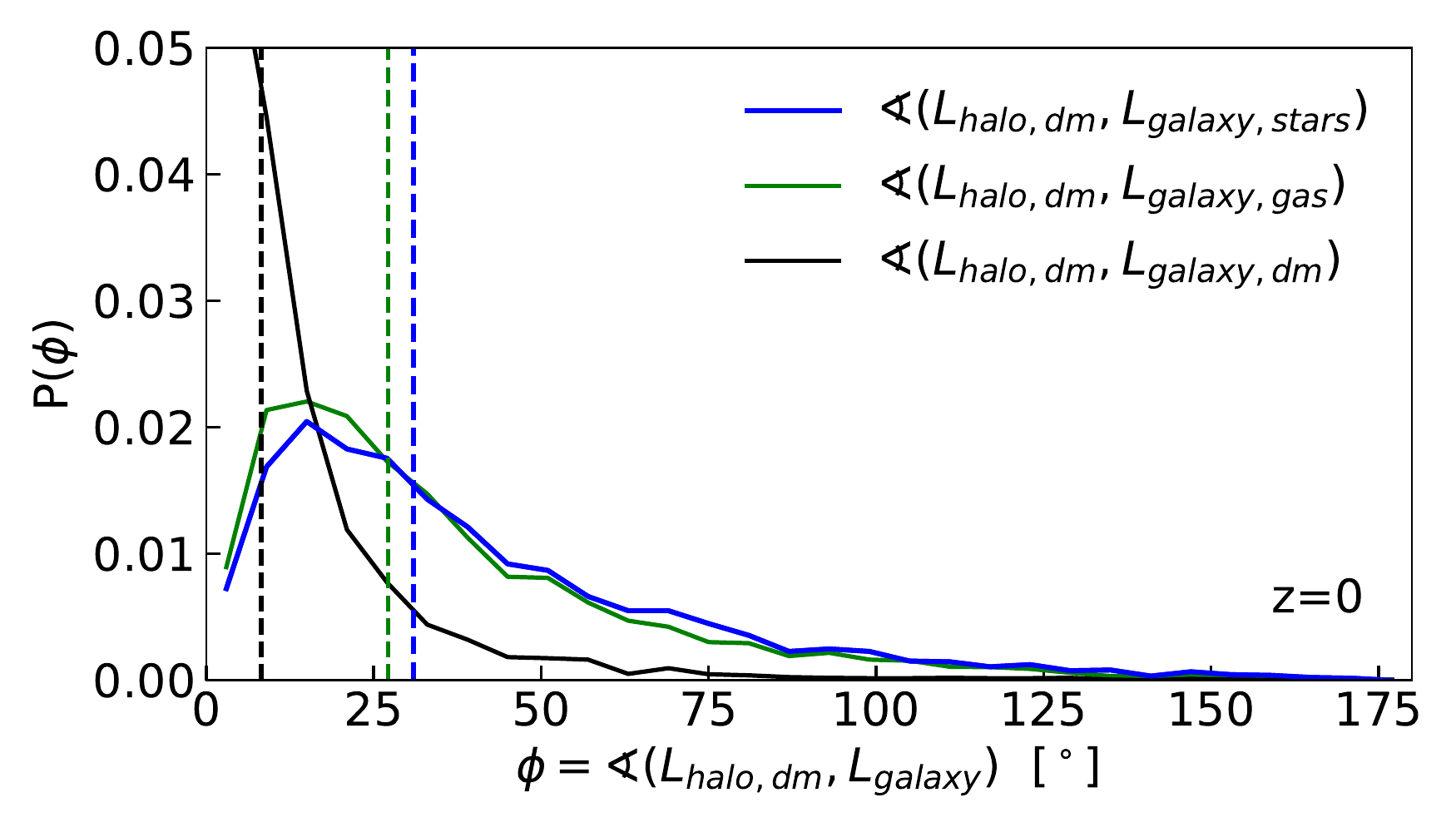}
\includegraphics[width=0.49\textwidth,trim= 10 10 0 10,clip]{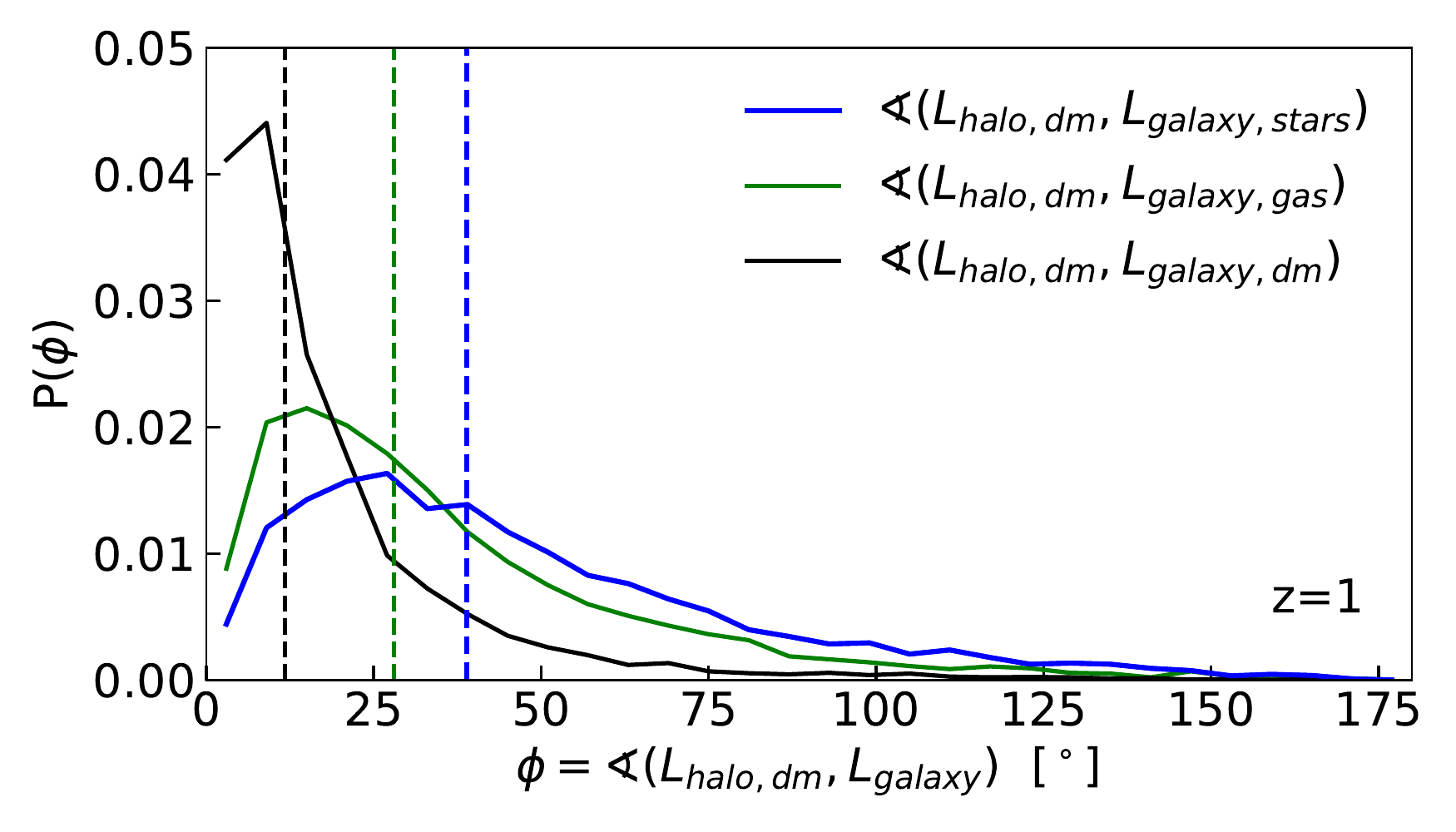}
\caption{Probability density function of the misalignment angle between the host dark matter halo angular momentum vector and the stellar (blue line), gas (green line), and dark matter (black line) angular momentum vector of central spirals at $z=0$ (\textit{top panel}) and $z=1$ (\textit{bottom panel}). Median values of each distribution are indicated by vertical dashed lines of corresponding colour. The statistical misalignment between stellar disc and host halo erases the correlation assumed in the quadratic model.}
\label{fig_angl_z0}
\end{figure}

\begin{figure}
\centering
\includegraphics[width=0.49\textwidth,trim= 10 0 -10 5,clip]{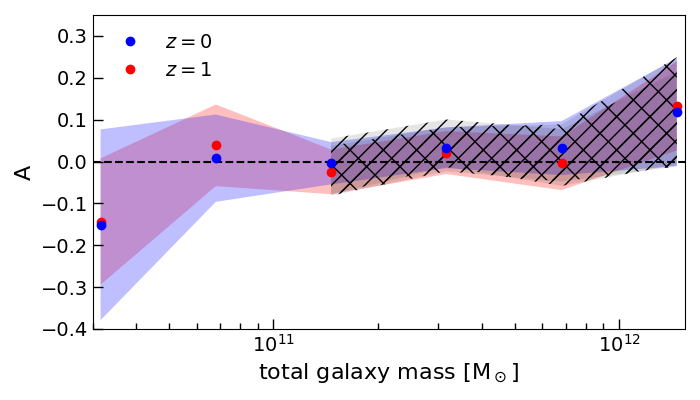}
\caption{Intrinsic alignment parameter $A$ derived from galaxy ellipticities as predicted by $\vec L_\text{halo,dm}$ as a function of total galaxy mass at $z=0$ and $1$, with the standard error shown as shaded area. Hashed areas show results for central spirals only. This figure demonstrates that at the source galaxy's redshift we do not find any correlation (being preserved) between the host halo's angular momentum vector and the gravitational tidal field, which is a fundamental assumption of the quadratic model.}
\label{fig_A_mass_tf}
\end{figure}

\paragraph*{Galaxy--halo angular momentum correlation.} \ Another way of seeing the same effect is by directly testing assumption (2), which states that the central galaxy inherits its host halo's angular momentum direction, in combination with assumption (4) that says that any correlation is preserved with time. In the top panel of Fig.~\ref{fig_angl_z0} we show the distribution of the misalignment angle between $\vec L_\text{halo,dm}$ and the angular momentum vector of the individual galaxy components for central galaxies at $z=0$. The distribution is obtained binning all measured misalignment angles in 30 eqidistant bins between $0^\circ$ and $180^\circ$, normalising to the total number of galaxies and the bin size. As vertical dashed lines we indicate the median values for all displayed distributions. The galaxy's dark matter component has a median misalignment angle of $\sphericalangle(\vec L_\text{halo,dm},\vec L_\text{galaxy,dm})_{z=0} \simeq 8^\circ$ with the host dark matter halo, which is considerably larger for the gas and stellar components and amounts to $ \sphericalangle(\vec L_\text{halo,dm}, \vec L_\text{galaxy,gas})_{z=0} \simeq 27^\circ$ and $\sphericalangle(\vec L_\text{halo,dm},\vec L_\text{galaxy,stars})_{z=0} \simeq 31^\circ$, respectively. This effect was also seen indirectly by \citet{Velliscig15ias} through a better alignment of the satellite system, tracing the host halo shape, with the dark matter of the central galaxy than its stellar component. The gas and stellar components of central galaxies do not only exhibit very similar median misalignments, but also remarkably similar distributions of their misalignment angles. 

\begin{figure*} 
\centering
\includegraphics[width=0.49\textwidth,trim= 10 10 10 10,clip]{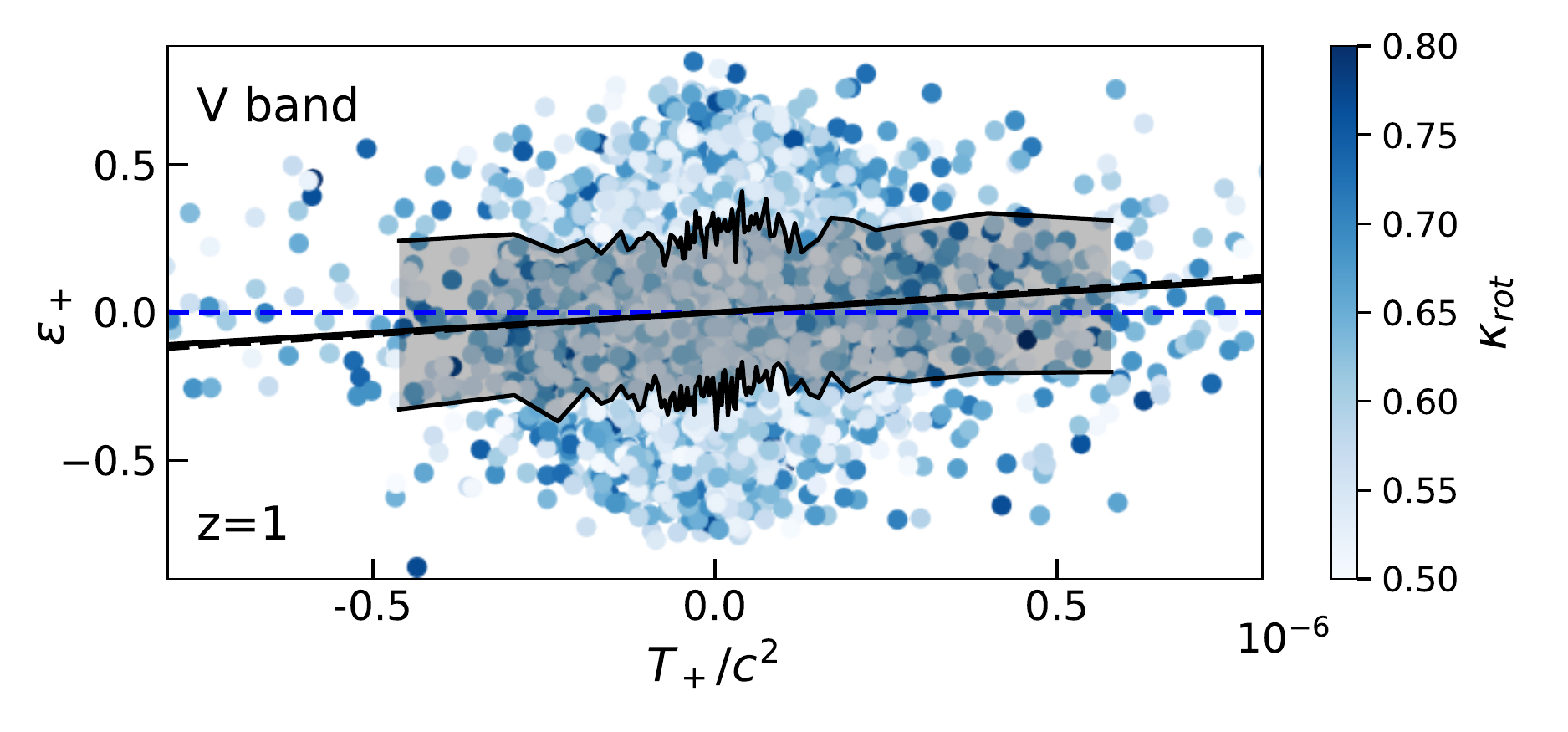}
\includegraphics[width=0.49\textwidth,trim= 10 10 10 10,clip]{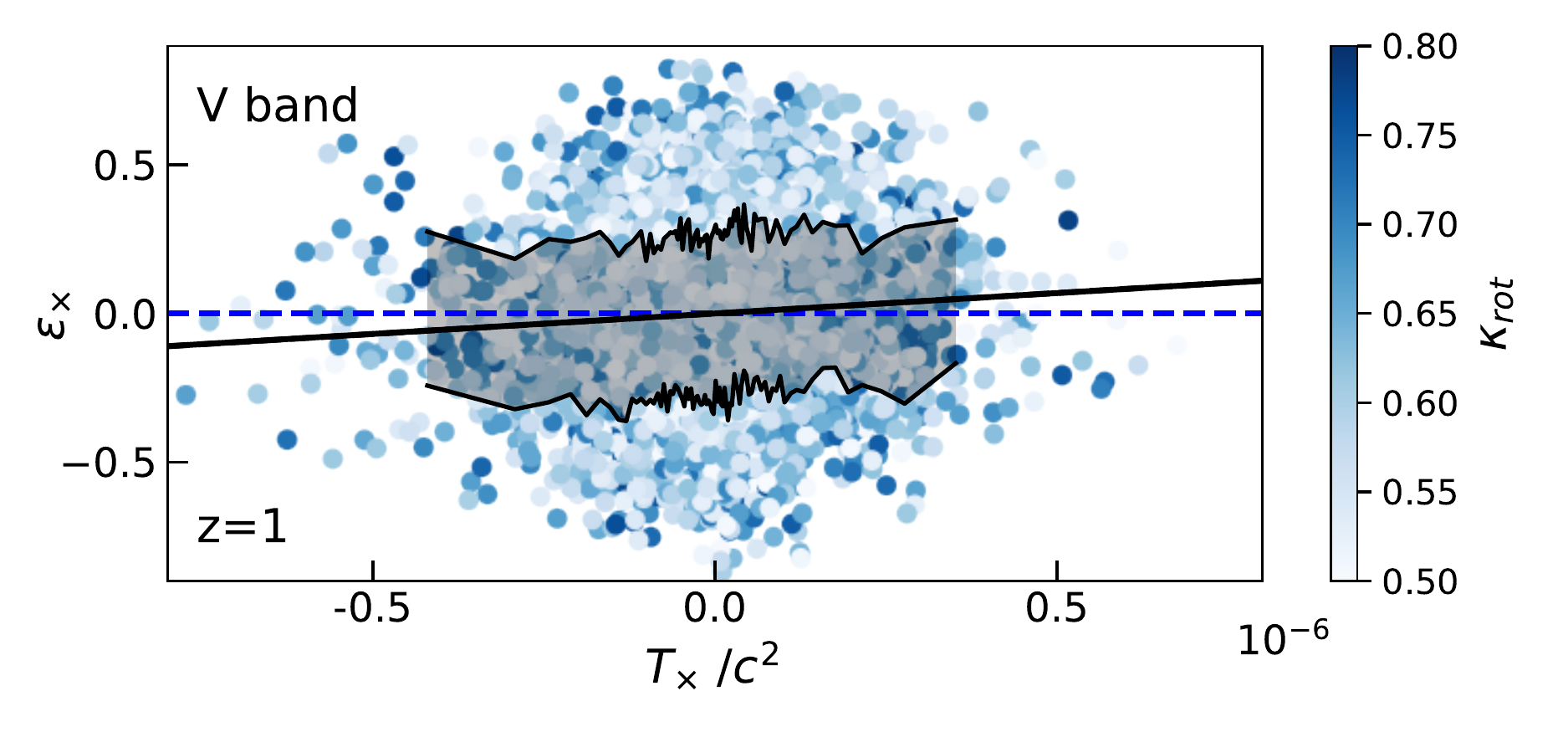}
\caption{Same as Fig.~\ref{fig_D_z1} verifying the linear model for elliptical galaxies, however, on our sample of spiral galaxies at $z=1$. In this figure, we show all our spiral galaxies as individual dots, colour-coded by their $\kappa_{\rm rot}$ parameter indicating rotational support. No correlation is shown as a blue dashed line, while the direct fit is shown as a black dashed line, and the anistropy correceted result as a black solid line that yields a significant alignment signal with a parameter of $D_{z=1} = (1.53 \pm 0.18) \times 10^{-6} c^2$. At $z=0$, however, we find no significant alignment of spirals according to the linear model (thus not explicitly shown).}
\label{fig_linsp}
\end{figure*}

Of primary relevance for the quadratic model, is that both Fig.~\ref{fig_eL} and~\ref{fig_angl_z0} show that at $z=0$ the angular momentum vector of the galaxy's stellar component $\vec L_\text{galaxy,stars}$ is considerably, and statistically randomly misaligned with $\vec L_\text{halo,dm}$. Therefore, even though $\vec L_\text{galaxy,stars}$ is suitable for predicting the observed galaxy ellipticity, this is by no means the case any more for $\vec L_\text{halo,dm}$. The misalignment between the central galaxy and its host halo is a consequence of ongoing galaxy evolution physics, and in particular feedback processes not accounted for in the quadratic model that stirr up the gas and subsequently the stellar populations that forms out of it. As a consequence we also expect the average misalignment between the baryonic components of the central galaxy and its host halo to grow in time, explaining the slightly higher alignment strength $A$ measured at $z=1$ compared to $z=0$. Therefore, in the bottom panel of Fig.~\ref{fig_angl_z0} we show the misalignment angle distribution for $5624$ central spiral galaxies at $z=1$. Interestingly, we find even somewhat larger misalignment angles of the different galactic components with $\vec L_\text{halo,dm}$. The galactic dark matter component exhibits a median misalignment angle of $\sphericalangle(\vec L_\text{halo,dm}, \vec L_\text{galaxy,dm})_{z=1} \simeq 12^\circ$, the gas and stellar components a median angle of $\sphericalangle(\vec L_\text{halo,dm}, \vec L_\text{galaxy,gas})_{z=1} \simeq 28^\circ$ and $\sphericalangle(\vec L_\text{halo,dm},\vec L_\text{galaxy,stars})_{z=1} \simeq 39^\circ$, respectively. We suspect that directed accretion of both dark matter and baryons from filaments, through both mergers and smooth accretion, onto galaxies serves to establish a statistical equilibrium of possible misalignments between the dark matter halo and the stellar disc across the epoch of active galaxy evolution. We defer tracing the evolution of those misalignments (for galaxies of distinct morphological type) through cosmic time and identifying their astrophysical origin to future work. 

As the alignment strength at $z=1$ is determined by the $3\sigma$ signal from massive, $\sim1- 2\times 10^{12} M_\odot$ spirals, we would expect to see smaller misalignment angles for galaxies in this specific narrow mass range. However, on the contrary, we find misalignment distributions that do not differ at all from the ones shown in Fig.~\ref{fig_angl_z0}. This gives rise to the question why the same degree of misalignment between host halo and the stellar disc seems to preserve a $3\sigma$ IA signal for a specific class of galaxies while erasing it otherwise. 

\paragraph*{Halo quadratic alignment.} \ To answer this question, we first test assumptions (1) and (4) of the quadratic model, namly $\vec L_\text{halo,dm}$ being set by tidal torques from the surrounding large-scale structure at high redshift and being preserved troughout cosmic time. To this end, we assume that the galaxy ellipticity is set by $\vec L_\text{galaxy,stars}$ according to eq.~(\ref{eq_eLz}) reflecting assumption (3), and that $\vec L_\text{galaxy,stars}$ is given by  $\vec L_\text{halo,dm}$ according to assumption (2), and such derive theoretical values for $\epsilon_+$ and $\epsilon_\times$ from $\vec L_\text{halo,dm}$. We correlate those analytically derived ellipticity values with the respective tidal fields at the same redshift, as in Fig.~\ref{fig_A_mass}, to obtain $A$ as a function of total galaxy mass, shown in Fig.~\ref{fig_A_mass_tf}. As this ansatz is only physically meaningful for galaxies residing in the centre of their host halo, we exclude satellites, and show in addition to the full sample results for central galaxies only in terms of their $1\sigma$ environment displayed as a backwards ($z=0$) and forwards ($z=1$) hashed area, in mass bins with a sufficient number of central spials. As can be seen in Fig.~\ref{fig_cntsat}, the spiral galaxy sample is dominated by satellites for $M_{\rm tot} \lesssim 2 \times 10^{11} M_\odot$, and by centrals for $M_{\rm tot} \gtrsim 2 \times 10^{11} M_\odot$, explaining why centrals yield almost identical results to the full sample in the respective high mass bins.

The striking result revealed by Fig.~\ref{fig_A_mass_tf} is that neither at $z = 0$ nor at $z = 1$, nor in any mass bin, including those that did exhibit a $3\sigma$ intrinsic alignment signal in Fig.~\ref{fig_A_mass}, we find a significant correlation between $\vec L_\text{halo,dm}$ and the tidal field, that is fundamental to the quadratic model. Note, that according to tidal torque theory the correlation between $\vec L_\text{halo,dm}$ and tidal field is set before turnround (spherical collapse model), and is preserved in linear theory. Verifying the initial correlation between $\vec L_\text{halo,dm}$ and tidal field, requires tracing all halo constituents back in time to very high redshift, and is complicated by the fact that the Lagrangian region of the proto-halo typically will be much more extended than the characteristic scale of $1$~Mpc we employ for smoothing of the tidal field. We leave such considerations for future work. Fig.~\ref{fig_A_mass_tf} states that at the source galaxy's redshift the tidal torque theory induced correlation between $\vec L_\text{halo,dm}$ and tidal field is not detectable, either due to not being present in the first place, violating assumption (1), or far more likely due to being erased over time due non-linear evolution and galaxy physics, violating assumption (4). Nevertheless, correlating with the tidal field at the same redshift, the quadratic model yields a signal that can be detected with $3\sigma$ in massive spirals at $z=1$. An alternative, likely rather instantaneous, mechanism thus must be present that sets intrinsic alignments of massive, high redshift spirals, which would also explain why the misalignment between the angular momentum vectors of the stellar disc and the host halo has no impact on the IA signal measured for spiral galaxies.

\subsection{Spiral galaxies and the linear alignment model}

Recently, \cite{Yao20a} measured a significant IA signal in the redshift range $0 \lesssim z \lesssim 1$ directly from observations of $\sim 23 \times 10^6$ galaxies from {\sc DESI}, while \cite{Yao20b} repeated the analysis on a total of $\sim 25 \times 10^6$ galaxies from the {\sc KiDS} and {\sc VIKING} surveys, and showed the obtained signal to be consistent with predictions by the non-linear alignment model. Furthermore, \cite{Yao20a} devide their galaxy sample by colour and find in their lowest redshift bin $0.1\lesssim z\lesssim 0.3$ a significant signal from blue galaxies, which the authors argue cannot be explained by a contamination of the blue galaxy sample by red galaxies, while no signal is detected at higher redshift. This motivates us to apply the linear model to our sample of simulated spiral galaxies from TNG100. Doing so, we find no signal at $z=0$, however an alignment strength of
\begin{equation}
D_V = (1.53 \pm 0.18) \times 10^{-6} c^2
\end{equation}
at $z=1$ which is significantly different from zero at a $8.5\sigma$ level. We show the corresponding fits in in Fig.~\ref{fig_linsp}, derived analogously to Fig.~\ref{fig_D_z0} and Fig.~\ref{fig_D_z1}, whereas here we show as individual dots all spiral galaxies from our sample. 

We thus find that spiral galaxies, though to a smaller degree than ellipticals, are affected by the tidal alignment mechanism described by the linear model, as also already seen by \citet{Hilbert17}. The fact that we obtain a significant signal only at $z=1$ is in accordance with the growing alignment signal of elliptical galaxies with increasing redshift, but also with our previous finding that spiral galaxies exhibit a stronger IA signal according to the the quadratic model at higher redshift. This trend is however opposite to the finding by \cite{Yao20a}, who find a significant signal only close to $z\simeq 0$. We suspect some discrepancy to originate from the different classification methods of galaxies into blue/spiral and red/elliptical. We leave detailed comparisons to observations, that could provide valuable constrains to numerical galaxy evolution models, for future analysis. 

Finally, note that this finding is a strong argument in favour of having one combined model applicable to all galaxy types that incorporates all relevant alignment mechanisms, as also advocated by \cite{Blazek19}. Whereby, above we have shown that tidal torquing (as estimated form same-epoch tidal fields) is not one of the relevant mechanisms for intrinsic alignments of spiral galaxies, in agreement with \citet{Samuroff20}, who find zero values for the tidal torquing term in their framework. Our measurements support a picture in which all galaxy types align according to the linear model (whether due to tidal shearing or any other mechanism that yields the same scaling behaviour), alleviating the need for a more complex treatment of intrinsic alignments, and strengthening the argument for the applicability of the (non-)linear model to account for the IA contamination in observational shear data.

\begin{figure} 
\centering
\includegraphics[width=0.49\textwidth,trim= 10 5 10 10,clip]{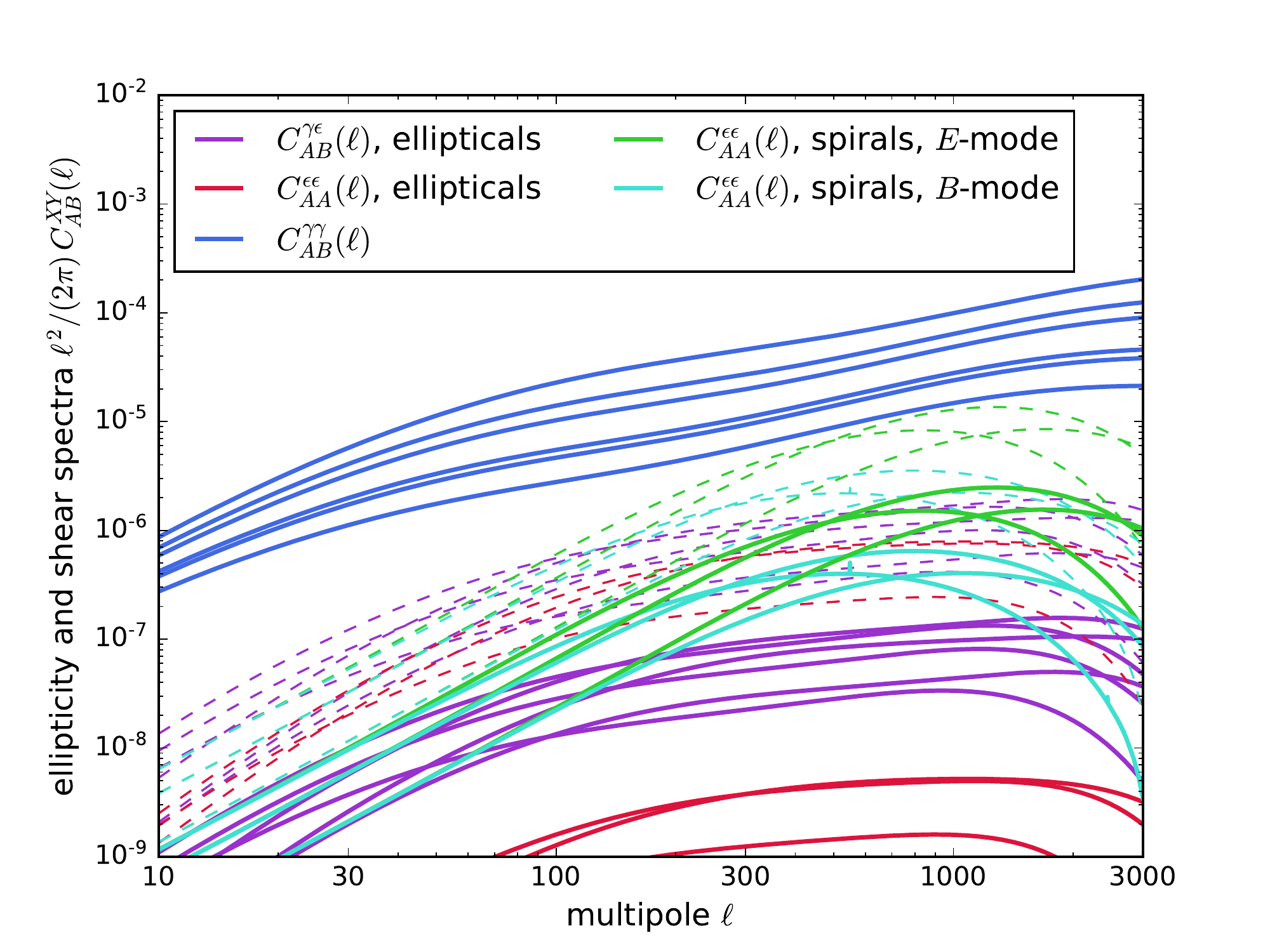}
\caption{Intrinsic alignment spectra $C^{\epsilon\epsilon}_{AA}(\ell)$ for elliptical and spiral galaxies, as well as the IA-lensing cross-spectrum $C^{\gamma\epsilon}_{AB}(\ell)$ for elliptical galaxies in comparison to weak lensing spectra $C^{\gamma\gamma}_{AB}(\ell)$, for values of the alignment parameters used in analytical computations (dashed lines) and for those derived from {\sc IllustrisTNG} (solid lines), for 3-bin tomography with {\sc Euclid}, with an assumed morphological mix of two spirals for one elliptical.}
\label{lensing_spectra}
\end{figure}

\section{Lensing predictions}\label{sect_lens}

Isolating alignment parameters for spiral and elliptical galaxies as they are predicted from the tidal torquing and tidal shearing models led us to measure the correlation between galaxy ellipticities and the ambient gravitational tidal field in {\sc IllustrisTNG} data. The intrinsic alignment parameters derived in this way are smaller than those used previously in analytical computations, in particular $D\simeq -10^{-5}$ and $A = 0.25^2$ \citep{Tugendhat18}, whereby the second factor of $0.25$ multiplied to the default $A \simeq 0.25$ from \citet{LeePen01} combines the impact from a stellar disc of finite thickness ($\simeq 0.85$ according to \citealt{Crittenden01}), but also effects such as the assumed statistical misalignment between the galactic disc and host halo angular momentum vector. Furthermore, here we quote $D$ with a minus sign, that was explicitly separated out in eq.~(\ref{eq_D}), to emphasize that the tidal shear field distorts `intrinsic' galaxy shapes in the opposite direction than gravitational lensing caused by the same tidal fields. 
As a consequence, predictions based on our measurements for IA spectra from surveys such as {\sc Euclid} are smaller accordingly, with implications on the measurability of ellipticity correlations, and the contaminating effect of alignments in weak lensing data. 
To achieve consistency between simulation and analytics, we impose a smoothing scale of $1~\mathrm{Mpc}/h$, corresponding to a mass scale of $\sim 10^{12}~M_\odot/h$ and a virial velocity of $\sigma^2 = (10^5~\mathrm{m}/\mathrm{s})^2$. Angular ellipticity spectra, as final results of the analytical calculation, only depend on a specific value of $h$ through the CDM-shape parameter $\Gamma\simeq \Omega_m h$, as all other scalings of $h$ drop out in Limber-projection. Fixing the cosmology in the analytical part to values from \citet{Planck15params} used in {\sc IllustrisTNG} thus allows a direct extrapolation of the IA amplitudes measured in TNG100 onto predictions for {\sc Euclid} spectra.

\begin{figure} 
\centering
\includegraphics[width=0.49\textwidth,trim= 10 5 10 10,clip]{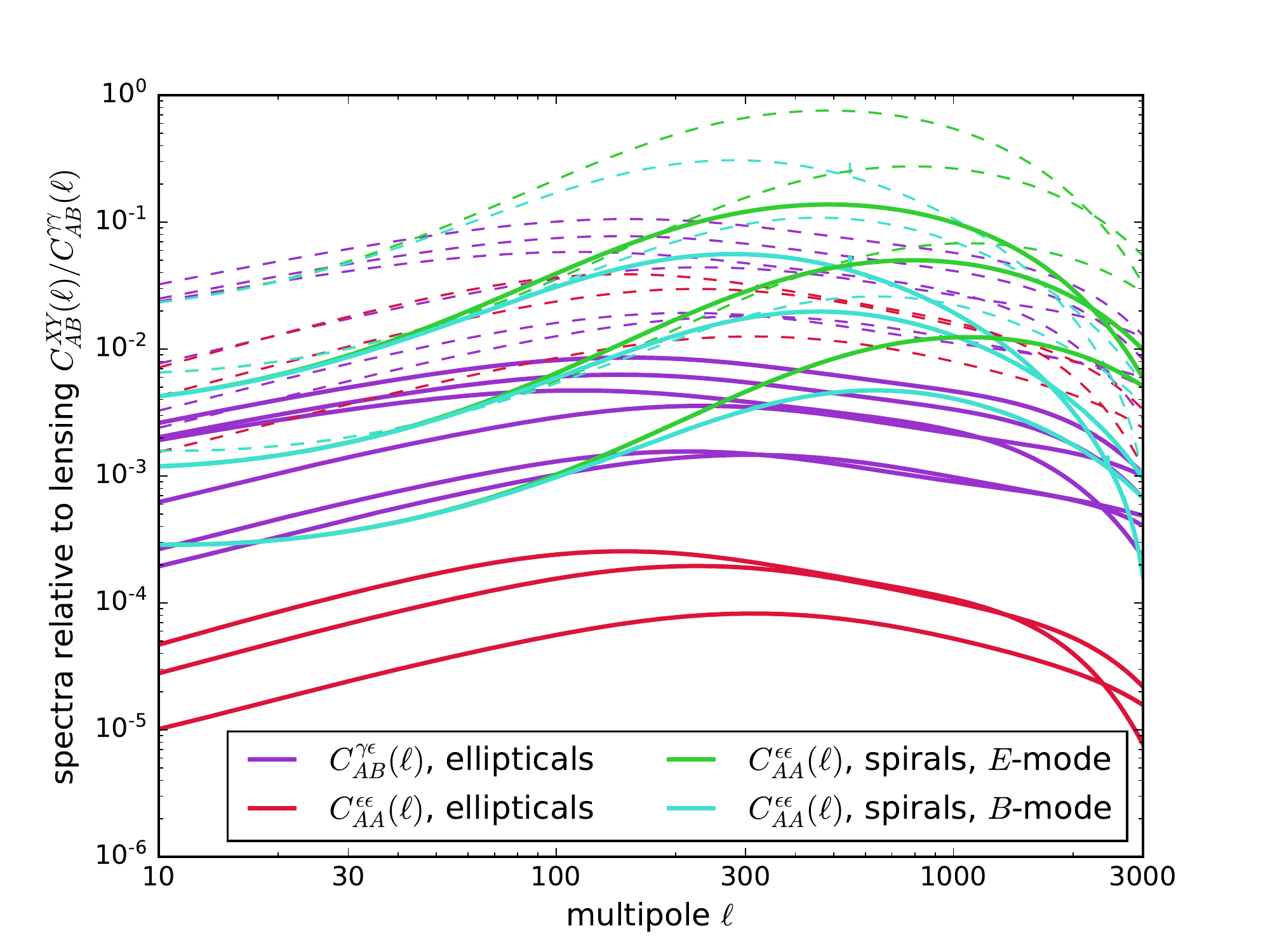}
\caption{Ratios $C^{\epsilon\epsilon}_{AA}(\ell)/C^{\gamma\gamma}_{AA}(\ell)$ between the IA-spectra and the weak lensing spectra, as well as the ratios $C^{\gamma\epsilon}_{AB}(\ell)/C^{\gamma\gamma}_{AB}(\ell)$ between the IA-lensing cross-spectra and the weak lensing spectra, for values of the alignment parameter used in analytical computations (dashed lines) and derived from {\sc IllustrisTNG} (solid lines), for 3-bin tomography with {\sc Euclid}, with an assumed morphological mix of two spirals for one elliptical.}
\label{lensing_suppression}
\end{figure}

\begin{figure} 
\centering
\includegraphics[width=0.49\textwidth,trim= 10 5 10 10,clip]{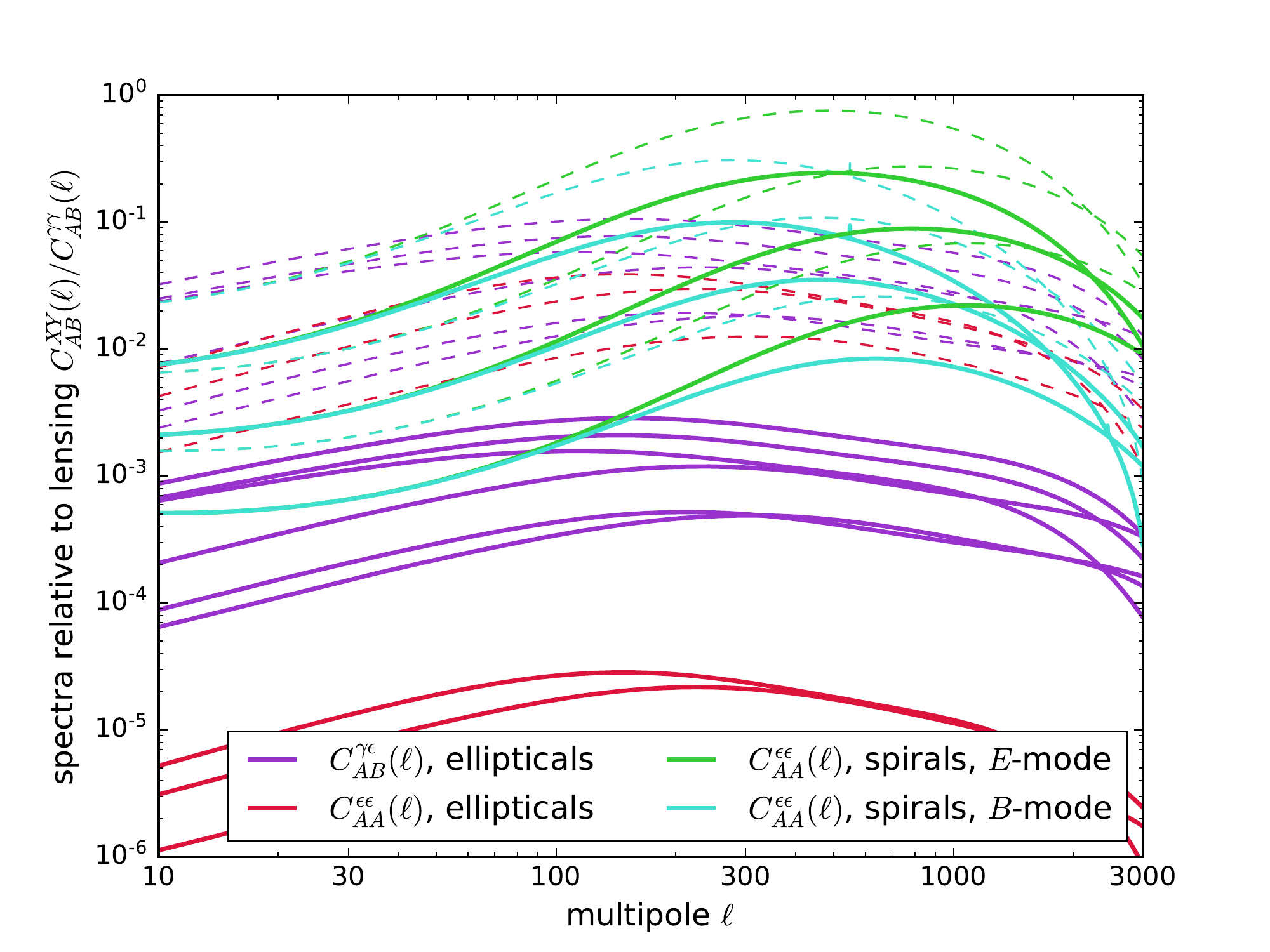}
\caption{Ratios $C^{\epsilon\epsilon}_{AA}(\ell)/C^{\gamma\gamma}_{AA}(\ell)$ between the IA-spectra and the weak lensing spectra, as well as the ratios $C^{\gamma\epsilon}_{AB}(\ell)/C^{\gamma\gamma}_{AB}(\ell)$ between the IA-lensing cross-spectra and the weak lensing spectra, for values of the alignment parameter used in analytical computations (dashed lines) and derived from {\sc IllustrisTNG} (solid lines), for 3-bin tomography with {\sc Euclid} with the morphological mix found in {\sc IllustrisTNG} with our selection criteria.}
\label{lensing_suppression_rich}
\end{figure}

Fig.~\ref{lensing_spectra} illustrates intrinsic alignment spectra for elliptical and spiral galaxies as they would be measured from the {\sc Euclid} survey, where we are using 3-bin tomography for simplicity. The tomographic bins are chosen to contain equal fractions of the total galaxy population, and we use classical fractions of $q = 1/3$ elliptical galaxies and $p = 1 - q = 2/3$ spiral galaxies \citep{Dressler84} to down-weigh the intrinsic alignment correlations relative to weak lensing, which operates equally on all galaxy types: This is motivated by the fact that only a fraction of $q^2$ of all galaxies would align according to the linear tidal shearing model for elliptical galaxies, only a fraction of $p^2$ of all galaxies would show alignments as predicted by the quadratic tidal torquing model for spiral galaxies, and only a fraction of $q$ of all galaxies would exhibit cross-alignments with the weak lensing signal, as typical for the linear alignment model for ellipticals. With the fundamental assumption that the tidal gravitational field that aligns the galaxies is Gaussian, there should be no cross-correlations between the intrinsic ellipticities of spiral and elliptical galaxies, and no cross-correlations between the shapes of spiral galaxies and weak gravitational lensing, because both correlation functions amount to a third moment of a Gaussian random field, which is equal to zero. Likewise, under identical assumptions, there would be no cross-correlations in the intrinsic shapes of spiral and elliptical galaxies.
As most galaxies of {\sc Euclid}'s sample are found at redshift $z\simeq 1$, we use the alignment parameters isolated at that particular redshift, noting that for instance spiral galaxies do not show significant alignments at $z=0$ in {\sc IllustrisTNG}. Typically, the alignments of elliptical galaxies are smaller by a factor of $\sim 4^2$, and those of spiral galaxies by a factor of $\sim 1.5^2$, while the cross-alignment between lensing and the shape of elliptical galaxies is smaller by a factor of $\sim 4$. Comparing to previous projections for the measurability of shape alignments would correct significances by the same factors, because the covariance is dominated by weak lensing and shape noise, but implying that {\sc Euclid} should still be able to perform detections of some IA-spectra and of the lensing-alignment cross-correlation.

Similarly, Fig.~\ref{lensing_suppression} shows the ratio of the IA spectra and weak lensing, in order to illustrate how the significantly smaller alignment parameters derived from {\sc IllustrisTNG} cause a much weaker IA-contamination of the weak lensing data, most notably for elliptical galaxies, but also to some extent for the spiral galaxies. The alignment contamination, that amounts to a few $10^{-2}$ for the IA-lensing cross-alignment and a few $10^{-3}$ for the ellipticity correlation of elliptical galaxies, and which reaches values of about $10^{-1}$ for spiral galaxies on very high, shape-noise dominated multipoles, are corrected towards much smaller values, alleviating systematical errors in weak lensing data analysis.
With Fig.~\ref{lensing_suppression_rich} we comment on the morphological mix that we extracted with our selection criteria from the {\sc IllustrisTNG} simulation, where in our galaxy catalogue spiral galaxies are much more abundant, with a ratio of about eight spirals to one elliptical at $z=1$. With these values, ellipticity spectra involving elliptical galaxies are significantly down-weighed to the point where spiral galaxies are responsible for the intrinsic ellipticity correlation, even dominating over the $GI$-spectra, in contrast to results from e.g. the CFHTLenS survey, which reports ellipticity correlations from elliptical but not for spiral galaxies.

\section{Summary and discussion}\label{sect_disc}

We use a sample of realistic galaxies from the {\sc IllustrisTNG} (TNG100) simulation at $z=0$ and $z=1$ to study the tidal shearing and tidal torquing alignment mechanisms for elliptical and spiral galaxies, respectively. We measure the corresponding intrinsic alignment parameters for the linear and quadratic alignment models, used to quantify the bias, caused by intrinsic galaxy formation processes, in weak lensing shear measurements. We furthermore measure intrinsic alignment parameters as a function of total and stellar galaxy mass, as well as environmental properties. We use the obtained alignment parameters, as well as information about the morphological mix of galaxies from {\sc IllustrisTNG}, to calculate angular ellipticity spectra and quantify the expected contamination in the weak lensing signal obtained by {\sc Euclid}.

From all well-resolved galaxies in TNG100 ($>$1000 stellar particles, $M_{\rm tot} > 10^{10}M_\odot$, which  are above the {\sc Euclid} sensitivity limit of 24.5 mag at $z\simeq 1$) we select galaxies with a distinct morphological type based on their star-formation properties and kinematics. We select red ellipticals based on a specific star-formation rate cut of sSFR $< 0.04\ Gyr^{-1}$, and a cut for (the lack of) rotational support $\kappa_\mathrm{rot} < 0.5$, and blue spirals based on sSFR $> 0.04\ Gyr^{-1}$ and $\kappa_\mathrm{rot} > 0.5$. For all our sample galaxies we derive observed ellipticities on the projected sky in the 8 observational bands available for {\sc IllustrisTNG} data. We do not find any dependence of our results on observational band, given the absence of dust modeling. We expect the presence of dust to affect observed galaxy ellipticities especially in the $U$-band and in the infrared. Thus we present our main results for the $V$-band. Including the impact of dust attenuation on observed galaxy ellipticities is a next crucial step for accurate intrinsic alignment modeling. Furthermore, we derive for each galaxy the surrounding tidal field, responsible for the various alignment mechanisms, at five different smoothing lengths, 250~kpc, 500~kpc, 1~Mpc, 2~Mpc, and 5~Mpc. We quote our main results for tidal fields smoothed at 1~Mpc, which corresponds to the characteristic length scale of typical Milky Way-sized galaxies.

Note, that when deriving the alignment parameters $D$ and $A$ for the linear and quadratic models, respectively, we have accounted for the large-scale anisotropies in TNG100's cosmologically small $(75$ Mpc$/h)^3$ volume, by randomising the reference frame in which we measure each galaxy's ellipticity and tidal field. We explicitly caution that any analysis of cosmological probes performed in a volume similar or smaller than the one of TNG100 is affected by large-scale structure anisotropy. 
Alternatively, the same analysis can be performed in the larger $(205$ Mpc$/h)^3$ volume of the TNG300 simulation, that we expect to fullfil large-scale isotropy. However, the {\sc IllustrisTNG} galaxy formation subgrid models have been calibrated to the TNG100 default resolution, whereas TNG300 was run at lower spatial resolution than TNG100. As a consequence basic galaxy properties such as stellar mass and sSFR for galaxies with total mass below $\sim 10^{12} M_\odot$ are not fully converged (see figure A2 in \citealt{Pillepich18}), and can only be used applying further recalibration techniques. We leave this alternative approach, that would also increase sample size, for future work. 

For elliptical galaxies we find a significant correlation between the tidal field and the observed ellipticity as predicted by the linear alignment model with $V$-band alignment strengths of $D_{z=0} = (6.53 \pm 0.58) \times 10^{-7} c^2$ and $D_{z=1} = (2.43 \pm 0.36) \times 10^{-6} c^2$ at $z=0$ and $1$, respectively. The value of $D$, however, depends on both galaxy as well as environmental properties. We find $D$ to strongly increase with tidal field smoothing scale, both stellar and total galaxy mass, and ultimately with redshift, while we find a decrease with larger local overdensity. The mass trends have also been previously observed by \citet{Chisari15}, \citet{Tenneti15}, and \citet{Velliscig15ias} in various galaxy formation simulations through other means, while an increase of the IA signal of elliptical galaxies with redshift has recently been detected by \citet{Yao20a} in observational data. Trends of increasing/decreasing $D$ can be traced back to a scaling with the local tidal field on a designated scale, and to relate to the galaxy's size and mass. The intrinsic alignment contamination therefore depends on the specifications of a given observation survey, and can be estimated for a galaxy sample with known properties.

For spiral galaxies we do not find significant intrinsic alignments as predicted by the quadratic model. We measure alignment parameters of $A_{z=0}= 0.026 \pm 0.017$ and $A_{z=1}= 0.039 \pm 0.016$ at $z=0$ and $1$, that are compatible with zero at a $2\sigma$ and $3\sigma$ level, respectiely, and are considerably smaller than previous theoretical estimates, even taking into account a finite disc thickness that further reduces the expected alignment signal. We also observe both the magnitude of $A$ as well as its significance to grow with redshift. On closer inspection, we find massive $\sim 10^{12} M_\odot$ spirals to exhibit a significant signal at high redshift, $z=1$. We show, however, that the quadratic model breaks down at its fundamental assumptions, namely, the angular momentum of the stellar disc pointing in the same direction as the host halo angular momentum, and the halo angular momentum being related to the same-epoch tidal fields through tidal torquing. It is not excluded that the observed alignments of spiral galaxies are a remainder of correlations set by the primordial tidal field, whereby the latter has then decorrelated with the present-epoch tidal field due to non-linear evolution and due to the displacement of galaxies caused by their peculiar velocities. Alternatively, another, rather instantaneous, mechanism can be at play that sets intrinsic alignments of spiral galaxies, which is supported by the fact that we find spirals to exhibit an alignment according to the linear model.

Motivated by observational findings from \citet{Yao20a}, who observe a low redshift alignment signal from spiral galaxies, and show it to be in agreement with a tidal shearing origin, we also test intrinsic alignments of spiral galaxies with the linear model, and find a significant signal at $z=1$, but not $z=0$. We report a measurement of this particular alignment on a galaxy by galaxy-basis, supplementing direct measurements of correlations in the projected light distribution by \citet{Hilbert17} and \citet{Samuroff20}.
A more detailed comparison to recent and upcoming observational studies is needed for a more profound understanding of the particular redshift trends. Such a comparison foremost requires a more sophisticalted framework for the classification of galaxy morphologies on both the theoretical and observational side that allows for direct comparison of the alignment signal from specific galaxy types. 
Discrepancies between observational findings and theory predictions, hints of which we possibly see in the opposite redshift behaviour of IAs of spiral galaxies, are a powerful probe for galaxy formation physics that can be used to constrain the next generation of subgrid models for galaxy formation simulations.

The fact that spiral galaxies tend to align according to the linear model could strongly simplify the treatment of intrinsic alignments, alleviating the need to incorporate a multitude of different alignment mechanisms into one framework applicable to all galaxy types, as proposed by e.g. \citet{Blazek19}. This finding is in support of the (non-)linear model being generally applicable, whereas however the alignment strength of $D_{z=1} = (1.53 \pm 0.18) \times 10^{-6} c^2$ for spirals at $z=1$ is substantially lower than for elliptical galaxies, while even indistinguishable from zero at $z=0$, such that a distinction between the different galaxy types still exists. Note, that though we measure spiral galaxies to align according to the linear model, we make no statement about the physical nature of the alignment mechanism (being tidal shearing or an alternative mechanism) that yields this particular scaling behaviour. Furthermore, for both spiral and elliptical galaxies the IA signal could arguably be caused by an integrated effect of all tidal interactions over the full history of galaxy assembly, opposed to the instantaneous correlations studied here, which also would take into account further effects such as peculiar velocities of galaxies (see \citealt{Schmitz18}).

With numerical values for the the alignment parameters $A$ and $D$ for spiral and elliptical galaxies, respectively, we compute angular ellipticity spectra as they would be measured from the {\sc Euclid} survey. Specifically, we determine ellipticity spectra for a tomographic setup with three equipopulated redshift bins and use the alignment parameters determined at $z=1$, corresponding to the median redshift of the {\sc Euclid} galaxy sample. As the alignment parameters from {\sc IllustrisTNG} are smaller than those used in analytical studies, we see much smaller amplitudes for the intrinsic ellipticity spectra, with a typical contamination of the lensing spectra on the percent-level on intermediate scales by the $GI$-term from elliptical galaxies, and about $10\%$ on very small scales due to the $II$-term from spiral galaxies, with an overall reduction of the alignment contamination by about an order of magnitude, with accordingly smaller estimation biases in the measurement of cosmological parameters. Compared to \citet{Blazek19}, our methodology differs substantially, as we isolate the alignment parameters by analysing the relation between shape and tidal field for individual galaxies and using those in an analytical model for the spectra relying on a Gaussian field in comparison to establishing a parameterised relationship between shape and a perturbatively evolved random field. Nevertheless, the final results for angular ellipticity correlations show comparable amplitudes, in particular at high redshift. 

The result that spiral galaxies align with the tidal field according to a linear relation is puzzling and is unexplained by the classical tidal shearing and tidal torquing models. In the lensing spectra this linear alignment would be reflected in a tidal shearing contribution that is up-weighted as it refers to all galaxies and not only ellipticals. The precise extent of this effect is to be determined in future work, complicated by the fact that, it is yet unclear to what extent galaxy formation simulations converge in predicting intrinsic shape correlations correctly, as shown by \citet{Samuroff20}. This uncertainty arrises due to a multitude of physical processes and their non-linear interplay being potentially relevant in determining the shape of galaxies and their relation to a tidal gravitational field. Whether galaxy shapes can be a valid test for galaxy formation and evolution models is an interesting question for further study.

\section*{Acknowledgments}
JZ thanks Ue-Li Pen for useful discussions. JZ and OH acknowledge funding from the European Research Council (ERC) under the European Union's Horizon 2020 research and innovation programme (Grant Agreement No.\,679145, project `COSMO-SIMS'). BMS is grateful for the hospitality of the Observatoire of the C{\^o}te d'Azur at Nice.

\subsection*{Data availability}
{\sc IllustrisTNG} data is publicly available at \href{https://tng-project.org}{tng-project.org} \citep{Nelson19TNGDR}.

\bibliography{refs/citations}
\bibliographystyle{mnras}

\appendix
\begin{appendix}

\section{Anisotropy of the TNG100 large-scale structure as seen by massive ellipticals} \label{sec:app_anisotropy_lss}

\begin{figure*} 
\centering
\includegraphics[width=0.49\textwidth,trim= 10 10 10 10,clip]{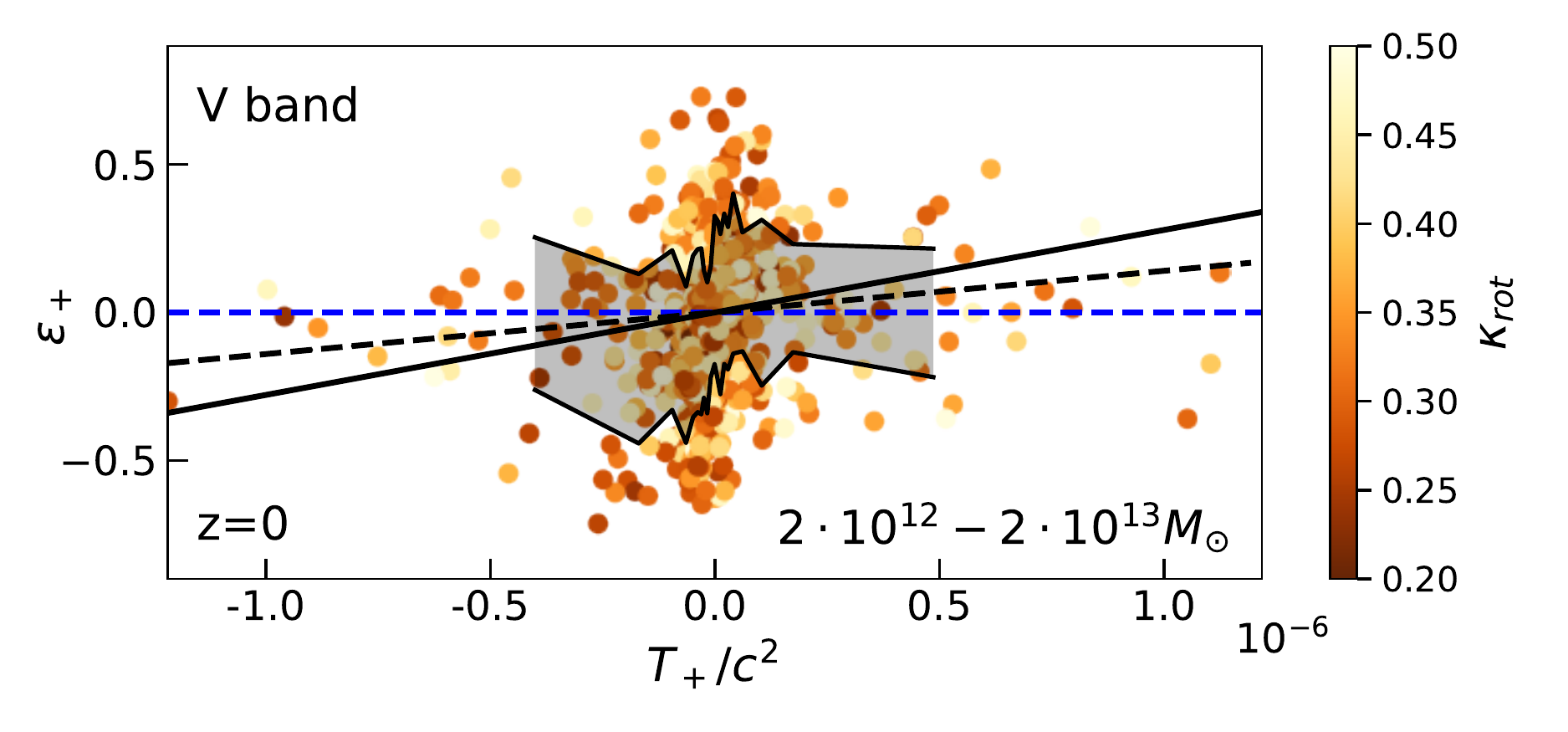}
\includegraphics[width=0.49\textwidth,trim= 10 10 10 10,clip]{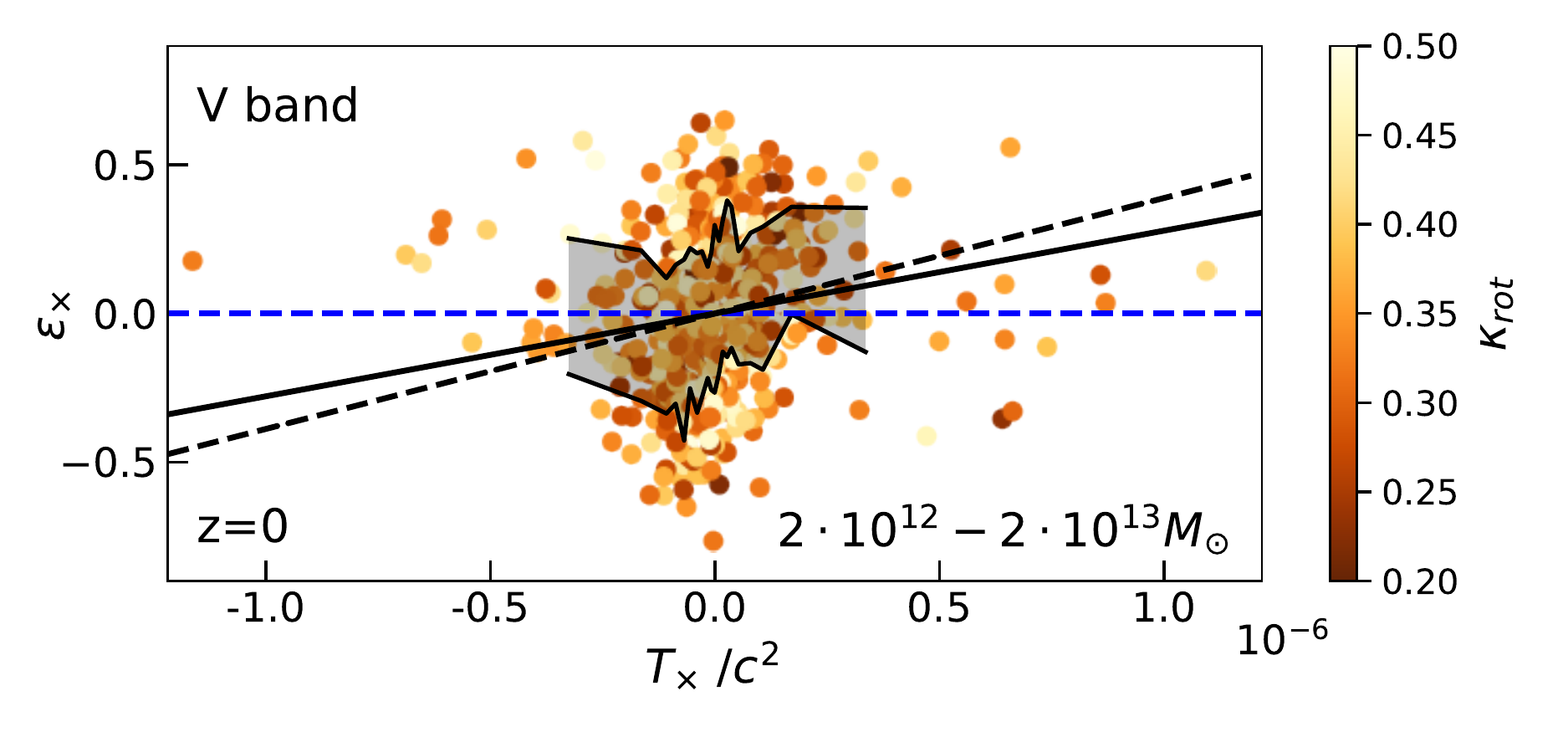}
\caption{Same as Fig.~\ref{fig_D_z0}, but for $z=0$ ellipticals in the mass range $2\times 10^{12} - 2\times 10^{13} M_\odot$. A linear fit to the binned data is shown as a black dashed line for the real (\textit{left panel}) and imaginary part (\textit{right panel}) of the ellipticity tensor directly measured from TNG100 along the cartesian $z$-axis, and yields intrinsic alignment strengths $D_+$ and $D_\times$ that are significantly different from each other. $1\sigma$-errors on the binned data points are are shown as a grey band, and the black line indicates the anisotropy corrected alignment parameter $D$ after randomisation of the reference frame. No correlation is indicated by a blue dashed line. The discrepancy in the raw measurements originates from an anistropic large-scale structure in TNG100 caused by its cosmologically small volume.\label{fig_D_aniso}}
\end{figure*}

In this appendix, we investigate the subset of massive ellipticals from the $z=0$ sample studied in section~\ref{sect_e}, which will allow us to demonstrate the reason for the necessary randomization of the reference frame (see section~\ref{sec:measurement_parameters}). In Fig.~\ref{fig_D_aniso}, we show $\epsilon_{+}(T_+)$ and $\epsilon_{\times}(T_\times)$ for elliptical galaxies with total mass between $2\times 10^{12} M_\odot$ and $2\times 10^{13} M_\odot$ at $z=0$ without frame randomisation. Due to the smaller sample size, we bin all galaxies in just 20 bins along $T_{+}$ and $T_{\times}$ with approximately the same number of $\sim 35$ galaxies per bin, and perform the same linear regression to the mean ellipticity as outlined before, which is shown as a black dashed line. This procedure yields $D_{+,V}= ( 1.56 \pm 0.64) \times 10^{-6} c^2$ and $D_{\times,V}= ( 4.32 \pm 0.78) \times 10^{-6} c^2$. These two values are clearly in tension with one another. The origin for this discrepancy lies in the statistical anisotropy of the large-scale structure in the cosmologically small TNG100 volume, as we will demonstrate below. Note that the E/B-decomposition of $D$ is sensitive to such an anisotropy, since this decomposition is not invariant under rotations in the $x$-$y$-plane.

Elliptical galaxies in the mass range $2\times 10^{12} - 2\times 10^{13} M_\odot$, which exhibit a significant difference in the measured alignment strength $D$, are dominated by central galaxies, as can be seen in Fig.~\ref{fig_cntsat} where this mass range is indicated by a grey band. Fig.~\ref{fig_D_aniso} further reveals massive central ellipticals to have on average lower $\kappa_\mathrm{rot}$ values than the full sample, which is reflected in the darker orange colour as compared to Fig.~\ref{fig_D_z0}.
Those most massive and therefore rarest objects form in knots and the most prominent filaments in the cosmic web. Strong intrinsic alignments have been observed for massive galaxies, between massive BCGs \citep[e.g.][]{Binggeli:1980,West:2017} or LRGs in SDSS \citep[e.g.][]{Mandelbaum:2006}, and are impacted by anisotropic mergers. Massive, merger dominated objects are thus particularly affected by the presence of large-scale anisotropies. This hints at the fact that the discrepancy between the $D_+$ and $D_\times$ values measured from the directly observed ellipticities in TNG100 stems from the $\sim 100$~Mpc simulation volume being not large enough to yield a statistically isotropic large-scale structure distribution. While this anisotropy would be much suppressed in the larger TNG300 volume, we consider the better resolved galaxy population (down to smaller masses) a larger benefit, since the intrinsic anisotropy can be easily corrected for as describe in section~\ref{sec:measurement_parameters}.

\begin{figure} 
\centering
\includegraphics[width=0.46\textwidth,trim= 0 5 8 5,clip]{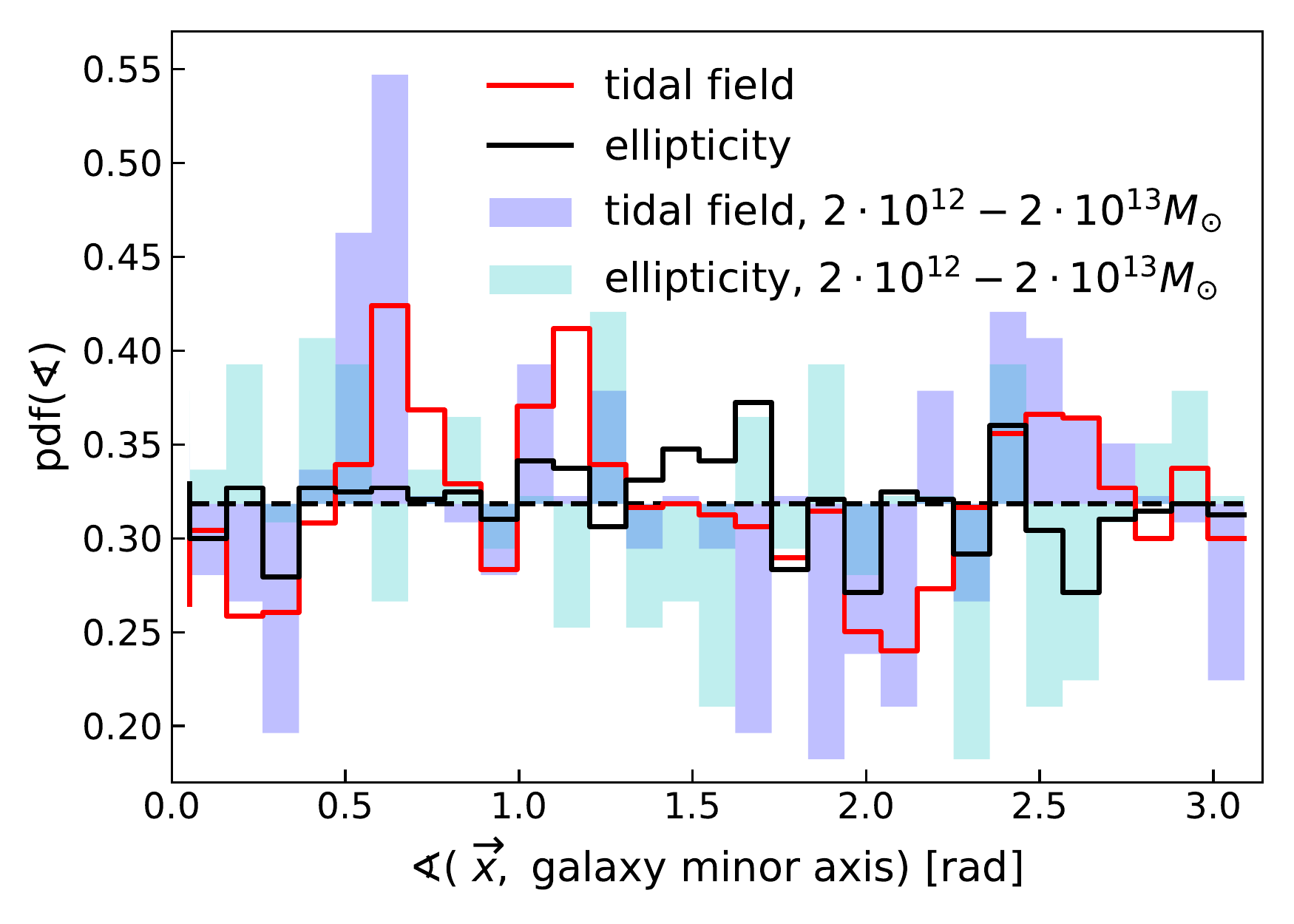}
\includegraphics[width=0.47\textwidth,trim= 12 10 0 0,clip]{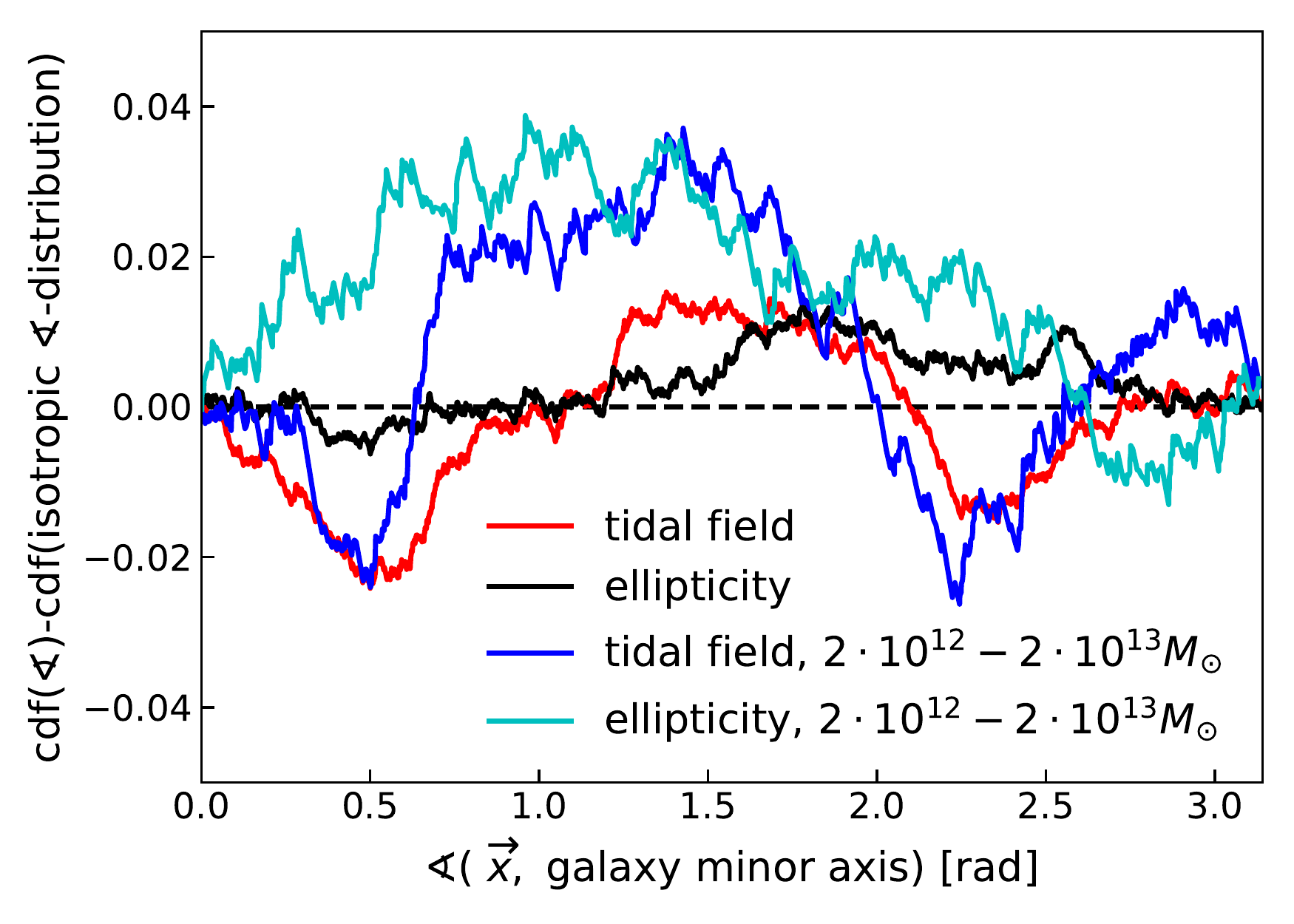}
\caption{Probability distributions (pdf, \textit{top panel}) and cumulative distribution function (cdf, \textit{bottom panel}) relative to the cdf of the uniform distribution of the angle between the strongest tidal field direction in projection, as well as the corresponding minor axis of the observed galaxy ellipticity and the $x$-axis of the simulation box, for all elliptical galaxies at $z=0$, and for the $2\times 10^{12} - 2\times 10^{13} M_\odot$ mass range, coloured relative to the expectation value of a uniform distribution (black dashed line). Preferred orientations are already ingrained in the tidal field, and are more prominent for the selected mass range. Obeserved ellipticities exhibit similar, slightly smaller deviations from the uniform distribution than the corresponding tidal fields, highlighting large-scale structure anisotropies in TNG00 caused by its limited box size which bias the direct measurement of galaxy ellipticities.}
\label{fig_tidalfield_aniso}
\end{figure}

To demonstrate that the $3\sigma$ discrepancy between the directly measured values of $D_{+}$ and $D_{\times}$ of massive ellipticals indeed originates from the anisotropy of the large-scale structure of TNG100, we calculate for each elliptical galaxy in our sample two angles indicative of their orientation in the simulation volume. We calculate the angle between the $x$-axis of the simulation box and the minor axis of the observed galaxy ellipticity, to test whether the galaxy ellipticities are distributed in a non-uniform way. According to the linear alignment model the minor axis of the observed galaxy ellipticity is set by the direction of the strongest tidal field. We thus then calculate the angle between the $x$-axis and the strongest tidal field direction in the plane perpendicular to the line-of-sight, to see whether large-scale anisotropies are already ingrained in the tidal gravitational field, and then transmit to the observed galaxy ellipticities. The direction of the strongest tidal field is given by the direction of the eigenvector corresponding to the largest eigenvalue of the two dimensional tidal field tensor $\Phi_{,ij}$ ($i,j\in\{x,y\}$, given the line-of-sight chosen along the $z$-direction). In case of the observed ellipticity, the direction of the minor axis is obtained from the eigenvector corresponding to the largest eigenvalue of the tensor $q_{ij}$ of second brightness moments.

We show the probability distribution function of the two angles both for the full sample of elliptical galaxies (solid lines), as well as only for elliptical galaxies in the relevant mass range $2\times 10^{12} - 2\times 10^{13} M_\odot$ (coloured areas) in the top panel of Fig.~\ref{fig_tidalfield_aniso}. The distribution is obtained binning the derived angles in 30 bins, and normalising to the number of elliptical galaxies in given sample as well as to the bin size. The black dashed line indicates a distribution derived from perfectly isotropically distributed angles, and the distribution for the selected mass range is coloured in such a way as to indicate its deviation from the isotropic distribution. Fig.~\ref{fig_tidalfield_aniso} reveals deviations from the isotropic distribution for all displayed quantities. It becomes apparent that anisotropies are already present in the distribution of the strongest tidal field directions and are larger than the anisotropies in the observed ellipticities for both the full elliptical galaxy sample, as well as in the selected mass range. This can be taken as evidence for tidal field anisotropies causing an anisotropic distribution of ellepticities on the projected sky. 

An alternative for seeing the same effect is through the cumulative distribution function of the two angles calculated above, displayed in the bottom panel of Fig.~\ref{fig_tidalfield_aniso} relative to the cumulative distribution of an isotropic distribution. Clearly, there is a deviation of all quantities from the isotropic prediction, with a much stronger deviation in the selected mass range. Elliptical galaxies in the above selected mass range are massive central objects of their host haloes thus preferentially tracing the large-scale filamentary structure of the cosmic web. Anisotropies in the large-scale matter distribution will therefore be most strongly reflected in properties of massive elliptical galaxies. This explains why for elliptical galaxies in the $2\times 10^{12} - 2\times 10^{13} M_\odot$ mass range the alignment strengths $D_{+}$ and $D_{\times}$, derived independently from the real and imaginary parts of the observed ellipticity, exhibit a significant discrepancy, while for the full sample the derived alignment strengths are still compatible with each other. This highlights the absolute necessity to apply correcting measures (as outlined in section~\ref{sec:measurement_parameters}) to alignment parameters derived from cosmological simulation of TNG100 volume. 

\end{appendix}

\label{lastpage}

\end{document}

%% file: journals.tex
\def\araa{ARA\&A}%
\def\apj{ApJ}%
\def\apjl{ApJ}%
\def\apjs{ApJS}%
\def\ao{Appl.~Opt.}%
\def\aap{A\&A}%
\def\mnras{MNRAS}%
\def\prd{Phys.~Rev.~D}%
\def\nat{Nature}%